\documentclass[twocolumn,prb,showpacs,floatfix]{revtex4}
\usepackage{amssymb}
\usepackage{amsmath}
\usepackage{epsfig}
\usepackage{graphics}
\usepackage[dvips]{color}

\begin{document}

\title{Shot noise and spin-orbit coherent control of entangled and spin polarized
electrons}
\author{J. Carlos \surname{Egues}$^{1,2}$}
\author{Guido Burkard$^{1}$}
\author{D. S. Saraga$^{1}$}
\author{John Schliemann$^1$}
\author{Daniel Loss$^1$}
\affiliation{$^1$Department of Physics and Astronomy, University of Basel,
Klingelbergstrasse 82, CH-4056 Basel, Switzerland}
\affiliation{$^2$Departamento de F\'{\i}sica e Inform\'{a}tica, Instituto de F\'{\i}sica
de S\~{a}o Carlos, Universidade de S\~{a}o Paulo, 13560-970 S\~{a}o Carlos, S%
\~{a}o Paulo, Brazil}

\begin{abstract}

We extend our previous work on shot noise for entangled and spin
polarized electrons in a beam-splitter geometry with spin-orbit
(\textit{s-o}) interaction in one of the incoming leads (lead 1).
Besides accounting for both the Dresselhaus and the Rashba
spin-orbit terms, we present general formulas for the shot noise
of singlet and triplets states derived within the scattering
approach. We determine the full scattering matrix of the system
for the case of leads with \textit{two} orbital channels coupled
via weak \textit{s-o} interactions inducing channel anticrossings.
We show that this interband coupling coherently transfers
electrons between the channels and gives rise to an additional
modulation angle -- dependent on both the Rashba and Dresselhaus
interaction strengths -- which allows for further independent
coherent control of the electrons traversing the incoming leads.
We derive explicit shot noise formulas for a variety of correlated
pairs (e.g., Bell states) and lead spin polarizations.
Interestingly, the singlet and \textit{each} of the triplets
defined along the quantization axis perpendicular to lead 1 (with
the local \textit{s-o} interaction) and in the plane of the beam
splitter display distinctive shot noise for injection energies
near the channel anticrossings; hence, one can tell apart all the
triplets, in addition to the singlet, through noise measurements.
We also find that spin-orbit induced backscattering within lead 1
reduces the visibility of the noise oscillations, due to the
additional partition noise in this lead. Finally, we consider
injection of two-particle wavepackets into leads with multiple
discrete states and find that two-particle entanglement can still
be observed via noise bunching and antibunching.
\end{abstract}

\date{\today }
\maketitle
\newpage

\section{Introduction}

Spin-related effects underlie promising possibilities in the
emerging field of semiconductor spintronics and spin-based quantum
computing \cite{als,rmp}. Spin-entangled electron pairs in
unconventional geometries, e.g., electron beam splitters
\cite{liu}, offer a unique setting in which to investigate
fundamental \emph{non-local} electron correlations in solids
\cite{review}. Several schemes for creating and injecting
entangled pairs in mesoscopic systems have recently been proposed
involving quantum dots, superconductors, and interference in the
electron flow
\cite{div-loss-jmmm,ble,choi-bruder-loss,recher-suk-loss,leso-mart-blat,falci-fein-hekk,
merlin,costa-bose,oliver-yam-yam,bose-home,recher-loss,bena-vish-bal-fish,saraga-loss,
bouchiat,recher-loss-prl2003,beenakker-03,samuelsson-prl03,saraga-al-loss-wes,
samuelsson-prl04, sauret-mart-fein,dup-hur}. Detection, coherent
manipulation, and transfer of spin entanglement (\textquotedblleft
flying qubits\textquotedblright ) in nanostructures are crucial
ingredients for quantum-information processing and communication.
Non-equilibrium noise, shot noise, is a useful probe for detecting
entanglement\cite{ble,taddei}.

More recently, the Rashba spin-orbit interaction present in
confined electron systems lacking structural inversion
asymmetry\cite{rashba} has been proposed as a convenient means to
spin rotate entangled pairs \cite{egd}. Interestingly, it was
found that a local Rashba spin-orbit interaction acting upon a
\begin{figure}[t!]
\begin{center}
\epsfig{file=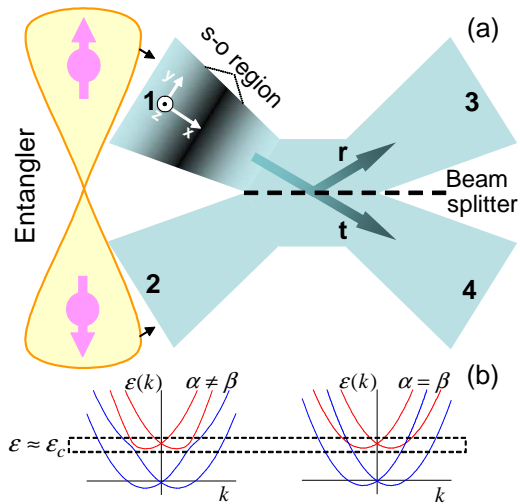, width=0.40\textwidth}
\end{center}
\vspace{-0.7cm} \caption{(a) Spin-entangled electrons injected
into a beam-splitter setup with spin-orbit interactions, Rashba
and Dresselhaus, within a finite region $L$ of lead 1. The
strength $\protect\alpha $ of the Rashba interaction can, in
principle, be controlled via a top gate so as to be equal or
unequal to the Dresselhaus coupling $\protect\beta $. For two
orbital channels in lead 1 and $\protect\alpha =\protect\beta $,
no \textit{s-o} induced band mixing occurs, right panel (b). For
$\protect\alpha \neq \protect\beta $ (or when either
$\protect\alpha =0$ or $\protect\beta =0$) the bands anti cross,
left panel (b). Only a single spin rotation $\protect\theta
_{so}=2m\protect\sqrt{
\protect\alpha ^{2}+ \protect\beta ^{2}}L/\hbar ^{2}$ is present for $%
\protect\alpha =\protect\beta $, while an additional
\textquotedblleft mixing\textquotedblright\ spin rotation
$\protect\theta _{d}$ modulates the electron transport in lead 1
for $\protect\alpha \neq \protect\beta $
and impinging energies near the crossing $\protect\varepsilon %
\approx \protect\varepsilon _{c}$. This modulation appears in the
current fluctuations (shot noise) measured in lead 3. In
particular, each of the triplets -- for a quantization axis along
the $y$ direction -- exhibits a
distinctive noise as a function of $(\protect\theta _{so},\protect\theta %
_{d})$.} \label{fg1}
\end{figure}
non-local portion of spatially-separated entangled electron pairs
injected into a beam splitter gives rise to sizable modulation of
the shot noise in the outgoing leads\cite{egd}. The use of the
Rashba interaction to controllably rotate the electron spin was
first proposed by Datta and Das \cite{datta-das}. Motivated by
this earlier proposal and its potential impact on semiconductor
spintronics, many researchers are actively investigating
spin-orbit related physics in a variety of semiconductor
nanostructures\cite{moroz-barnes,
mireles-kirc,gover-zu,streda-seba,erasmo,wang,winkler,shen,mish-brat-tse,
usaj-bal,rom-ull-tamb,kno-thom,deba-kram,yu}.

Here we extend our previous investigation on the coherent
\textit{s-o} control of entangled and spin-polarized electrons and
their shot noise for transport in a beam-splitter configuration
(Fig. 1) with local spin-orbit interactions, i.e., interactions
acting within only a finite region of one of the two
one-dimensional incoming leads\cite{egd}. We include both the
Rashba \cite{rashba} and the Dresselhaus \cite{dress} spin-orbit
terms\cite{ganichev}. Since the Rashba part of the \emph{s-o}
coupling is gate-tunable\cite{nitta}, one can controllably spin
rotate the incoming correlated spinor pairs thus changing the
degree of symmetry of the \emph{spin} part of pair wave function.
The stringent requirement of antisymmetry for fermions -- the
Pauli principle -- intrinsically links the spin and the orbital
(charge) degrees of freedom \cite{div-loss-jmmm}. Thus the
spin-orbit induced spin rotation affects the spatial charge
distribution of the pair which can be probed via
current-fluctuation measurements: charge shot noise.

We consider a beam-splitter with quasi-one dimensional incoming
leads with one and two channels. (i) For single-moded leads and
within the scattering approach we generalize our previous results
\cite{egd} by deriving general expressions for the shot noise of
singlet and triplet pairs injected into the beam splitter. We
present explicit formulas for the particular beam-splitter
scattering matrix of the experiment in Ref. \onlinecite{liu} and a
variety of incoming electron pairs: singlet and entangled and
unentangled triplet states defined along distinct quantization
axis. (ii) The case with two channels is particularly interesting
as the \emph{s-o} terms give rise to inter-channel coupling which
results in anticrossings of the bands. For incoming energies near
these avoided crossings, we find similarly to Ref.
\onlinecite{egd} an additional spin phase due to the coherent
transfer of carriers between the \emph{s-o} coupled bands. Here,
however, this modulation angle depends on both the Rashba and the
Dresselhaus coupling strengths. Interestingly, for singlet and
triplets defined along the $y$ quantization axis (Fig. 1) and
injected into only one of the two channels of the incoming leads,
we find that \emph{each} of the triplet (besides the singlet)
pairs displays distinctive noise modulations. This provides a way
of distinguishing all of these triplet pairs via noise
measurements. The interband coupling controlling the extra phase
can, in principle, be varied via independent side gates which
changes the width of the incoming channels\cite{egd}; this
provides an additional mechanism for electric spin control.
Moreover, for tuned \textit{s-o} couplings (i.e., equal strengths)
the Rashba and Dresselhaus terms partially cancel themselves out,
thus giving rise to parabolic-band crossings for arbitrary
strength of the \emph{s-o} interaction\cite{john}, Fig. 1 (b).
This allows for the propagation of electron spins protected
against non-magnetic scattering, i. e., robust entangled or
unentangled spin pairs.

We also consider spin-polarized injection\cite{spin-pol,egues}
into the beamsplitter. Here we find that noise measurement can
probe the spin-polarization of the Fermi liquid leads along
distinct quantization directions. We also discuss the effects of
backscattering in the incoming leads, due to, e.g., the potential
discontinuities at the entrance and exit of the \textit{s-o}
active region in lead 1 (see Appendix \ref{app-john} for an
explicit evaluation of the transmission coefficient for electrons
crossing a 1D lead with \textit{s-o} interaction). Backscattering
reduces the visibility of the shot noise oscillations, because of
the additional partition noise in the incoming leads. Finally, we
investigate transport of injected two-particle wavepackets into
leads with multiple discrete states but without \textit{s-o}
interaction. Similarly to our previous results \cite{ble} with
ordinary two-particle pairs (i.e., ``plane waves''), we find that
two-particle entanglement can also be detected via noise
measurements (bunching and antibunching) even with incoming
wavepackets.

This paper is organized as follows. In Sec. II we introduce the
spin-orbit Hamiltonian in 1D channels. We consider both the Rashba
and the Dresselhaus \emph{s-o} terms. \ We present exact and
approximate solutions for wires with, respectively, equal and
unequal (Rashba and Dresselhaus) \emph{s-o} coupling strengths.
The full \emph{s-o} transfer matrix for wires with one and two
(coupled) channels is also derived. The boundary conditions for
the two coupled channel case is discussed in detail. In Sec. III
we present the basics of the scattering formalism for current and
shot noise of spin-entangled electron pairs and spin-polarized
electrons. We derive general formulas for the shot noise of
singlet and triplet pairs injected into a beam splitter with an
arbitrary scattering matrix (III-B). The effect of backscattering
is also discussed (III-C) for electron pairs in single-moded
incoming leads. We present many specific formulas for the noise of
Bell pairs, electron pairs defined along distinct quantization
axes for both single- and double-moded wires. Noise for
spin-polarized injection in discussed in (III-D).  We also
consider (III-E) the injection of entangled and unentangled
wavepackets into leads with multiple energy levels. We summarize
our results and conclusions in Sec. IV. Many technical details of
our calculation are discussed in the Appendices A-E.

\section{Spin-orbit coupling in 1D channels: Rashba and Dresselhaus}

Quantum wires can be defined from two-dimensional electron gases
by further constraining the electron motion to one spatial
direction via, for instance, gate electrodes. When the underlying
2DEG has spin-orbit interactions of the Dresselhaus\cite{dress}
and Rashba\cite{rashba} types, due to bulk inversion asymmetry
(BIA) and structural inversion asymmetry (SIA), respectively, the
1D channel so formed will also present such interaction terms
\cite{Hau04}. The Hamiltonian of a 2DEG with
spin orbit interaction and an additional gate-induced confining potential $%
V(y)$ reads
\begin{eqnarray}
H &=&-\frac{\hbar ^{2}}{2m}\left( \frac{\partial ^{2}}{\partial x^{2}}+\frac{%
\partial ^{2}}{\partial y^{2}}\right) +V(y)+  \notag \\
&&i\alpha \left( \sigma _{y}\partial _{x}-\sigma _{x}\partial _{y}\right)
+i\beta \left( \sigma _{y}\partial _{y}-\sigma _{x}\partial _{x}\right) ,
\label{eq1}
\end{eqnarray}%
where\ $\partial _{i}\equiv \partial /\partial i$, $i=x,y$ and the
third and fourth terms are the usual Rashba (strength $\alpha $)
and the linearized Dresselhaus (strength $\beta $) \textit{s-o}
terms, respectively.

\subsection{Exact solution: $\protect\alpha =\protect\beta $ case} \label{exactsol}

Similarly to the two-dimensional case treated in Ref.
\onlinecite{john}, the \textit{s-o} wire problem here is exactly
solvable for tuned couplings $|\alpha |=\beta $. Let us first
consider the general case of a two-dimensional electron gas with
an arbitrary scalar potential $V(\vec{r})$ which can, e.g,
describe static non-magnetic impurities, or further confinements
creating a quantum wire or a quantum dot. At the symmetry points
$\alpha =\pm \beta $ the operator $\Sigma =(\sigma ^{x}\mp \sigma
^{y})/\sqrt{2}$
provides an additional conserved quantity, and a general eigenstate of $%
\mathcal{H}$ and $\Sigma $ reads (for $\alpha =-\beta $)
\begin{equation}
\psi _{\pm }(\vec{r})=\frac{1}{\sqrt{2}}\left(
\begin{array}{c}
1 \\
\pm e^{i\pi /4}%
\end{array}%
\right) \varphi (\vec{r})e^{\mp i\sqrt{2}\alpha m(x-y)/\hbar ^{2}}
\label{eigenstate}
\end{equation}%
in the $\sigma_z$ basis.
The function $\varphi (\vec{r})$ fulfills the usual spin-independent
Schr\"{o}dinger equation
\begin{equation}
\left( -\frac{\hbar ^{2}}{2m}\nabla ^{2}+V(\vec{r})\right) \varphi (\vec{r}%
)=\left( \varepsilon +\frac{2\alpha ^{2}m}{\hbar ^{2}}\right) \varphi (\vec{r%
}),  \label{schroedinger}
\end{equation}%
and $\varepsilon $ is the energy eigenvalue of the wave function $\psi _{\pm
}(\vec{r})$ with $\Sigma =\pm 1$. Now consider a quantum wire
along the $x$-direction, i.e. $V(\vec{r})=V(y)$. At $\alpha =-\beta $ the
wave functions are of the form (\ref{eigenstate}) with $\varphi _{n}(\vec{r}%
)=\phi_{n}(y)\exp (i(k\pm \sqrt{2}\alpha m/\hbar ^{2})x)/\sqrt{L_x}$, $L_x$ is
a normalizing length, such that the full wave function reads
\begin{equation}
\psi _{n,\pm }(\vec{r})=\frac{1}{\sqrt{2}}\left(
\begin{array}{c}
1 \\
\pm e^{i\pi /4}%
\end{array}%
\right) \frac{e^{ikx}}{\sqrt{L_x}}\phi_{n}(y)e^{\pm i\sqrt{2}\alpha my/\hbar
^{2}},  \label{generalwire}
\end{equation}%
where $\phi_{n}(y)$ obeys the usual Schr\"{o}dinger equation for
the transverse variable $y$ with quantized eigenvalues
$\tilde{\varepsilon}_{n}$ . The eigenstates (\ref{generalwire})
are characterized by the subband index $ n $ and the wave number
$k$, and the corresponding eigenenergies are given by $\varepsilon
_{n}^{\pm }(k)=\tilde{\varepsilon}_{n}+(\hbar ^{2}/2m)(k\pm
\sqrt{2}\alpha m/\hbar ^{2})^{2}-2\alpha ^{2}m/\hbar ^{2}$. Note
that, similarly to the two-dimensional case discussed earlier
\cite{john}, the wire energy dispersions here are also parabolic
-- for \textit{any} strength of the $|\alpha |=\beta $ coupling, see \
Fig. 1(b).

\subsection{Approximate solutions: $\protect\alpha \neq \protect\beta $ case}

For unequal couplings we first solve the quantum wire problem in
the absence of spin orbit coupling and then use this solution as a basis to
write down the Hamiltonian matrix with the \textit{s-o} terms.
Here we neglect any additional \emph{ s-o} terms arising from the
further confinement\cite{moroz-barnes} $V(y)$.

\subsubsection{Quantum wire eigenstates}

The solution to Eq. (\ref{eq1}) without the \textit{s-o} terms is%
\begin{equation}
\varphi _{k,n,\sigma _{z}}(x,y)=\frac{e^{ikx}}{\sqrt{L_x}}\phi _{n}(y)|\sigma
_{z}\rangle ,  \label{eq2}
\end{equation}%
where $|\sigma _{z}\rangle \in \{|\uparrow \rangle _{z}\,
|\downarrow \rangle _{z}\}$ is the electron spin state in the
$\sigma _{z}$ basis, with
eigenvalues%
\begin{equation}
\varepsilon _{k,n,\sigma _{z}}=\frac{\hbar ^{2}k^{2}}{2m}+\epsilon _{n},%
\text{ }  \label{eq3}
\end{equation}%
and $n=a,b...$ denoting the transverse modes with energies
$\epsilon _{n}$ (note that $\tilde{\varepsilon}_{n} = \epsilon
_{n}$ in the absence of \textit{s-o}). The transverse confining
eigenfunctions $\phi _{n}(y)$ obey the 1D Sch\"odinger equation%
\begin{equation}
-\frac{\hbar ^{2}}{2m}\frac{d^{2}\phi _{n}(y)}{dy^{2}}+V(y)\phi
_{n}(y)=\epsilon _{n}\phi _{n}(y).  \label{eq4}
\end{equation}%
The confining potential in Eq. (\ref{eq4}) is arbitrary. Later on
we consider an explicit form (obtained for hard-wall confinement)
so as to obtain simple estimates.

\subsubsection{Rashba-Dresselhaus wire}

We\ can derive a reduced Hamiltonian for our quantum wire with \emph{s-o} by
expanding the solution of Eq. (\ref{eq1}) in the basis of the wire without
\emph{s-o},
$\left\{ \varphi _{k,a,\uparrow},\varphi _{k,a,\downarrow},
\varphi _{k,b,\uparrow},\varphi _{k,b,\downarrow}\right\}$.
Here we
consider only two wire modes. We then find
\begin{widetext}%
\begin{equation}
H=\left[
\begin{array}{cccc}
\frac{\hbar ^{2}k^{2}}{2m}+\epsilon _{a} & \left( i\alpha +\beta \right) k &
0 & \left( -i\alpha +\beta \right) d_{ab} \\
\left( -i\alpha +\beta \right) k & \frac{\hbar ^{2}k^{2}}{2m}+\epsilon _{a}
& \left( -i\alpha -\beta \right) d_{ab} & 0 \\
0 & \left( i\alpha -\beta \right) d^{\ast}_{ab} & \frac{\hbar ^{2}k^{2}}{2m}%
+\epsilon _{b} & \left( i\alpha +\beta \right) k \\
\left( i\alpha +\beta \right) d^{\ast}_{ab} & 0 & \left( -i\alpha
+\beta \right) k
& \frac{\hbar ^{2}k^{2}}{2m}+\epsilon _{b}%
\end{array}%
\right].  \label{eq5}
\end{equation}%
\end{widetext}$\allowbreak $The matrix element
\begin{equation}
d_{ab}=-d^{\ast}_{ba}\equiv \langle \phi _{a}|\partial /\partial
y|\phi _{b}\rangle \label{eq5a}
\end{equation}%
in Eq. (\ref{eq5}) defines the \emph{s-o} induced interband mixing
between the wire modes arising from the \textit{s-o} terms
proportional to $p_y$ in Eq. (\ref{eq1}). For hard-wall
confinement $d_{ab}=8/3w$, where $w$ is the wire width. It is
convenient to rewrite the above matrix in the basis of the
eigenstates corresponding to $d_{ab}=0.$ For null interband
coupling the
Hamiltonian decouples into two sets of \emph{s-o} bands%
\begin{equation}
\varepsilon _{n}^{s}(k)=\frac{\hbar ^{2}k^{2}}{2m}+\epsilon _{n}-sk\sqrt{%
\alpha ^{2}+\beta ^{2}},\text{ }  \label{eq6}
\end{equation}%
where $n=a,b$ and $s=\pm $, and eigenvectors%
\begin{equation}
\varphi _{k,n,s}(x,y)=\frac{e^{ikx}}{\sqrt{L_x}}\phi _{n}(y)|s\rangle ,
\label{eq7}
\end{equation}%
with
\begin{equation}
|s\rangle =\frac{1}{\sqrt{2}}\left(
\begin{array}{c}
1 \\
-s\xi%
\end{array}%
\right) =\frac{1}{\sqrt{2}}\left( |\uparrow \rangle _{z}-s\xi
|\downarrow \rangle _{z}\right) ,  \label{eq8}
\end{equation}%
where
\begin{equation}
\xi =\sqrt{\alpha ^{2}+\beta ^{2}}/\left( i\alpha +\beta \right).
\label{eq-gamma}
\end{equation}
We define the transformed Hamiltonian matrix as $\bar{H}=U^{\dag
}HU$, with
\begin{equation}
U=\frac{1}{\sqrt{2}}\left(
\begin{array}{cccc}
1 & 1 & 0 & 0 \\
-\xi & \xi & 0 & 0 \\
0 & 0 & 1 & 1 \\
0 & 0 & -\xi & \xi
\end{array}%
\right) .  \label{eq9}
\end{equation}%
We find
\begin{widetext}%
\begin{equation}
\bar H=\left(
\begin{array}{cccc}
\varepsilon _{a}^{+} & 0 & 2id_{ab}\frac{\alpha \beta }{\sqrt{\alpha
^{2}+\beta ^{2}}} & -d_{ab}\frac{\alpha ^{2}-\beta ^{2}}{\sqrt{\alpha
^{2}+\beta ^{2}}} \\
0 & \varepsilon _{a}^{-} & d_{ab}\frac{\alpha ^{2}-\beta ^{2}}{\sqrt{\alpha
^{2}+\beta ^{2}}} & -2id_{ab}\frac{\alpha \beta }{\sqrt{\alpha ^{2}+\beta
^{2}}} \\
-2id^{\ast}_{ab}\frac{\alpha \beta }{\sqrt{\alpha ^{2}+\beta
^{2}}} & \allowbreak d^{\ast}_{ab}\frac{\alpha ^{2}-\beta
^{2}}{\sqrt{\alpha ^{2}+\beta ^{2}}} &
\varepsilon _{b}^{+} & 0 \\
-d^{\ast}_{ab}\frac{\alpha ^{2}-\beta ^{2}}{\sqrt{\alpha
^{2}+\beta ^{2}}} & 2id^{\ast}_{ab}\frac{\alpha \beta
}{\sqrt{\alpha ^{2}+\beta ^{2}}} & 0 &
\varepsilon _{b}^{-}%
\end{array}%
\right).  \label{eq10}
\end{equation}%
\end{widetext}The diagonalization of (\ref{eq10}) is straightforward; the
eigenenergies are
\begin{widetext}%
\begin{eqnarray}
\varepsilon _{s,s^{\prime }}(k) &=&\frac{\hbar ^{2}k^{2}}{2m}+\frac{1}{2}%
\left( \epsilon _{b}+\epsilon _{a}\right) +  \notag \\
&&s\frac{1}{2}\sqrt{\left( \epsilon _{b}-\epsilon _{a}\right) ^{2}+4\left(
|d_{ab}|^{2}+k^{2}\right) \left( \alpha ^{2}+\beta ^{2}\right) +s^{\prime }4k%
\sqrt{\left( \alpha ^{2}+\beta ^{2}\right) \left( \epsilon
_{b}-\epsilon _{a}\right) ^{2}+16|d_{ab}|^{2}\alpha ^{2}\beta
^{2}}}  \label{eq11}
\end{eqnarray}%
\end{widetext}where $s,s^{\prime }=\pm $ $.$ The corresponding
eigenfunctions are too lengthy to be shown here. Figure 1(b) shows
the above energy dispersions for $\alpha \neq \beta $ and $\alpha
=\beta $ for nonzero interband coupling $d_{ab}$. In general, the
energy dispersions present avoided crossings for $\alpha \neq
\beta$. In contrast, the \emph{s-o} tuned $\alpha =\beta $ case
has eigenvalues which are quadratic in $k$ with \emph{no} avoided
crossings.
This $k$ dependence is easily seen by setting $\alpha =\beta $ in Eq. (\ref{eq11}%
)

\begin{eqnarray}
\varepsilon _{s,s^{\prime }}(k) &=&\frac{\hbar ^{2}k^{2}}{2m}+\frac{1}{2}%
\left( \epsilon _{b}+\epsilon _{a}\right) +s\sqrt{2}k\alpha +  \notag \\
&&s^{\prime }\frac{1}{2}\sqrt{(\epsilon _{b}-\epsilon
_{a})^{2}+8\alpha ^{2}|d_{ab}|^{2}}.  \label{eq12}
\end{eqnarray}

In what follows we discuss in more detail the cases $d_{ab}=0$ and $%
d_{ab}\neq 0$ corresponding to the uncoupled and interband-coupled
channels, respectively. We emphasize\ again that the interband
coupling described by the matrix element $d_{ab}$ is purely
induced by the \emph{s-o}. As we will see below, the uncoupled
case gives rise to a single spin-rotation modulation. The
interband coupled case, on the other hand, will have two
independent modulation angles for injected electrons with energies
near the band crossings.

\subsection{Uncoupled 1D channels ($d_{ab}=0$): single spin rotation $%
\protect\theta _{R}$}

Here we have in mind a two-terminal geometry with the source and drain
connected by a Rashba-Dresselhaus wire. For simplicity, we neglect the band
offsets between the various interfaces. That is, we assume a unity
transmission through the \emph{s-o} region\cite{mismatch}. Finite offsets
give rise to\ Fabry-Perot type oscillations which further modulate the
transport properties\cite{mireles-kirc} of the system. The uncoupled case ($%
d_{ab}=0$) considered here should be a good approximation also for finite $%
d_{ab}$, provided that $\alpha |d_{ab}|$ be much smaller than the
interband energy separation ($\alpha |d_{ab}|\ll \epsilon
_{b}-\epsilon _{a}$). The solution for $d_{ab}=0$ is
straightforward (see Ref. \onlinecite{mireles-kirc}
for the case where only the Rashba coupling is active). From Eq. (\ref{eq10}%
), which is diagonal for $d_{ab}=0$, we immediately obtain the two sets of
\emph{s-o} bands [Eq. (\ref{eq6})] which we rewrite as
\begin{equation}
\varepsilon _{a,b}^{(s)}(k)=\frac{\hbar ^{2}}{2m}\left(
k-sk_{so}\right) ^{2}+\epsilon _{a,b}-\frac{\hbar
^{2}k_{so}^{2}}{2m} \text{, \ }s=\pm ,  \label{eq13}
\end{equation}
where
\begin{equation}
k_{so}\equiv m\sqrt{\alpha ^{2}+ \beta^{2}}/\hbar ^{2}
\label{eq-kso}
\end{equation}
 is the \textit{s-o} wave vector. The
corresponding eigenvectors are given in Eqs. (\ref{eq7}) and
(\ref{eq8}). For \ $d_{ab}=0$ the \emph{s-o} bands cross at
\begin{equation}
k_{c}=\frac{\epsilon _{b}-\epsilon _{a}}{2\sqrt{\alpha ^{2}+\beta ^{2}}},
\label{eq14}
\end{equation}%
which is obtained by setting $\varepsilon _{a}^{-}(k_{c})=\varepsilon
_{b}^{+}(k_{c})$ (see thin solid line in the inset of Fig. 2); a symmetric
crossing also occurs for at $k=-k_{c}$.

As first pointed out by Datta and Das\cite{datta-das}, injected
electrons moving down the 1D channel will spin precess due to the
action of the \textit{s-o} interaction. Here the spin rotation is
due to the combined effects of the Rashba and Dresselhaus terms.
In analogy to the case discussed by Datta and Das, here we find
that a spin-up electron, say in channel \emph{a}, crossing the
length $L$ of the \emph{s-o} active region will emerge in the
state

\begin{equation}
|\uparrow \rangle _{z}\rightarrow \cos (\theta _{so}/2)|\uparrow \rangle
_{z}-\sin (\theta _{so}/2)|\downarrow \rangle _{z}\text{,}  \label{eq15}
\end{equation}%
where
\begin{equation}
\theta _{so}=2m\sqrt{\alpha ^{2}+\beta ^{2}}L/\hbar ^{2}
\label{eq-theta-so}
\end{equation}
is the spin rotation angle about the $y$ axis. Similarly, a spin
down electron evolves into
\begin{equation}
|\downarrow \rangle _{z}\rightarrow \sin (\theta _{so}/2)|\uparrow \rangle
_{z}+\cos (\theta _{so}/2)|\downarrow \rangle _{z}\text{.}  \label{eq16}
\end{equation}%
The same reasoning applies to impinging electrons in channel \emph{b}.
Hence, we can described the \emph{s-o} region in the absence of \emph{s-o}
induced channel coupling (uncoupled channels) by the\ 4x4 \textquotedblleft
transfer\textquotedblright\ matrix $\mathbf{U}_{so}^{u}$%
\begin{equation}
\mathbf{U}_{so}^{u}=\left(
\begin{array}{cc}
\mathbf{U}_{so}^{a} & 0 \\
0 & \mathbf{U}_{so}^{b}%
\end{array}%
\right) ,  \label{eq16a}
\end{equation}%
where%
\begin{equation}
\mathbf{U}_{so}^{a}=\mathbf{U}_{so}^{b}=\left(
\begin{array}{cc}
\cos (\theta _{so}/2) & \sin (\theta _{so}/2) \\
-\sin (\theta _{so}/2) & \cos (\theta _{so}/2)%
\end{array}%
\right) ,  \label{eq17}
\end{equation}%
defines the single-channel transfer matrix for the uncoupled channels \emph{a%
} and \emph{b}. Later on we introduce the scattering matrix approach to
calculate current and noise in a beam-splitter geometry. The \emph{s-o}
rotation matrix above (and its generalization for two channels) will prove
very convenient in accounting for \emph{s-o} effects on the transport
properties of the beam-splitter within the scattering approach. Note that
only the Rashba coupling constant appearing in the rotation angle $\theta
_{so}$ can be varied externally via a gate electrode, while the Dresselhaus
coupling $\beta $ is a material property. As a final point, we note that the
above \emph{s-o} rotated states satisfy the proper boundary conditions for
the wavefunction at $x=0$ and $x=L$. This is discussed in some detail in
appendix A for both the one- and two-channel cases.

\subsection{Coupled 1D channels ($d_{ab}\neq 0$): additional spin rotation $%
\protect\theta _{d}$ for $\protect\alpha \neq \protect\beta $}

For nonzero \emph{s-o} induced interband coupling $d_{ab}$, the subbands
anticross for distinct coupling strengths $\alpha \neq \beta $. Similarly to
the one-channel case, here we also have to find out how incoming spin up (or
down) electrons emerge after traversing the \emph{s-o} active region of
length $L$. Here we have in mind incoming electrons with energies near the $%
d_{ab}=0$ crossing of the bands at $k_{c}$, i.e., $\varepsilon
\sim \varepsilon _{a}^{-}(k_{c})=\varepsilon _{b}^{+}(k_{c})$.
This is the relevant energy range where \emph{s-o} induced
interband crossing should play a role (unless $\alpha=\beta $). In
what follows we present a simple analysis of this injection
problem by using a perturbative approach (\textquotedblleft near
free electron model\textquotedblright\ \cite{am}) to describe the
\emph{s-o} states near the crossings.

For injection energies near the $d_{ab}=0$ crossings, we can approximate the
Hamiltonian in Eq. (\ref{eq10}) by
\begin{equation}
H_{app}=\left(
\begin{array}{cccc}
\varepsilon _{a}^{-} & 0 & 0 & 0 \\
0 & \varepsilon _{a}^{+} & \allowbreak d_{ab}\frac{\alpha ^{2}-\beta ^{2}}{%
\sqrt{\alpha ^{2}+\beta ^{2}}} & 0 \\
0 & \allowbreak d^\ast_{ab}\frac{\alpha ^{2}-\beta
^{2}}{\sqrt{\alpha ^{2}+\beta
^{2}}} & \varepsilon _{b}^{-} & 0 \\
0 & 0 & 0 & \varepsilon _{b}^{+}%
\end{array}%
\right)  \label{eq20}
\end{equation}%
i.e., we drop all the off-diagonal matrix elements except those
directly coupling the states near the crossing. From the form of
$H_{app}$ it is obvious that the crossing states [middle block of
Eq. (\ref{eq20})] will split due to the $d_{ab}$ coupling. The new
eigenvalues are
\begin{equation}
\varepsilon _{\pm }(k)=\frac{\hbar ^{2}k^{2}}{2m}+\frac{1}{2}\left( \epsilon
_{b}+\epsilon _{a}\right) \pm |d_{ab}|\frac{\alpha ^{2}-\beta ^{2}}{\sqrt{%
\alpha ^{2}+\beta ^{2}}}\sqrt{1+x}  \label{eq21}
\end{equation}%
where
\begin{equation}
x=\frac{\left[ (\epsilon _{b}-\epsilon _{a})-2\sqrt{\alpha ^{2}+\beta ^{2}}k%
\right] ^{2}}{4\left( |d_{ab}|\frac{\alpha ^{2}-\beta
^{2}}{\sqrt{\alpha ^{2}+\beta ^{2}}}\right) ^{2}}  \label{eq22}
\end{equation}%
can be viewed as an expansion parameter near $k_{c}$ [Eq. (\ref{eq14})].
Expanding $\varepsilon _{\pm }(k)$ near $k_{c}$ (we should keep only the
lowest order in $x$ since the third term of Eq. (\ref{eq21}) is already
proportional to $d_{ab}$), we find to zeroth order in $x$%
\begin{equation}
\varepsilon _{\pm }(k)=\frac{\hbar ^{2}k^{2}}{2m}+\frac{1}{2}\left( \epsilon
_{b}+\epsilon _{a}\right) \pm |d_{ab}|\frac{\alpha ^{2}-\beta ^{2}}{\sqrt{%
\alpha ^{2}+\beta ^{2}}}.\   \label{eq23}
\end{equation}%
The corresponding eigenvectors are
\begin{equation}
|\psi _{\pm }\rangle =\frac{1}{\sqrt{2}}\left( |-\rangle _{a}\pm |+\rangle
_{b}\right) ,  \label{eq24}
\end{equation}%
where $|-\rangle _{a}\rightarrow $ $\varphi _{k,a,-}(x,y)$ and $|+\rangle
_{b}$ $\rightarrow \varphi _{k,b,+}(x,y)$ are the eigenstates in Eq. (\ref%
{eq7}). The new eigenstates $|\psi _{\pm }\rangle $ are zeroth-order linear
combinations of the crossing states (remember that the energies are linear
in $\alpha d_{ab}$). More explicitly,
\begin{equation}
|\psi _{\pm }\rangle =\left[ \frac{1}{2}\left(
\begin{array}{c}
1 \\
\xi%
\end{array}%
\right) \phi _{a}(y)\pm \frac{1}{2}\left(
\begin{array}{c}
1 \\
-\xi
\end{array}%
\right) \phi _{b}(y)\right] \frac{e^{ikx}}{\sqrt{L_x}}.  \label{eq25}
\end{equation}%
In a \textquotedblleft four-vector notation\textquotedblright\ we can write%
\begin{equation}
|\psi _{\pm }\rangle =\frac{1}{2}\left(
\begin{array}{c}
1 \\
\xi \\
\pm 1 \\
\mp \xi%
\end{array}%
\right) \frac{e^{ikx}}{\sqrt{L_x}}.  \label{eq26}
\end{equation}%
As Fig. 2 clearly shows, $\varepsilon _{\pm }(k)$ [Eq. (\ref{eq23})]
approximate well the exact energy dispersions $\varepsilon _{s,s^{\prime
}}(k)$ [Eq. \ref{eq11}] of the problem near $k_{c}$. By using Eq. (\ref{eq23}%
) we can analytically determine the wave vectors $k_{c1}$ and $k_{c2}$
relevant for the spin injection problem. This is easily done by imposing $%
\varepsilon _{F}=\varepsilon _{+}(k_{c1})=\varepsilon _{-}(k_{c2})$ which
yields%
\begin{equation}
\frac{\hbar ^{2}k_{c2}^{2}}{2m}-\frac{\hbar ^{2}k_{c1}^{2}}{2m}=2|d_{ab}|\frac{%
\alpha ^{2}-\beta ^{2}}{\sqrt{\alpha ^{2}+\beta ^{2}}}.  \label{eq27}
\end{equation}%
For small \textit{s-o} induced interband coupling we look for
symmetric solutions
around $k_{c}$ [Eq. (\ref{eq14})]: $k_{c1}=k_{c}-\Delta /2$ and $%
k_{c2}=k_{c}+\Delta /2$. Equation (\ref{eq27}) then gives%
\begin{equation}
\Delta =\frac{2m|d_{ab}|}{k_{c}\hbar ^{2}}\frac{\alpha ^{2}-\beta ^{2}}{\sqrt{%
\alpha ^{2}+\beta ^{2}}}.  \label{eq28}
\end{equation}%
Having determined the wave vectors $k_{c1}$ and $k_{c2}$, we can
now solve the injection problem. The idea is to expand the
incoming electron state, say spin up in channel \emph{a}, in terms
of the eigenstates of the \textit{s-o} region.\ The expansion has
to satisfy the boundary conditions (continuity of the wavefunction
and flux conservation) at both the entrance and the exit of the
\emph{s-o} region.
\begin{figure}[th]
\begin{center}
\epsfig{file=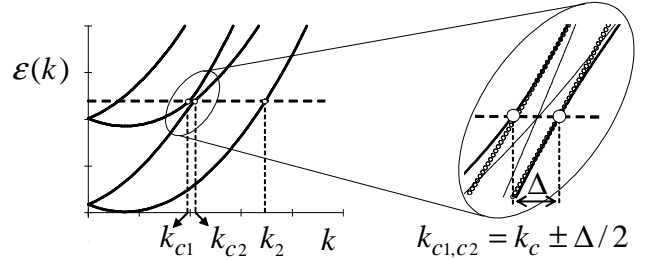, width=0.55\textwidth}
\end{center}
\par
\vspace{-0.7cm}
\caption{Schematic of the quantum wire energy dispersions $\protect%
\varepsilon _{s,s^{\prime }}(k)$ (Eq. \protect\ref{eq11}) for $\protect%
\alpha \neq \protect\beta $. The blowup shows the band anticrossing for $%
d\neq 0$ in more detail. The crossing thin solid lines represent the
uncoupled case, $d_{ab}=0$. The curves with circles are obtained from Eq. (%
\protect\ref{eq23}) ($\protect\varepsilon _{\pm }(k)$) and are good
approximation for the actual dispersions near crossing point $k_{c}^{0}$.
The wave-vectors $k_{c1}$, $k_{c2}$, and $k_{2}$, used to expand an incoming
plane wave within the \textit{s-o} region [Eq. (\protect\ref{eq29})], are
also shown in the inset. }
\label{fig22}
\end{figure}

\subsubsection{Boundary conditions}

Here we show that spin injection with energies near the band anticrossing is
possible in our system, provided that the \emph{s-o} interband coupling be
small compared to the Fermi energy. Details are given in Appendix A.

\paragraph{Continuity of the wave function.}

A spin-up electron in channel $a$ entering the \emph{s-o} region at $x=0$
with an energy $\varepsilon _{F}\sim \varepsilon _{+}(k_{c1})=\varepsilon
_{-}(k_{c2})$ has to satisfy
\begin{widetext}%
\begin{equation}
\left(
\begin{array}{c}
1 \\
0 \\
0 \\
0%
\end{array}%
\right) e^{ikx}|_{_{x\rightarrow 0^{-}}}=\left\{ \frac{1}{4}\left(
\begin{array}{c}
1 \\
\xi \\
1 \\
-\xi
\end{array}%
\right) e^{ik_{c1}x}+\frac{1}{4}\left(
\begin{array}{c}
1 \\
\xi  \\
-1 \\
\xi
\end{array}%
\right) e^{ik_{c2}x}+\frac{1}{2}\left(
\begin{array}{c}
1 \\
-\xi  \\
0 \\
0%
\end{array}%
\right) e^{ik_{2}x}\right\} _{x\rightarrow 0^{+}}.  \label{eq29}
\end{equation}%
\end{widetext}The above condition is clearly fulfilled; a similar condition
holds at $x=L$ (see Appendix A).

\paragraph{Continuity of the current flow.}

The continuity of the (non-diagonal) velocity operator\cite{bc}
acting on the wave function at $x=0$ which assures current
conservation yields
\begin{widetext}%
\begin{eqnarray}
\left(
\begin{array}{c}
\frac{\hbar k_{F}}{m} \\
0 \\
0 \\
0%
\end{array}%
\right) e^{ikx}|_{x\rightarrow 0^{-}} &=&\left\{ \frac{1}{4}\left(
\begin{array}{c}
\frac{\hbar }{m}\left( k_{c}-\Delta /2+k_{so}\right)  \\
\xi \frac{\hbar }{m}\left( k_{c}-\Delta /2+k_{so}\right)  \\
\frac{\hbar }{m}\left( k_{c}-\Delta /2-k_{so}\right)  \\
-\xi \frac{\hbar }{m}\left( k_{c}-\Delta /2-k_{so}\right)
\end{array}%
\right) e^{ik_{c1}x}+\frac{1}{4}\left(
\begin{array}{c}
\frac{\hbar }{m}\left( k_{c}+\Delta /2+k_{so}\right)  \\
\xi \frac{\hbar }{m}\left( k_{c}+\Delta /2+k_{so}\right)  \\
-\frac{\hbar }{m}\left( k_{c}+\Delta /2-k_{so}\right)  \\
\xi \frac{\hbar }{m}\left( k_{c}+\Delta /2-k_{so}\right)
\end{array}%
\right) e^{ik_{c2}x}\right. +  \notag \\
&&\left. \frac{1}{2}\left(
\begin{array}{c}
1 \\
-\xi \\
0 \\
0%
\end{array}%
\right) \frac{\hbar }{m}\left( k_{2}-k_{so}\right) e^{ik_{2}x}\right\}
_{x\rightarrow 0^{+}}.  \label{eq30}
\end{eqnarray}%
\end{widetext}which simplifies to
\begin{widetext}%
\begin{equation}
\left(
\begin{array}{c}
\frac{\hbar k_{F}}{m} \\
0 \\
0 \\
0%
\end{array}%
\right) =\frac{1}{2}\left(
\begin{array}{c}
\frac{\hbar }{m}\left( k_{c}+k_{2}\right)  \\
\xi \frac{\hbar }{m}\left( k_{c}-k_{2}-2k_{so}\right)  \\
-\frac{\hbar }{m}\Delta /2 \\
\xi \frac{\hbar }{m}\Delta /2%
\end{array}%
\right) =\frac{\hbar k_{F}}{m}\left(
\begin{array}{c}
\frac{\left( k_{c}+k_{2}\right) }{2k_{F}} \\
0 \\
-\frac{\Delta }{4k_{F}} \\
\xi \frac{\Delta }{4k_{F}}%
\end{array}%
\right) ,  \label{eq32}
\end{equation}%
\end{widetext}where we have used $k_{2}-k_{c}=2k_{so}$ [Eq. (\ref{eq13})].
From\ Eq. (\ref{eq32}) we see that the matching of the derivative is
fulfilled provided that $\Delta \ll 4k_{F}$. As we show later on, this is
the case for realistic parameters. The velocity operator matching at $x=L$
holds similarly (see Appendix A).

\subsubsection{General spin-rotated state at $x=L$}

After traversing the \emph{s-o} region a, say, spin-up electron in channel
\emph{a }is described by the state
\begin{widetext}%
\begin{equation}
\Psi _{\uparrow ,a}=\frac{1}{4}\left(
\begin{array}{c}
1 \\
\xi  \\
1 \\
-\xi
\end{array}%
\right) e^{ik_{c1}L}+\frac{1}{4}\left(
\begin{array}{c}
1 \\
\xi  \\
-1 \\
\xi
\end{array}%
\right) e^{ik_{c2}L}+\frac{1}{2}\left(
\begin{array}{c}
1 \\
-\xi  \\
0 \\
0%
\end{array}%
\right) e^{ik_{2}L}.  \label{eq33}
\end{equation}

\end{widetext}Straightforward manipulations lead to
\begin{equation}
\Psi _{\uparrow ,a}=\frac{1}{2}e^{i\left( k_{c}+k_{so}\right) L}\left(
\begin{array}{c}
\cos \left( \theta _{d}/2\right) e^{-i\theta _{so}/2}+e^{i\theta _{so}/2} \\
\xi \left[ \cos \left( \theta _{d}/2\right) e^{-i\theta
_{so}/2}-e^{i\theta _{so}/2}\right] \\
-i\sin \left( \theta _{d}/2\right) e^{-i\theta _{so}/2} \\
i\xi \sin \left( \theta _{d}/2\right) e^{-i\theta _{so}/2}%
\end{array}%
\right) .  \label{eq34}
\end{equation}%
where we have introduced the additional modulation angle
\begin{equation}
\theta _{d}=\Delta L=\left( |d_{ab}|/k_{c}\right) \theta
_{so}(\alpha ^{2}-\beta ^{2})/(\alpha ^{2}+\beta ^{2})
\label{eq-theta_d}
\end{equation}
due to \emph{s-o} induced interband mixing. We show in the
Appendix A that for $x\geqslant L$, the state%
\begin{widetext}%
\begin{equation}
\Psi (x,y)=\left(
\begin{array}{c}
\cos \left( \theta _{d}/2\right) e^{-i\theta _{so}/2}+e^{i\theta _{so}/2} \\
\xi\left[ \cos \left( \theta _{d}/2\right) e^{-i\theta
_{so}/2}-e^{i\theta _{so}/2}\right]
\end{array}%
\right) \frac{1}{2}e^{i\left( k_{c}+k_{so}\right) x}\phi _{a}(y)+\left(
\begin{array}{c}
-i\sin \left( \theta _{d}/2\right) e^{-i\theta _{so}/2} \\
i\xi \sin \left( \theta _{d}/2\right) e^{-i\theta _{so}/2}%
\end{array}%
\right) \frac{1}{2}e^{i\left( k_{c}-k_{so}\right) x}\phi _{b}(y),
\label{eq35}
\end{equation}%
\end{widetext}satisfies the proper boundary condition for the velocity
operator [note that setting $x=L$ in (\ref{eq35}) gives (\ref{eq34}), thus
fulfilling the continuity of the wave function at this interface]. Equation (%
\ref{eq35}) shows that upon traversing the \emph{s-o} active
region of length $L,$ a spin-up electron in the incoming channel
\emph{a}, acquires a spin-down component in the same channel and,
more importantly, coherently transfers into channel \emph{b}. This
coherent transfer from channel \emph{a} to channel \emph{b} is
solely due to the \emph{s-o} induced interband coupling near
$k_{c}$, described by the mixing angle $\theta _{d}$. Hence, a
\emph{weak} \emph{s-o} induced interchannel mixing -- rather than
being detrimental to transport -- offers a unique possibility for
further spin modulating the electron flow.

\subsubsection{s-o transfer matrix: coupled channels}

Similarly to the uncoupled-channel case, here we can also define a
\textit{s-o} transfer matrix $U_{so}^{c}$ describing the effect of
the \emph{s-o}
interaction on electrons impinging near the band crossing at $k_{c}$%
. This transfer matrix is readily contructed in terms of the column vectors
similar to the one in Eq. (\ref{eq34}), which describes how a spin up
electron in channel \emph{a} evolves upon crossing the \emph{s-o} region. We
obtain
\begin{widetext}%
\begin{equation}
\mathbf{U}_{so}^{cc}=\frac{1}{2}e^{ik_{c}L}\left(
\begin{array}{cccc}
\cos \left( \theta _{d}/2\right) +e^{i\theta _{so}} & \xi ^{\ast
}\left[ \cos (\theta _{d}/2)-e^{i\theta _{so}}\right]  & -i\sin
(\theta _{d}/2) &
i\xi^{\ast }\sin \left( \theta _{d}/2\right)  \\
\xi \left[ \cos \left( \theta _{d}/2\right) -e^{i\theta
_{so}}\right]  & \cos (\theta _{d}/2)+e^{i\theta _{so}} & -i\xi
\sin (\theta _{d}/2) &
i\sin \left( \theta _{d}/2\right)  \\
-i\sin \left( \theta _{d}/2\right)  & -i\xi ^{\ast }\sin \left(
\theta _{d}/2\right)  & \cos (\theta _{d}/2)+e^{-i\theta _{so}} &
-\xi ^{\ast }
\left[ \cos \left( \theta _{d}/2\right) -e^{-i\theta _{so}}\right]  \\
i\xi \sin \left( \theta _{d}/2\right)  & i\sin \left( \theta
_{d}/2\right)  & -\xi \left[ \cos (\theta _{d}/2)-e^{-ik\theta _{so}}%
\right]  & \cos \left( \theta _{d}/2\right) +e^{-i\theta _{so}}%
\end{array}%
\right) {\small, }  \label{eq36}
\end{equation}
\end{widetext} where the modulation angles
$\theta_{so}$ and $\theta_d$ are given in Eqs. (\ref{eq-theta-so})
and (\ref{eq-theta_d}), respectively. We should keep in mind that
Eq. (\ref{eq36}) describes electrons traversing the \emph{s-o}
region with energies near the crossing energy. As we discuss later
on, the \emph{s-o} transfer matrix above is also useful for
spin-rotating entangled and/or unentangled electron pairs injected
into a four-terminal geometry (beam splitter). The idea is that
$U^{cc}$ operates on the member of the pair traversing the
\textit{s-o} region. Note that the transfer matrix in Eq.
(\ref{eq36}) reduces to that of the uncoupled case, Eq.
(\ref{eq16a}), for $\alpha =\beta $ ($\theta _{d}=0$). Next we
estimate the magnitude of the spin rotations we have described
here.

\subsection{Single spin rotation $\protect\theta _{so}$ for coupled channels
($d_{ab}\neq 0$) with $\protect\alpha =\protect\beta $ \ }

Here the calculation is simpler since the bands \emph{do not} anticross even
for non-zero $d_{ab}$ as we discussed earlier. The crossing wave vector $%
\bar{k_{c}}$ for $\alpha =\beta $ is determined from Eq. (\ref{eq12}%
). For instance, the $k>0$ crossing is obtained by setting
$\varepsilon _{+,-}(\bar{k_{c}})=\varepsilon _{-,+}(\bar{k_{c}})$
which gives
\begin{equation}
\bar{k_{c}}=\frac{1}{2\sqrt{2}\alpha }\sqrt{(\epsilon
_{b}-\epsilon _{a})^{2}+8\alpha ^{2}|d_{ab}|^{2}}.  \label{eq19}
\end{equation}
For $d_{ab}=0$ and $\alpha =\beta $ the above wave vector reduces to\ $k_{c}$
defined in Eq. (\ref{eq14}).

By expanding the incoming electron states into the exact eigenstates derived
in Sec. IIA we can obtain the modulation angle $\theta _{so}=2\sqrt{2}%
mL/\hbar ^{2}$. Note that $\theta _{d}=0$ for $\alpha =\beta $.
Interestingly, the matching of the boundary conditions here and the general
state at $x=L$ can be straightforwardly obtained from the $\alpha \neq \beta
$ case by setting $\Delta =0$ (or equivalently, $\theta _{d}=0$). However,
it is important to note that the crossing wave-vector is now $%
\bar{k_{c}}$ (not $k_{c}$)\ and that $k_{2}-\bar{k_{c}}=2k_{so}$,
where $k_{so}$ is calculated for $\alpha =\beta $. Note also that
only one modulation angle $\theta _{so}$ is present for the
tuned-coupling case $\alpha =\beta $. Hence, this case is similar
to the uncoupled channel problem treated by Datta and
Das\cite{datta-das}, even though here $d_{ab}\neq 0$. The
identical coupling strengths makes the problem similar to that of
the uncoupled channels; however the rotation angle is now
renormalized.

\subsection{Estimates for the modulating angles $\protect\theta _{so}$ and $%
\protect\theta _{d}$}

Simple estimates for the spin-rotation angle $\theta _{so}$ and
the mixing angle $\theta _{d}$ can be obtained by assuming a
hard-wall transverse confinement of width $w$. Using the
well-known analytical results for the wire problem, we find
$d_{ab}=8/3w$ for the interband mixing and $\epsilon _{b}=3\pi
^{2}\hbar ^{2}/2mw^{2}$\ (assuming $\epsilon _{a}\equiv 0$). The
quantity $\epsilon _{so}\equiv \hbar ^{2}k_{so}^{2}/2m=$ $m\left(
\alpha ^{2}+\beta ^{2}\right) /2\hbar ^{2}$ sets an energy scale
in our problem. For the sake of concreteness, let us choose
$\epsilon _{b}=16\epsilon _{so}$ which leads to $\sqrt{\alpha
^{2}+\beta ^{2}}=(\sqrt{3/2}\pi /4)\hbar ^{2}/mw=2.44\times
10^{-2}$ eVnm (and $\epsilon _{so}\sim 0.2$ meV) for $m=0.05m_{0}$
(see Ref. [\onlinecite{nitta}]) and $w=60$ nm. The energy at the
band crossing points is then $\varepsilon
_{a}^{-}(k_{c})=24\epsilon _{so}\sim 4.\,\allowbreak 8$ meV; note
that for Fermi energies close to this
value, \emph{s-o} induced channel-mixing effects are important. From Eq. (%
\ref{eq14}) we find $k_{c}=8\epsilon _{so}/\sqrt{\alpha ^{2}+\beta
^{2}}$. Assuming an active \emph{s-o} region of length $L=69$ nm
we can estimate the spin-rotation angles; we find $\theta
_{so}=\pi $. To obtain $\theta _{d}=\left( d_{ab}/k_{c}\right)
\theta _{so}(\alpha ^{2}-\beta ^{2})/(\alpha ^{2}+\beta ^{2})$ we
need an estimate for $\beta $. To estimate the Dresselhaus
coefficient in a quantum well geometry we use $\beta =
\tilde{\gamma} \langle k_{z}^{2}\rangle $, where $\langle
k_{z}^{2}\rangle $ denotes the expectation value of the wave
vector component along the growth direction. For the lowest
infinite-well eigenstate we find $\langle k_{z}^{2}\rangle =\left(
\pi /w\right) ^{2}$. The coefficient $\tilde{\gamma}$ is typically
$\approx 25$
$\mathrm{eV\mathring{A}}^{3}$\cite{Lommer88,Jusserand92,Jusserand95}
which yields $\beta \approx 10^{-5}$ eVnm. Hence, for such III-V
materials we can neglect $\beta$ and use $\theta _{d}=\left(
d_{ab}/k_{c}\right) \theta _{so}$, which gives $\theta_{d}=\pi /2$
since $d_{ab}/k_{c}=2/(3k_{so}w)\sim 0.5$.

In order to obtain comparable Rashba and Dresselhaus coupling
strengths, we could use a setup with wider wires and materials
with a larger effective mass \cite{john}. In addition, we could
consider an inhomogeneous beam-splitter with a different material
with larger Dresselhaus coupling in one of the incoming arms. Note
that the possibility of tuned couplings $\alpha=\beta$ is very
attractive since in this case the spin of the electron propagating
in the \textit{s-o} coupled channels is insensitive to
non-magnetic impurity scattering (\ref{exactsol}), i.e., the
spinor is $k$ independent for $\alpha=\beta$.

We stress that the modulation angles $\theta _{so}$ and $\theta _{d}$ can, in
principle, be tuned independently via a proper gating structure. This could
involve, for instance, both side (top) and back gates to induce changes in
the channel width $w$ (confining potential) and the Rashba constant. The
above conservative estimates suggests that the spin rotations we are
considering here are sizable.
Finally, we note that for the
above parameters $\Delta /4k_{F}\sim 0.05\ll 1$, which justifies the
approximation made in the velocity operator matching [Eq. (\ref{eq32})].

\section{Tranport properties: current and noise}

In what follows we calculate the current and its dynamic
fluctuations (shot noise) for electrons traversing a
beam-splitter. We use the scattering approach of Landauer and
B\"{u}ttiker \cite{butt}. We consider injection of (i) electron
pairs (singlet and triplets) from an \textquotedblleft
entangler\textquotedblright\ tunnel-coupled to the incoming leads
of the beam-splitter and (ii) spin-polarized electrons from
Fermi-liquid leads which are assumed to be thermal reservoirs each held at a given
chemical potential. For a calculation of shot
noise for entangled electrons in a beam-splitter where a Berry
phase provides an additional modulation, see Ref.
\onlinecite{Zha05}.

\subsection{Scattering approach: basics}

Here we briefly outline the scattering-matrix formulation for
current and noise \cite{butt}.

\subsubsection{Current}

Within the Landauer-B\"{u}ttiker approach, the transport properties of a
mesoscopic system are expressed in terms of the scattering matrix $\mathbf{s}%
_{\gamma \mu }$ connecting the many incoming and outgoing attached
leads. The current operator in lead $\gamma $ is
\begin{equation}
\hat{I}_{\gamma }(t)=\frac{e}{h \nu }\sum_{\alpha \beta \sigma
\sigma ^{\prime }} \sum_{\varepsilon \varepsilon^{\prime} }
A_{\alpha ,\beta }^{\sigma ,\sigma ^{\prime }}(\gamma ;\varepsilon
,\varepsilon ^{\prime }) e^{i(\varepsilon -\varepsilon ^{\prime
})t/\hbar } a_{\alpha \sigma }^{\dagger }(\varepsilon )a_{\beta
\sigma ^{\prime }}(\varepsilon ^{\prime }), \label{eq37}
\end{equation}%
with%
\begin{equation}
A_{\alpha ,\beta }^{\sigma ,\sigma ^{\prime }}(\gamma ;\varepsilon
,\varepsilon ^{\prime })=\delta _{\sigma ,\sigma ^{\prime }}\delta
_{\gamma ,\alpha }\delta _{\gamma ,\beta }-\sum\limits_{\sigma
^{\prime \prime }}s_{\gamma \alpha ;\sigma \sigma ^{\prime \prime
}}^{\ast }(\varepsilon )s_{\gamma \beta ;\sigma ^{\prime \prime
}\sigma ^{\prime }}(\varepsilon ^{\prime }) \label{eq38}
\end{equation}%
where  $\sigma =\uparrow
,\downarrow $ is the relevant spin component along a proper quantization
direction (\textquotedblleft \emph{x}, \emph{y}, or \emph{z}%
\textquotedblright ).
We have introduced the creation (annihilation) fermionic operator
 $a_{\alpha \sigma }^{\dagger }(\varepsilon )$
$[a_{\alpha \sigma}(\varepsilon )]$ for an electron with energy
$\varepsilon $ in lead $\alpha $, which satisfy the
anticommutation relation $\{ a_{\alpha \sigma }^{\dagger
}(\varepsilon ), a_{\alpha' \sigma'}(\varepsilon' )\}
=\delta_{\alpha \alpha'}\delta_{\sigma \sigma'}\delta_{\varepsilon
\varepsilon'}$ We have considered beam-splitter leads with
discrete longitudinal energy levels $\varepsilon,\varepsilon '$.
This yields the factor $\nu=(L_x/2\pi \hbar)\sqrt{m/2 E_F}$ in
(\ref{eq37}), which actually is the 1D density of states for only
forward propagating states (positive momenta). In the standard
expression for the current with continuous energies \cite{butt},
this factor cancels with the density of states appearing when
transforming
discrete states into continuous ones. For a study of noise in a
beam-splitter with continuous energies, see Refs.
\onlinecite{Samu04} and \onlinecite{Das04}. We discuss in more
details the transition from the discrete case to the continuous
one in Sec. \ref{multiple-inj}.

\subsubsection{Shot noise}

At a time \emph{t}, the current fluctuation about its average in
lead $ \gamma $ is $\delta \hat{I}_{\gamma }(t)=$ $\hat{I}_{\gamma
}(t)-\langle \hat{I}_{\gamma }\rangle $. In a multiple-lead
configuration, the shot noise between leads $\gamma $ and $\mu $
is defined as the Fourier transform of the symmetrized
current-current autocorrelation function
\begin{equation}
S_{\gamma \mu }(\omega )=\frac{1}{2}\int \langle \delta
\hat{I}_{\gamma }(t)\delta \hat{I}_{\mu }(t^{\prime })+\delta
\hat{I}_{\mu }(t^{\prime })\delta \hat{I}_{\gamma }(t)\rangle
e^{i\omega t}dt. \label{eq39}
\end{equation}
The angle brackets in Eq. (\ref{eq39}) stand for either (i) a \emph{quantum
mechanical} expectation value between two-particle states or (ii) a standard
\emph{ensemble} average (thermal reservoirs). Note that the non-equilibrium
current noise defined above arises physically from the discrete nature of
the charge flow in the system.\ This is strictly true only at zero
temperatures; at finite temperatures Eq. (\ref{eq39}) contains also thermal
noise.

\subsubsection{Beam-splitter scattering matrix}

To calculate the noise from Eq. (\ref{eq39}) we need to specify
the beam splitter scattering matrix. For a symmetric beam splitter
without {\it s-o} interaction and single-mode channels,  we have
the scattering matrix \cite{liu}

\begin{equation}
\mathbf{s}=\left(
\begin{array}{cccc}
0 & 0 & s_{13} & s_{14} \\
0 & 0 & s_{23} & s_{24} \\
s_{31} & s_{32} & 0 & 0 \\
s_{41} & s_{42} & 0 & 0%
\end{array}%
\right) =\left(
\begin{array}{cccc}
0 & 0 & r & t \\
0 & 0 & t & r \\
r & t & 0 & 0 \\
t & r & 0 & 0%
\end{array}%
\right) ,  \label{eq40}
\end{equation}%
that is, the beam splitter transmits electrons between leads 1 and 3 and
leads 2 and 4 with amplitude $r$ and between leads 2 and 3 and leads 2 and 4
with amplitude $t$. Note that backscattering is neglected in $\mathbf{s}$;
see Sec. III-C and Appendix \ref{app-back-bs} for a beamsplitter including backscattering
effects. If the incoming or outgoing leads have more than one mode (i.e.,
many quantized channels) we can, in a first approximation, assume that the
beamsplitter does not mix the orbital channels so that Eq. (\ref{eq40})
holds true for each of the modes separately.

Interestingly, in the presence of \emph{s-o} interaction in lead 1, Fig. 1,
we can define an extended beam-splitter scattering matrix to incorporate\
the spin rotation described by the \emph{s-o} transfer matrix $\mathbf{U}%
_{so}$. Since an electron in lead 1 undergoes a spin rotation described by $%
\mathbf{U}_{so}$, we can redefine the matrix elements $s_{13}=s_{31}$ and $%
s_{14}=s_{41}$ as 4x4 matrices
\begin{equation}
\mathbf{s}_{13}^{so}=\mathbf{s}_{31}^{so}=s_{13}\mathbf{U}_{so},
\label{eq41}
\end{equation}%
and
\begin{equation}
\mathbf{s}_{14}^{so}=\mathbf{s}_{41}^{so}=s_{14}\mathbf{U}_{so},
\label{eq42}
\end{equation}%
to incorporate the effects of the \textit{s-o} interaction. Note that $\mathbf{U}%
_{so} $ is given by Eqs. (\ref{eq16a}) and (\ref{eq36}) for the uncoupled-
and the coupled two-channel cases, respectively. The other elements in $%
\mathbf{s}$ remain unaltered except that they are now $4 \times 4$ matrices, e.g., $%
\mathbf{s}_{24}=\mathbf{s}_{42}=s_{14}\mathbf{1}$, where
$\mathbf{1}$ denotes the $4 \times4$ unit matrix. Note that the
new beam splitter scattering matrix $\mathbf{s}^{so}$
incorporating the \textit{s-o} effects in lead 1 as defined above
is a $16 \times 16$ object as opposed to the $4 \times 4 $ matrix
in\ Eq. (\ref{eq40}).

\subsection{Shot noise for electron pairs: singlet and triplet states}

We assume that an entangler
\cite{div-loss-jmmm,ble,choi-bruder-loss,recher-suk-loss,leso-mart-blat,falci-fein-hekk,
merlin,costa-bose,oliver-yam-yam,bose-home,recher-loss,bena-vish-bal-fish,saraga-loss,
bouchiat,recher-loss-prl2003,beenakker-03,samuelsson-prl03,saraga-al-loss-wes,
samuelsson-prl04, sauret-mart-fein,dup-hur} is placed just before
leads 1 and 2, Fig. 1(a) \cite{delay}. Below we calculate the
noise for the states
\begin{eqnarray}
\left.
\begin{array}{c}
|S\rangle \\
|T_{e_{i}}\rangle%
\end{array}%
\right\} &=&\frac{1}{\sqrt{2}}[a_{1\uparrow }^{\dagger }(\varepsilon
_{1})a_{2\downarrow }^{\dagger }(\varepsilon _{2})\mp a_{1\downarrow
}^{\dagger }(\varepsilon _{1})a_{2\uparrow }^{\dagger }(\varepsilon
_{2})]|0\rangle ,  \notag \\
|T_{u_{\sigma ,i}}\rangle &=&a_{1\sigma }^{\dagger }(\varepsilon
_{1})a_{2\sigma }^{\dagger }(\varepsilon _{2})|0\rangle ,\text{ }\sigma
=\uparrow ,\downarrow \text{,}  \label{eq43}
\end{eqnarray}%
where $|0\rangle $ denotes the ground state (filled) Fermi sea of the leads
and $i=x,y,z$ any particular quantization axis. The states $|S\rangle $ and $%
|T_{e_{i}}\rangle $ are \emph{entangled} singlet and triplet, respectively,
while $|T_{u_{\uparrow ,i}}\rangle $ and $|T_{u_{\downarrow ,i}}\rangle $
are \emph{unentangled} triplets. Here we consider zero temperatures, zero
applied voltages and zero frequencies. In this limit the Fermi sea is
completely inert (\emph{noiseless}) and the noise in the system is solely due
to the injected pairs above the Fermi surface \cite{ble}. To determine the
shot noise we essentially evaluate matrix elements of the\ general form
\begin{eqnarray}
&&\langle 0|a_{\mu ,\sigma _{\mu }}(\varepsilon _{\mu })a_{\nu ,\sigma _{\nu
}}(\varepsilon _{\nu })a_{\alpha ,\sigma }^{\dagger }(\varepsilon )a_{\beta
,\sigma ^{\prime }}(\varepsilon ^{\prime })\times  \notag \\
&&a_{\alpha ^{\prime },\sigma ^{\prime \prime }}^{\dagger }(\varepsilon
^{\prime \prime })a_{\beta ^{\prime },\sigma ^{\prime \prime \prime
}}(\varepsilon ^{\prime \prime \prime })a_{\gamma ,\sigma _{\gamma
}}^{\dagger }(\varepsilon _{\gamma })a_{\eta ,\sigma _{\eta }}^{\dagger
}(\varepsilon _{\eta })|0\rangle ,  \label{eq44}
\end{eqnarray}%
appearing in the noise definition, Eq. (\ref{eq39}). This is most
systematically done via Wick's theorem since the object in (\ref{eq41})
resembles a four-particle Green function (see Appendix D for details).

\subsubsection{General noise formulas: single-channel case}

For the injected singlet and the triplets in Eq. (\ref{eq43}) we find the
following expressions for the \emph{zero-frequency} noise between leads $%
\gamma $ and $\mu $
\begin{widetext}%
\begin{eqnarray}
S_{\gamma \mu }^{S/T_{e_{i}}}=\frac{e^{2}}{2h\nu }&&\Big[\sum\limits_{%
\alpha =1,2,\beta =1..4,\sigma ^{\prime },\sigma }A_{\alpha ,\beta }^{\sigma
,\sigma ^{\prime }}(\gamma ;\varepsilon _{\alpha },\varepsilon _{\alpha
})A_{\beta ,\alpha }^{\sigma ^{\prime },\sigma }(\mu ;\varepsilon _{\alpha
},\varepsilon _{\alpha })+\sum\limits_{\alpha \neq \beta =1,2,\sigma }[\pm
A_{\alpha ,\beta }^{\sigma ,\sigma }(\gamma ;\varepsilon _{\alpha
},\varepsilon _{\alpha })A_{\beta ,\alpha }^{-\sigma ,-\sigma }(\mu
;\varepsilon _{\beta },\varepsilon _{\beta })  \notag \\
- &&A_{\alpha ,\beta }^{\sigma ,-\sigma }(\gamma ;\varepsilon _{\alpha
},\varepsilon _{\alpha })A_{\beta ,\alpha }^{-\sigma ,\sigma }(\mu
;\varepsilon _{\beta },\varepsilon _{\beta })]\delta _{\varepsilon _{\alpha
},\varepsilon _{\beta }}\mp \sum\limits_{\alpha \neq \beta =1,2;\sigma
}A_{\alpha ,\alpha }^{\sigma ,-\sigma }(\gamma ;\varepsilon _{\alpha
},\varepsilon _{\alpha })A_{\beta ,\beta }^{-\sigma ,\sigma }(\mu
;\varepsilon _{\beta })  \notag \\
&&+\frac{1}{2}\sum\limits_{\alpha \neq \beta =1,2;\sigma }A_{\alpha ,\alpha
}^{\sigma ,\sigma }(\gamma ;\varepsilon _{\alpha },\varepsilon _{\alpha
})A_{\beta ,\beta }^{-\sigma ,-\sigma }(\mu ;\varepsilon _{\beta
},\varepsilon _{\beta })-\frac{1}{2}\sum\limits_{\alpha ,\beta =1,2;\sigma
}A_{\alpha ,\alpha }^{\sigma ,\sigma }(\gamma ;\varepsilon _{\alpha
},\varepsilon _{\alpha })A_{\beta ,\beta }^{\sigma ,\sigma }(\mu
;\varepsilon _{\beta })  \notag \\
&&-\frac{1}{2}\sum\limits_{\alpha =1,2;\sigma }A_{\alpha ,\alpha
}^{\sigma ,\sigma }(\gamma ;\varepsilon _{\alpha },\varepsilon
_{\alpha })A_{\alpha ,\alpha }^{-\sigma ,-\sigma }(\mu
;\varepsilon _{\alpha },\varepsilon _{\alpha })\Big],\text{ }
\label{eq45}
\\  {\noindent  \rm and} \\
S_{\gamma \mu }^{T_{u_{\sigma ,i}}}=\frac{e^{2}}{h\nu }&&\Big[%
\sum\limits_{\alpha =1,2,\beta =1..4,\sigma ^{\prime }}A_{\alpha ,\beta
}^{\sigma ,\sigma ^{\prime }}(\gamma ;\varepsilon _{\alpha },\varepsilon
_{\alpha })A_{\beta ,\alpha }^{\sigma ^{\prime },\sigma }(\mu ;\varepsilon
_{\alpha },\varepsilon _{\alpha })-\sum\limits_{\alpha ,\beta =1,2}A_{\alpha
,\beta }^{\sigma ,\sigma }(\gamma ;\varepsilon _{\alpha },\varepsilon
_{\alpha })A_{\beta ,\alpha }^{\sigma ,\sigma }(\mu ;\varepsilon _{\beta
},\varepsilon _{\beta })\delta _{\varepsilon _{\alpha },\varepsilon _{\beta
}}\Big].  \label{eq46}
\end{eqnarray}%
\end{widetext}
Note that Eqs. (\ref{eq45}) and (\ref{eq46}) do not depend on the
particular form (\ref{eq40}) of the beam-splitter scattering matrix
and the quantization axis chosen. In what follows we present
explicit formulas for the noise derived from Eqs. (\ref{eq45}) and (\ref%
{eq46}). We also determine the noise for entangled and unentangled states
defined along distinct quantization axes ($i=x,y$ and $\emph{z}$)\emph{\ }%
and for the Bell states. Later on we present similar results for the
two-channel case as well.

\subsubsection{Specific formulas: uncoupled-channel case}

In the absence of \emph{s-o} induced interband coupling (uncoupled-channel
case), the channels \emph{a} and \emph{b}\ are independent within lead 1.
That is, if electrons are injected only in the channel \emph{a} of lead 1,
they will remain in that channel while propagating through the length $L$ of
the \emph{s-o} region in that lead. If fact, the channel index remains
unaltered as the electrons traverse the beam splitter since we assume the
the beamsplitter does not mix the channels. We present below results for
electron pairs injected only in channel \emph{a} of the incoming leads. The
case with two pairs injected into channels \emph{a} and \emph{b} is
straighforward (factor of two) since no \emph{s-o} interband mixing is
considered here. However, as we discuss later on, in the coupled-channel
case injection into just one of the channels is significantly different from
injection into both channels (not just a factor of two as here). Below we
detail the calculation of the noise from Eqs. (\ref{eq45}) and (\ref{eq46}).
To calculate the noise in lead 3, i.e., $S_{33}^{S/T_{e_{i}}}$ and $%
S_{33}^{T_{u_{\sigma ,i}}}$, we first have to determine the relevant
elements $A_{\alpha ,\beta }^{\sigma ,\sigma ^{\prime }}(3;\varepsilon
_{\alpha },\varepsilon _{\alpha })$ [Eq. (\ref{eq38})] appearing in these
quantities. Since our scattering matrix is assumed to be independent of the
energy, so is $A_{\alpha ,\beta }^{\sigma ,\sigma ^{\prime }}(3;\varepsilon
_{\alpha },\varepsilon _{\alpha })=$ $A_{\alpha ,\beta }^{\sigma ,\sigma
^{\prime }}(3)$.

\paragraph{Quantization axis along z.}

For the specific forms
\begin{equation}
\mathbf{s}_{13}^{so}=\left(
\begin{array}{cc}
r\cos (\theta _{so}/2) & r\sin (\theta _{so}/2) \\
-r\sin (\theta _{so}/2) & r\cos (\theta _{so}/2)%
\end{array}%
\right)  \label{eq47}
\end{equation}%
and
\begin{equation}
\mathbf{s}_{14}^{so}=\left(
\begin{array}{cc}
t\cos (\theta _{so}/2) & t\sin (\theta _{so}/2) \\
-t\sin (\theta _{so}/2) & t\cos (\theta _{so}/2)%
\end{array}%
\right) ,  \label{eq48}
\end{equation}%
the only non-zero $A_{\alpha ,\beta }^{\sigma ,\sigma ^{\prime }}(3)$'s are
\begin{eqnarray}
A_{1,2}^{\uparrow ,\uparrow }(3) &=&-s_{31;\downarrow \uparrow }^{\ast
}s_{32;\downarrow \uparrow }-s_{31;\uparrow \uparrow }^{\ast }s_{32;\uparrow
\uparrow }  \notag \\
&=&-r^{\ast }t\cos (\theta _{so}/2)=A_{1,2}^{\downarrow ,\downarrow }(3)
\label{eq49} \\
A_{1,2}^{\uparrow ,\downarrow }(3) &=&-s_{31;\downarrow \uparrow }^{\ast
}s_{32;\downarrow \downarrow }-s_{31;\uparrow \uparrow }^{\ast
}s_{32;\uparrow \downarrow }  \notag \\
&=&r^{\ast }t\sin (\theta _{so}/2)=-A_{1,2}^{\uparrow ,\downarrow }(3)
\label{eq50} \\
A_{2,2}^{\uparrow ,\uparrow }(3) &=&-s_{32;\downarrow \uparrow }^{\ast
}s_{32;\downarrow \uparrow }-s_{32;\uparrow \uparrow }^{\ast }s_{32;\uparrow
\uparrow }  \notag \\
&=&-|t|^{2}=A_{2,2}^{\downarrow ,\downarrow }(3).  \label{eq51}
\end{eqnarray}%
Note that in the above we have chosen the $\sigma $ index in $\mathbf{s}%
^{so} $ to be that of the $z$ component of the spin: $\sigma \rightarrow
\sigma _{z}=\uparrow ,\downarrow $, i.e., we have set the quantization axis
to be \emph{z}, Fig. 1. Hence, the entangled and non-entangled triplet
states here refer to this basis: $S_{33}^{T_{e_{z}}}$and $%
S_{33}^{T_{u_{\sigma ,z}}}$; the noise for the singlet state $S_{33}^{S}$ is
the same for all quantization axes. Plugging in the above $A_{\alpha ,\beta
}^{\sigma ,\sigma ^{\prime }}(3)$'s into Eq. (\ref{eq45}) we find%
\begin{eqnarray}
S_{33}^{S/T_{e_{z}}}(\theta _{so}) &=&\frac{2e^{2}RT}{h\nu }[1\pm \cos
(\theta _{so}/2)\cos (\theta _{so}/2)\delta _{\varepsilon _{1},\varepsilon
_{2}}  \notag \\
&&-\sin (\theta _{so}/2)\sin (\theta _{so}/2)\delta _{\varepsilon
_{1},\varepsilon _{2}}],  \label{eq52}
\end{eqnarray}%
where we have defined the transmission and reflection probabilities $%
T=|t|^{2}$ and $R=|r|^{2}$, respectively. Since $R+T=1$, further
simplifications lead to
\begin{equation}
S_{33}^{S}(\theta _{so})=\frac{2e^{2}}{h\nu }T(1-T)\left[ 1+\cos \left(
\theta _{so}\right) \delta _{\varepsilon _{1},\varepsilon _{2}}\right] ,
\label{eq53}
\end{equation}%
\begin{equation}
S_{33}^{T_{e_{z}}}(\theta _{so})=\frac{2e^{2}}{h\nu }T(1-T)\left( 1-\delta
_{\varepsilon _{1},\varepsilon _{2}}\right) .  \label{eq54}
\end{equation}%
Similarly, we find%
\begin{eqnarray}
S_{33}^{T_{u_{\uparrow ,z}}}(\theta _{so})
&=&S_{33}^{T_{u_{\downarrow ,z}}}(\theta _{so})=  \notag \\
&&\frac{2e^{2}}{h\nu }T(1-T)[1 -\cos ^{2}(\theta _{so}/2)\delta
_{\varepsilon _{1},\varepsilon _{2}}], \label{eq55}
\end{eqnarray}%
for the unentangled triplets with spin polarization along \emph{z}.

The above formulas have been derived for the case where the
injected electrons in arm 1 and 2 have the infinitely-sharp
energies $\varepsilon_1$ and $\varepsilon_2$.  In Sec.
\ref{multiple-inj} we consider the case where the injected
electrons are described by Lorentzian wave-packets of width
$\gamma$ centered on $\varepsilon_1$ and $\varepsilon_2$. We then
find, in the continuous limit, that
$\delta_{\varepsilon_1,\varepsilon_2}$ in Eqs.
(\ref{eq53}-\ref{eq55}) is replaced by the function
$H(\Delta)=\gamma^2/(\Delta^2/4 +\gamma^2)$ where $\Delta =
\varepsilon_1-\varepsilon_2$. This function, which interpolates
between $H=1$ when $\Delta\ll \gamma$ and $H=0$ when $\Delta \gg
\gamma$, corresponds to the one appearing in Eq. (28) of Ref.
\onlinecite{Samu04}.

\paragraph{Quantization axis along y.}

To obtain the corresponding noise espressions for injected electron pairs
with spin polarization defined along the \emph{y} we have to first rewrite $%
\mathbf{s}^{so}$ in the basis of $\sigma _{y}$: $|\uparrow \rangle
_{y}=\left( |\uparrow \rangle _{z}+i|\downarrow \rangle _{z}\right) /\sqrt{2}
$, $|\downarrow \rangle _{y}=\left( |\uparrow \rangle _{z}-i|\downarrow
\rangle _{z}\right) /\sqrt{2}$. For instance, $\mathbf{s}_{13}^{so}$ becomes

\begin{equation}
\mathbf{s}_{13,y}^{so}=\left(
\begin{array}{cc}
r\exp (i\theta _{so}/2) & 0 \\
0 & r\exp (-i\theta _{so}/2)%
\end{array}%
\right) .  \label{eq56}
\end{equation}%
Now the only non-zero elements are $A_{1,2}^{\uparrow ,\uparrow
}(3)=-r^{\ast }t\exp (-i\theta _{so}/2)=\left[ A_{2,1}^{\downarrow
,\downarrow }(3)\right] ^{\ast }$ and $A_{2,1}^{\uparrow ,\uparrow
}(3)=-rt^{\ast }\exp (-i\theta _{so}/2)=\left[ A_{1,2}^{\downarrow
,\downarrow }(3)\right] ^{\ast }$. Substituting these terms into the general
Eqs. (\ref{eq45}) and (\ref{eq46}), we find%
\begin{equation}
S_{33}^{S/T_{e_{y}}}(\theta _{so})=\frac{2e^{2}}{h\nu }T(1-T)\left[ 1\pm
\cos \left( \theta _{so}\right) \delta _{\varepsilon _{1},\varepsilon _{2}}%
\right] ,  \label{eq57}
\end{equation}%
and%
\begin{equation}
S_{33}^{T_{u\uparrow ,y}}(\theta _{so})=S_{33}^{T_{u_{\downarrow
_{y}}}}(\theta _{so})=\frac{2e^{2}}{h\nu }T(1-T)\left( 1-\delta
_{\varepsilon _{1},\varepsilon _{2}}\right) .  \label{eq58}
\end{equation}

\paragraph{Bell states.}

For completeness we also calculate the noise for injected Bell
pairs \cite{science-2005}(maximally entangled states) described by

\begin{eqnarray}
|\Psi _{0}\rangle &=&|S\rangle  \label{eq59} \\
|\Psi _{1}\rangle &=&|T_{e_{i}}\rangle  \label{eq60} \\
|\Psi _{2}\rangle &=&\frac{1}{\sqrt{2}}\left( |T_{u\uparrow ,i}\rangle
+|T_{u\downarrow ,i}\rangle \right)  \label{eq61} \\
|\Psi _{3}\rangle &=&\frac{1}{\sqrt{2}}\left( |T_{u\uparrow ,i}\rangle
-|T_{u\downarrow ,i}\rangle \right) .  \label{eq62}
\end{eqnarray}%
The Bell states above are defined with respect to an arbitrary quantization
axis. The noise expressions for the first two Bell states are the same as
those of the singlet and the entangled triplet derived above. The noise for
the states\ $|\Psi _{2}\rangle $ and $|\Psi _{3}\rangle $ can be easily
determined in terms of the results for the unentangled triplet states for
particular quantization axes. Let us consider the quantization axis along
\emph{z}, for concreteness. The procedure is straightforward:

\noindent i) After traversing the {\it s-o} region, the injected unentangled triplet state
$|T_{u_{\uparrow ,z}}\rangle$ becomes
$
|T_{u_{\uparrow ,z}}\rangle^L = \mathbf{U}^a_{so} |T_{u_{\uparrow ,z}}\rangle
$
and can be decomposed as
\begin{equation}
|T_{u_{\uparrow ,z}}\rangle^L =\left[ \cos (\theta _{so}/2)|\uparrow \rangle
_{1a,z}-\sin (\theta _{so}/2)|\downarrow \rangle _{1a,z}\right] \otimes
|\uparrow \rangle _{2a,z},  \label{eq65}
\end{equation}%
or%
\begin{equation}
|T_{u_{\uparrow ,z}}\rangle^L =\cos (\theta _{so}/2)|\uparrow \uparrow \rangle
_{aa,z}-\sin (\theta _{so}/2)|\downarrow \uparrow \rangle _{aa,z},
\label{eq66}
\end{equation}%
where we have used the shorthand notation $|\uparrow \uparrow \rangle
_{aa,z}=|\uparrow \rangle _{1a,z}\otimes |\uparrow \rangle _{2a,z}=|T_{u_{\uparrow ,z}}\rangle$, to
denote the tensor product state with one
electron injected into the channel \emph{a} of lead 1 and another in the
channel \emph{a} of lead 2. Similarly,%
\begin{equation}
|T_{u_{\downarrow z}}\rangle^L =\sin (\theta _{so}/2)|\uparrow \downarrow
\rangle _{aa,z}+\cos (\theta _{so}/2)|\downarrow \downarrow \rangle _{aa,z}.
\label{eq67}
\end{equation}

\noindent ii) Now, we rewrite the rotated Bell state $|\Psi _{2}\rangle^L $ explicitly
\begin{eqnarray}
|\Psi _{2}\rangle^L &=&\cos (\theta _{so}/2)\frac{1}{\sqrt{2}}\left( |\uparrow
\uparrow \rangle _{aa,z}+|\downarrow \downarrow \rangle _{aa,z}\right)
\notag \\
&&+\sin (\theta _{so}/2)\frac{1}{\sqrt{2}}\left( |\uparrow \downarrow
\rangle _{aa,z}-|\downarrow \uparrow \rangle _{aa,z}\right) .  \label{eq68}
\end{eqnarray}

\noindent iii) The noise in lead 3 corresponding to $|\Psi _{2}\rangle^L $ is
determined from the expectation value of Eq. (\ref{eq39}) in this state.
Since the beam-splitter scattering matrix does not include spin-flip
processes, this expectation value can be expressed in terms of the shot
noise for the unrotated singlet $\left(
|\uparrow \downarrow \rangle _{aa,z}-|\downarrow \uparrow \rangle
_{aa,z}\right) /2$ and unentangled triplets $|\uparrow \uparrow \rangle
_{aa,z}$, $|\downarrow \downarrow \rangle _{aa,z}$
\begin{eqnarray}
S_{33}^{\Psi _{2}}\left( \theta _{so}\right) &=&\frac{1}{2}\cos ^{2}(\theta
_{so}/2)[\widetilde{S}_{33;aa}^{T_{u_{\uparrow ,z}}}+\widetilde{S}%
_{33;aa}^{T_{u_{\downarrow ,z}}}]  \notag \\
&&+\sin ^{2}(\theta _{so}/2)\widetilde{S}_{33;aa}^{S},  \label{eq69}
\end{eqnarray}%
where
\begin{equation}
\widetilde{S}_{33;aa}^{S}=\frac{2e^{2}}{h\nu }T(1-T)\left( 1+\delta
_{\varepsilon _{1},\varepsilon _{2}}\right) ,  \label{eq63}
\end{equation}%
and%
\begin{equation}
\widetilde{S}_{33;aa}^{T_{u_{\uparrow ,z}}}=\widetilde{S}_{33;aa}^{T_{u_{%
\downarrow ,z}}}=\frac{2e^{2}}{h\nu }T(1-T)\left( 1-\delta _{\varepsilon
_{1},\varepsilon _{2}}\right) .  \label{eq64}
\end{equation}%
Note that $\widetilde{S}_{33;aa}^{S}$ and $\widetilde{S}_{33;aa}^{T_{u_{%
\sigma ,z}}}$ denote the shot noise for the bare singlet and triplet states,
respectively \cite{ble}. Equations (\ref{eq63}) and (\ref{eq64}) represent
the noise expectation value for the bare singlet and triplet states \cite%
{ble} and follows from Eqs. (\ref{eq54}) and (\ref{eq55}) with $\theta
_{so}=0$. Substituting the Eqs. (\ref{eq63}) and (\ref{eq64}) into (\ref%
{eq69}), we find
\begin{equation}
S_{33}^{\Psi _{2}}\left( \theta _{so}\right) =\frac{2e^{2}}{h\nu }T(1-T)%
\left[ 1-\cos (\theta _{so})\delta _{\varepsilon _{1},\varepsilon _{2}}%
\right] .  \label{eq70}
\end{equation}%
Similarly,
\begin{equation}
S_{33}^{\Psi _{3}}\left( \theta _{so}\right) =\frac{2e^{2}}{h\nu }%
T(1-T)\left( 1-\delta _{\varepsilon _{1},\varepsilon _{2}}\right) .
\label{eq71}
\end{equation}%
It is interesting to note that $S_{33}^{\Psi _{2}}=S_{33}^{T_{e_{y}}}$ and $%
S_{33}^{\Psi _{3}}=S_{33}^{T_{u_{\uparrow _{y}}}}=S_{33}^{T_{u_{\downarrow
y}}}$. This follows from $|T_{e_{y}}\rangle =$ $\frac{1}{\sqrt{2}}(|\uparrow
\downarrow \rangle _{y}-|\downarrow \uparrow \rangle _{y})$ being equal to $%
|\Psi _{2}\rangle $ and \ $|\Psi _{3}\rangle =|T_{u_{\uparrow y}}\rangle $
for the quantization axis along \emph{z}.

\paragraph{Local Zeeman and spin-orbit induced rotations.}

It is interesting to note that the shot noise modulation induced by the
local \emph{s-o }interaction in the uncoupled-channel case discussed above
is formally identical to that due to a local Zeeman-interaction in lead 1
[e.g., due to a local magnetic field or a distinct \emph{g} factors in lead
1]. For a local Zeeman interaction, $\theta _{so}$ should be equal to $g\mu
_{B}BL/v$, i.e., the phase acquired by the electron upon traversing, with
velocity $v$, the length $L$ of the Zeeman-active region. Note, however,
that this formal correspondence between local Zeeman and \emph{s-o}
interactions holds only for the uncoupled-channel case (or in a strictly
single-channel lead) and when magnetic-field-induced orbital effects are
neglected.

\paragraph{Shot noise for electron pairs: physical picture.}

Before moving over to the more elaborate case with coupled channels, we
provide here a simple picture for the shot noise results we obtained for the
singlet and triplets. Let us first consider the case with no spin orbit ($%
\theta _{so}=0$) and one orbital channel. Our formulas yield
bunching for singlet and anti-bunching for the triplets, as
previously found in Ref. \onlinecite{ble}. This bunching and
antibunching behavior for the singlet and triplets pairs can be
readily understood if we write out the pair wave-functions in the
outgoing leads (3,4) similarly to Ref. \onlinecite{loudon}. The
injected singlet and entangled triplets states
$|S/T_{e_{z}}\rangle =(a_{1,\uparrow }^{\dagger }a_{2,\downarrow
}^{\dagger }\mp a_{1,\downarrow }^{\dagger }a_{2,\uparrow
}^{\dagger })|0\rangle /\sqrt{2}$ and $|T_{u_{\sigma ,i}}\rangle
=a_{1\sigma }^{\dagger }a_{2\sigma }^{\dagger }|0\rangle $ after
trasversing the beam-splitter evolve into
\begin{eqnarray}
|S\rangle  &\rightarrow &\sqrt{2}rtb_{3,\uparrow }^{\dagger }b_{3,\downarrow
}^{\dagger }|0\rangle +\sqrt{2}rtb_{4,\uparrow }^{\dagger }b_{4,\downarrow
}^{\dagger }|0\rangle +  \notag \\
&&\frac{1}{\sqrt{2}}\left( r^{2}+t^{2}\right) \left( b_{3,\uparrow
}^{\dagger }b_{4,\downarrow }^{\dagger }-b_{3,\downarrow }^{\dagger
}b_{4,\uparrow }^{\dagger }\right) |0\rangle ,  \label{s1}
\end{eqnarray}%
\begin{equation}
|T_{e_{z}}\rangle \rightarrow \frac{1}{\sqrt{2}}\left( r^{2}-t^{2}\right)
\left( b_{3,\uparrow }^{\dagger }b_{4,\downarrow }^{\dagger
}+b_{3,\downarrow }^{\dagger }b_{4,\uparrow }^{\dagger }\right) |0\rangle ,
\label{t1}
\end{equation}%
and%
\begin{equation}
|T_{u_{\sigma ,z}}\rangle \rightarrow \left( t^{2}-r^{2}\right) b_{4,\sigma
}^{\dagger }b_{3,\sigma }^{\dagger },  \label{t2}
\end{equation}%
where the $b_{i,\sigma }^{\dagger }$'s ($i=3,4$) represent the creation
operators in the outgoing leads, directly related to the $a_{i,\sigma
}^{\dagger }$'s via the beam-splitter scattering matrix \textbf{s} (note
that here the particular quantization axis is irrelevant; this is not true
in the presence of the spin-orbit interaction as described below).
Interestingly, the incoming electrons in the triplet states in leads 1 and 2
have \emph{zero} probability of emerging in the same outgoing lead (note
that $\left\vert r^{2}-t^{2}\right\vert =1$ due to $\mathbf{s}^{\dag }%
\mathbf{s}=\mathbf{1}$). This is the (full) antibunching we mentioned above
and it results in zero shot noise for the injected triplet pairs.
Physically, this vanishing of the shot noise means that there is no
randomness in the electron flow in the outgoing leads: each electron of the
incoming triplet pairs goes -- with unity probability -- to distinct
outgoing leads. For the singlet pairs, on the other hand, the probability
for the two electrons to emerge in the same outgoing arm is not zero as for
the triplets. This probability is larger by a factor of two than the
classical value $\left\vert r\right\vert ^{2}\left\vert t\right\vert ^{2}$.
Hence the incoming electrons in a singlet pair tend to bunch, i.e., to go to
the same outgoing leads. Full bunching occurs only for the particular case
of a 50:50 beam splitter.\ In this case we have $r=\pm it$ and opposite
spins have a 1/2 probability to emerge in one of the leads 3 or 4. This
randomness in the electron flow in the outgoing leads increases the shot
noise as compared to both the triplet and the classical case.

Note that the bunching and anti-bunching behaviors described above follow
from the stringent requirement for anti-symmetry (Pauli's principle) of the
total  wave function of the electron pair.\ In an unentangled triplet, for
instance, the spin part of the wave function is symmetric thus forcing its
spatial component to be anti-symmetric which results in a \ strong
correlation between the electrons in the triplet pair, i.e., they avoid each
other by always going into distinct outgoing leads. The same is true for the
entangled triplet pair. This correlation reduces the shot noise, in
agreement with the classical notion of shot noise suppression which occurs
when the \textquotedblleft discreteness\textquotedblright\ of the electron
flow is reduced. For the singlet, on the other hand, the spin part of the
pair wave function is anti-symmetric which makes the spatial part symmetric.
The electrons in a singlet pair can overlap freely in space and hence are
less correlated than the triplet case thus giving rise to non-zero shot
noise.  Interestingly, the shot noise for the singlet is even larger than
the classical analog (i.e., a distinguishable pair injected into leads 1 and
2). This also agrees with the classical notion of shot noise as due to the
discreteness of the charge flow: because the spatial part of the singlet
pair is maximally \textquotedblleft uncorrelated\textquotedblright\
(\textquotedblleft negative correlation\textquotedblright ) the discreteness
of the electron charge is larger than the classical case (the singlet
electrons can lie completely on top of each other in real space) thus
yielding a larger shot noise.

In the presence of spin-orbit coupling, the above ideas can be
straightforwardly generalized. In this case the singlet and
triplets injected into single-channel leads evolve into
\begin{eqnarray}
|S\rangle  &\rightarrow &\sqrt{2}rt\cos (\theta _{so}/2)b_{3,\uparrow
}^{\dagger }b_{3,\downarrow }^{\dagger }|0\rangle +  \notag \\
&&\sqrt{2}tr\cos (\theta _{so}/2)b_{4,\uparrow }^{\dagger }b_{4,\downarrow
}^{\dagger }|0\rangle +  \notag \\
&&\left( t^{2}+r^{2}\right) \cos (\theta _{so}/2)\frac{1}{\sqrt{2}}\left(
b_{3,\uparrow }^{\dagger }b_{4,\downarrow }^{\dagger }-b_{3,\downarrow
}^{\dagger }b_{4,\uparrow }^{\dagger }\right) |0\rangle +  \notag \\
&&\left( t^{2}-r^{2}\right) \sin (\theta _{so}/2)\frac{1}{\sqrt{2}}\left(
b_{3,\downarrow }^{\dagger }b_{4,\downarrow }^{\dagger }+b_{3,\uparrow
}^{\dagger }b_{4,\uparrow }^{\dagger }\right) |0\rangle ,  \label{sb}
\end{eqnarray}%
\begin{eqnarray}
|T_{e_{z}}\rangle  &\rightarrow &\left( r^{2}-t^{2}\right) \cos (\theta
_{so}/2)\frac{1}{\sqrt{2}}\left( b_{3,\uparrow }^{\dagger }b_{4,\downarrow
}^{\dagger }+b_{3,\downarrow }^{\dagger }b_{4,\uparrow }^{\dagger }\right)
|0\rangle +  \notag \\
&&\left( t^{2}-r^{2}\right) \sin (\theta _{so}/2)\frac{1}{\sqrt{2}}\left(
b_{3,\downarrow }^{\dagger }b_{4,\downarrow }^{\dagger }-b_{3,\uparrow
}^{\dagger }b_{4,\uparrow }^{\dagger }\right) |0\rangle ,  \label{t1b}
\end{eqnarray}%
and
\begin{eqnarray}
|T_{u_{\uparrow },z}\rangle  &\rightarrow &\cos (\theta _{so}/2)\left(
t^{2}-r^{2}\right) b_{4,\uparrow }^{\dagger }b_{3,\uparrow }^{\dagger
}|0\rangle +  \notag \\
&&\sin (\theta _{so}/2)rt\left( b_{3,\uparrow }^{\dagger }b_{3,\downarrow
}^{\dagger }+b_{4,\uparrow }^{\dagger }b_{4,\downarrow }^{\dagger }\right)
|0\rangle +  \notag \\
&&\sin (\theta _{so}/2)t^{2}b_{3,\uparrow }^{\dagger }b_{4,\downarrow
}^{\dagger }|0\rangle -  \notag \\
&&\sin (\theta _{so}/2)r^{2}b_{3,\downarrow }^{\dagger }b_{4,\uparrow
}^{\dagger }|0\rangle .  \label{tub}
\end{eqnarray}%
The unentangled triplet $|T_{u_{\downarrow ,z}}\rangle $ evolves
similarly to $|T_{u_{\uparrow },z}\rangle $. Due to the continuous
spin rotation induced by the \emph{s-o} coupling, intermediate
degrees of bunching and antibunching are possible as the
modulation angle $\theta _{so}$ is varied. Note that a variety of
entangled and unentangled electron pairs (along distinct
quantization axes as well) can be generated according to the above
states by sending electron pairs through a \emph{s-o} active
region. See also Ref. \onlinecite{signal-zulicke} for entanglement
generation using a Mach-Zehnder interferometer.

\subsubsection{Coupled-channel case}

Here we assume that electron pairs are injected into channel \emph{a} of leads
1 and 2 with energies near the crossing of the bands at $k_{c}$, Figs. 1(b)
and 2. The case with injection into both channels \emph{a} and \emph{b} is
discussed at the end of this section. General expressions similar to Eqs. (%
\ref{eq45}) and (\ref{eq46}) can, in principle, be derived for
this case. Here, however, we generalize the approach outlined at
the end of the previous section to the the case of two
\emph{s-o}--coupled channels, which is both simpler and more
intuitive. An additional heuristic derivation of the noise
properties using simple number operators is given in the Appendix
\ref{app-back-bs}.

\paragraph{Quantization axis along z.}

The idea is essentially the same as before: for instance, an injected
unentangled spin-up triplet along the \emph{z} direction,\ with one electron
in channel \emph{a} of lead 1 an the other in channel \emph{a} of lead 2
evolves into [Eq. (\ref{eq34})]
\begin{eqnarray}
|T_{u_{\uparrow ,z}}\rangle^L &=&\frac{1}{2}\left[ \cos \left( \theta
_{d}/2\right) e^{-i\theta _{so}/2}+e^{i\theta _{so}/2}\right] |\uparrow
\uparrow \rangle _{aa,z}  \notag \\
&+&\frac{1}{2}\xi \left[ \cos \left( \theta _{d}/2\right)
e^{-i\theta _{so}/2}-e^{i\theta _{so}/2}\right] |\downarrow
\uparrow \rangle _{aa,z}
\notag \\
&-&\frac{1}{2}i\sin \left( \theta _{d}/2\right) e^{-i\theta
_{so}/2}|\uparrow \uparrow \rangle _{ba,z}  \notag \\
&+&\frac{1}{2}i\xi \sin \left( \theta _{d}/2\right) e^{-i\theta
_{so}/2}|\downarrow \uparrow \rangle _{ba,z},  \label{eq72}
\end{eqnarray}%
upon traversing the \emph{s-o} region in lead 1 (we dropped an
overall phase $e^{i\left( k_{c}+k_{so}\right) L}$. Similarly to
the uncoupled-channel case, we note that only the portion of the
electron pair going through lead 1 is subject to the \textit{s-o}
effect. Note that the \emph{s-o} induced interband coupling in
lead 1 makes the injected states initially in channel \emph{a}
leak into channel \emph{b}. The expectation value of the noise in
the state $|T_{u_{\uparrow ,z}}\rangle^L$ is
\begin{eqnarray}
S_{33}^{T_{u_{\uparrow ,z}}}(\theta _{so},\theta _{d}) &=&\frac{1}{4}%
\left\vert \cos \left( \theta _{d}/2\right) e^{-i\theta _{so}/2}+e^{i\theta
_{so}/2}\right\vert ^{2}\widetilde{S}_{33;aa}^{T_{u_{\uparrow ,z}}}  \notag
\\
&+&\frac{\left\vert \xi \right\vert ^{2}}{4}\left\vert \cos \left(
\theta
_{d}/2\right) e^{-i\theta _{so}/2}-e^{i\theta _{so}/2}\right\vert ^{2}%
\widetilde{S}_{33;aa}^{\downarrow \uparrow ,z}  \notag \\
&+&\frac{1}{4}\sin ^{2}\left( \theta _{d}/2\right) \widetilde{S}%
_{33;ba}^{T_{u_{\uparrow ,z}}}  \notag \\
&+&\frac{1}{4}\sin ^{2}\left( \theta _{d}/2\right) \widetilde{S}%
_{33;ba}^{\downarrow \uparrow ,z}\text{,}  \label{eq73}
\end{eqnarray}%
where $\widetilde{S}_{33;ba}^{\downarrow \uparrow ,z}$
denotes the noise expectation value, Eq. (\ref{eq39}),
for a pair with a spin-down electron in channel \emph{b}
of lead 1 and a spin-up electron in channel \emph{a} of lead 2,
$\protect{|\downarrow \uparrow \rangle _{ba,z}}$. Similarly, $%
\widetilde{S}_{33;aa}^{\downarrow \uparrow ,z}$ corresponds to the\ two
opposite spin electrons in the channel \emph{a} of the respective lead. The
noise contribution of the pair state $|\downarrow \uparrow \rangle _{aa,z}$
can be determined from Eq. (\ref{eq55}) by setting $\theta _{so}=\pi $ which
makes\ $|T_{u_{\uparrow ,z}}\rangle^L =-|\downarrow \uparrow \rangle _{aa,z}$, see (\ref{eq66}),
\begin{equation}
\widetilde{S}_{33;aa}^{\downarrow \uparrow ,z}=\frac{2e^{2}}{h\nu }T(1-T).
\label{eq74}
\end{equation}%
Note that the pair state $|\downarrow \uparrow \rangle _{aa,z}$ is
distinguishable (spin down in lead 1 and spin up in lead 2, both
in channel \emph{a}), and therefore yields the classical shot
noise. Similarly, for the distinguishable state $|\downarrow
\uparrow \rangle _{ba,z}$ we should have
$\widetilde{S}_{33;ba}^{\downarrow \uparrow
,z}=\widetilde{S}_{33;aa}^{\downarrow \uparrow ,z}$. Hence we find
\begin{eqnarray}
S_{33}^{T_{u_{\uparrow ,z}}}(\theta _{so},\theta _{d}) &=&\frac{e^{2}}{h\nu }%
T(1-T)\left[ 1+\sin ^{2}\left( \theta _{d}/2\right) /2\delta _{\varepsilon
_{1},\varepsilon _{2}}\right.  \notag \\
&&\left. -\cos \left( \theta _{d}/2\right) \cos \left( \theta _{so}\right)
\delta _{\varepsilon _{1},\varepsilon _{2}}\right] ,  \label{eq75}
\end{eqnarray}%
and, similarly, $S_{33}^{T_{u_{\downarrow ,z}}}(\theta _{so},\theta
_{d})=S_{33}^{T_{u_{\uparrow ,z}}}(\theta _{so},\theta _{d})$. The noise for
the entangled triplet and the singlet are, respectively,%
\begin{equation}
S_{33}^{T_{e_{z}}}(\theta _{so},\theta _{d})=\frac{2e^{2}}{h\nu }T(1-T)\left[
1-\frac{1}{2}\left( \cos ^{2}\left( \theta _{d}/2\right) +1\right) \delta
_{\varepsilon _{1},\varepsilon _{2}}\right] ,  \label{eq76}
\end{equation}%
and
\begin{equation}
S_{33}^{S}(\theta _{so},\theta _{d})=\frac{2e^{2}}{h\nu }T(1-T)\left[ 1+\cos
\left( \theta _{d}/2\right) \cos \left( \theta _{so}\right) \delta
_{\varepsilon _{1},\varepsilon _{2}}\right] .  \label{eq77}
\end{equation}

\paragraph{Quantization axis along y.}

We can again straightforwardly find all the shot noise expressions
for the quantization axis along \emph{y} by following the same
procedure as above. However, here a spin up\ (along the \emph{y}
direction) electron injected into
the channel \emph{a} of lead 1 evolves into%
\begin{eqnarray}
|\uparrow \rangle _{a,y}^L &=&\frac{1}{2\sqrt{2}}\{[(1+i\xi ^{\ast
})\cos \left( \theta _{d}/2\right) e^{-i\theta _{so}/2}  \notag \\
&+&(1-i\xi ^{\ast })e^{i\theta _{so}/2}]|\uparrow \rangle _{a,z}
\notag
\\
&+&[\left( \xi +i\right) \cos \left( \theta _{d}/2\right)
e^{-i\theta _{so}/2}+\left( i-\xi \right) e^{i\theta
_{so}/2}]|\downarrow \rangle
_{a,z}  \notag \\
&+&\left( \xi ^{\ast }-i\right) \sin \left( \theta _{d}/2\right)
e^{-i\theta _{so}/2}|\uparrow \rangle _{b,z}  \notag \\
&+&\left( i\xi -1\right) \sin \left( \theta _{d}/2\right)
e^{-i\theta _{so}/2}|\downarrow \rangle _{b,z}\},  \label{eq78a}
\end{eqnarray}%
upon traversing the \emph{s-o} region in lead 1, while a spin down electron
(along the \emph{y}) evolves according to%
\begin{eqnarray}
|\downarrow \rangle _{a,y}^L &=&\frac{1}{2\sqrt{2}}\{[(1-i\xi
^{\ast
})\cos \left( \theta _{d}/2\right) e^{-i\theta _{so}/2}  \notag \\
&+&(1+i\xi ^{\ast })e^{i\theta _{so}/2}]|\uparrow \rangle _{a,z}
\notag
\\
&+&\left[ \left( \xi -i\right) \cos \left( \theta _{d}/2\right)
e^{-i\theta _{so}/2}-\left( i+\xi \right) e^{i\theta
_{so}/2}\right]
|\downarrow \rangle _{a,z}  \notag \\
&-&\left( i+\xi ^{\ast }\right) \sin \left( \theta _{d}/2\right)
e^{-i\theta _{so}/2}|\uparrow \rangle _{b,z}  \notag \\
&+&\left( i\xi +1\right) \sin \left( \theta _{d}/2\right)
e^{-i\theta _{so}/2}|\downarrow \rangle _{b,z}\}.  \label{eq78b}
\end{eqnarray}

For simplicity, we consider below the case with only the Rashba coupling,
i.e., $\alpha \neq 0$ and $\beta =0$. In this case, the above equations
simplify to
\begin{equation}
|\uparrow \rangle _{a,y}^L=\exp (i\theta _{R}/2)|\uparrow \rangle _{a,y},
\label{eq78}
\end{equation}%
and%
\begin{equation}
|\downarrow \rangle _{a,y}L=\exp (-i\theta _{R}/2)\left[ \cos \left(
\theta _{d}/2\right) |\downarrow \rangle _{a,y}-i\sin (\theta
_{d}/2)|\uparrow \rangle _{b,y}\right] ,  \label{eq79}
\end{equation}%
where $\theta _{so}\rightarrow \theta _{R}=2m\alpha L/\hbar ^{2}$
and $|\uparrow \rangle _{a,y}$ and $|\downarrow \rangle _{a,y}$
are the eigenspinors of $\sigma _{y}$ in channel \emph{a} (similar
definitions hold for channel \emph{b}). Note that the spin-up and
spin-down states in channel \emph{a} of lead 1 evolve quite
differently. In particular, the spin-up state remains in channel
\emph{a} and only picks up a overal phase, while the spin-down one
acquires a spin-up component in channel \emph{b}. This distinct
evolution of $|\uparrow \rangle _{a,y}^L$ and $|\downarrow \rangle
_{a,y}^L$ can be understood if we recall that the incoming
electron state $|\uparrow \rangle _{a,y}$ is essentially an
eigenstate of the system at $x=0$ (far away from any level
crossing) and hence evolves as such, as the electron traverses the
\emph{s-o }region.\emph{\ }In contrast, the state $|\downarrow
\rangle _{a,y}^L$ results from the incoming spin-down state in
channel \emph{a}\ being injected at the level
crossing corresponding to the states $|\downarrow \rangle _{a,y}$ and $%
|\uparrow \rangle _{b,y}$ and thus evolves as a coherent superposition of
these two states along lead 1. Hence, $|\uparrow \rangle _{a,y}^L$ is an
eigenstate of our Hamiltonian\ in the approximation we considered in Eq. (%
\ref{eq20}), while $|\downarrow \rangle _{a,y}^L$ is not, due to the weak
coupling between channels \emph{a} and \emph{b} at the crossing.

From the states (\ref{eq78}) and (\ref{eq79})\ we can construct
the electron pairs (entangled or not) that we are interested in,
e.g., the singlet and the triplet states along the \emph{y}
direction. Let us consider a pair injected into only the channel
\emph{a} of leads 1 and 2. For the  spin-up triplet we find
\begin{equation}
|T_{u_{\uparrow ,y}}\rangle ^L=\exp (i\theta _{R}/2)|\uparrow \uparrow
\rangle _{aa,y},  \label{eq80}
\end{equation}%
while for the spin down
\begin{eqnarray}
|T_{u_{\downarrow ,y}}\rangle ^L &=&\exp (-i\theta _{R}/2)[\cos \left(
\theta _{d}/2\right) |\downarrow \downarrow \rangle _{aa,y}  \notag \\
&-&i\sin (\theta _{d}/2)|\uparrow \downarrow \rangle _{ba,y}].  \label{eq81}
\end{eqnarray}%
The entangled triplet and singlet are found to be
\begin{eqnarray}
|S/T_{e_{y}}\rangle ^L &=&\frac{1}{\sqrt{2}}[\exp (i\theta
_{R}/2)|\uparrow \downarrow \rangle _{aa,y}  \notag \\
&\mp &\exp (-i\theta _{R}/2)\cos (\theta _{d}/2)|\downarrow \uparrow \rangle
_{ba,y}]  \notag \\
&\pm &\frac{1}{\sqrt{2}}i\exp (-i\theta _{R}/2)|\uparrow \uparrow \rangle
_{ba,y},  \label{eq82}
\end{eqnarray}%
where the upper (lower) sign corresponds to the singlet (triplet) state. The
noise expression corresponding to the above states are%
\begin{equation}
S_{33}^{T_{u_{\uparrow ,y}}}(\theta _{d})=\frac{2e^{2}}{h\nu }T(1-T)\left(
1-\delta _{\varepsilon _{1},\varepsilon _{2}}\right) ,  \label{eq83}
\end{equation}%
\begin{equation}
S_{33}^{T_{u_{\downarrow ,y}}}(\theta _{d})=\frac{2e^{2}}{h\nu }T(1-T)\left(
1-\cos ^{2}(\theta _{d}/2)\delta _{\varepsilon _{1},\varepsilon _{2}}\right)
,  \label{eq84}
\end{equation}%
and%
\begin{equation}
S_{33}^{S/T_{e_{y}}}(\theta _{R},\theta _{d})=\frac{2e^{2}}{h\nu }T(1-T)%
\left[ 1\pm \cos (\theta _{d}/2)\cos (\theta _{R})\delta _{\varepsilon
_{1},\varepsilon _{2}}\right] .  \label{eq85}
\end{equation}

Interestingly, the above results show that by measuring the noise
in lead 3 along the \emph{y} quantization axis one can distinguish
\emph{all} the electron pairs as they display distinct noise for
non-zero \emph{s-o} induced interchannel mixing angles ($\theta
_{d}\neq 0$).\ Figure 3 shows the reduced Fano factor
$f=S_{33}/\left[ 2e^{2}T(1-T)/h\nu \right] $ for the singlet
and triplets along the \emph{y} axis as a function of $\theta _{so}$ and $%
\theta _{d}$. It clearly shows that the singlet and triplets
display distinct shot noise in a wide range of angles. Note that
this result holds only for incoming electron pairs injected
initially into channel \emph{a} and with energies near the
crossing.

\begin{figure}[th]
\begin{center}
\epsfig{file=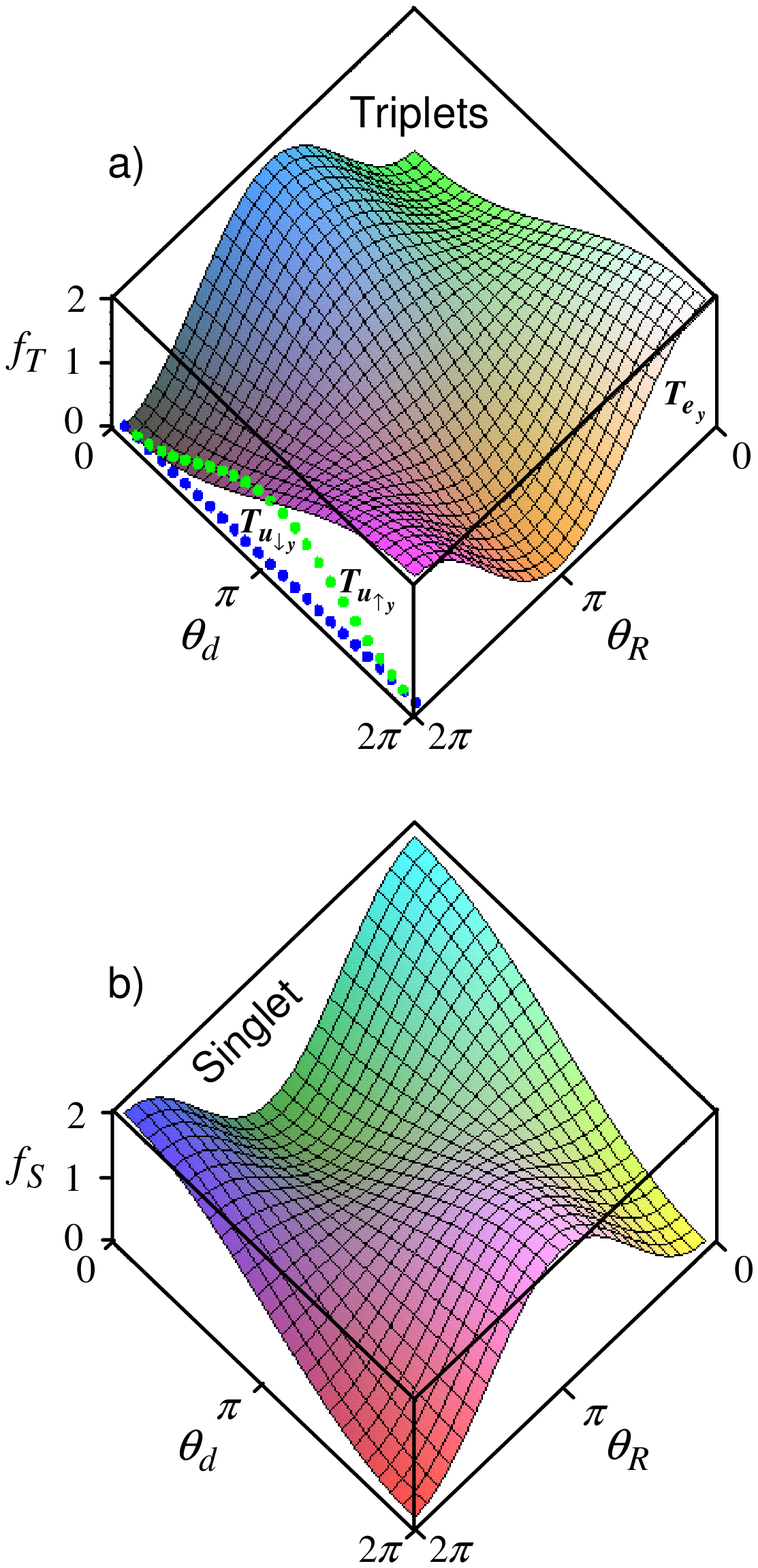, width=0.45\textwidth}
\end{center}
\caption{Fano factors as a function of $\protect\theta _{R}$ and $\protect%
\theta _{d}$ for triplet (a) and singlet (b) pairs (defined along the $y$
axis) with injection energies near the channel anticrossings. We assume the
injected pairs to be initially in channel \textit{a} of the non-Rashba
region in lead 1. In addition to the usual Rashba-induced spin rotation $%
\protect\theta _{R}$, the spin-orbit interaction induces a further
modulation on the Fano factors via the coherent transfer of electrons
between the channels (mixing angle $\protect\theta _{d}$). Interestingly,
the triplets display distinct Rashba\emph{\ s-o} modulations: the
unentangled spin-up pair $T_{u_{\uparrow }y}$ is not sensitive to spin-orbit
effects and shows full anti-bunching, the unentangled spin-down triplet $%
T_{u_{\downarrow }y}$ depends on only $\protect\theta _{d}$ and oscillates
between the full anti-bunching and the classical value for distinguishable
pairs, the entangled triplet $T_{e_{y}}$ displays sizable oscillations
between full bunching and anti-bunching as $\protect\theta _{d}$ and $%
\protect\theta _{R}$ are varied. For the sake of clarity, we have omitted in
(a) the $\protect\theta _{R}$ dependence of Fano factors corresponding to $%
T_{u_{\uparrow }y}$ and $T_{u_{\downarrow }y}$. }
\label{fig3}
\end{figure}

\subsection{Effect of backscattering in the Rashba lead\label{back-body}}

\begin{figure}[th]
\begin{center}
\epsfig{file=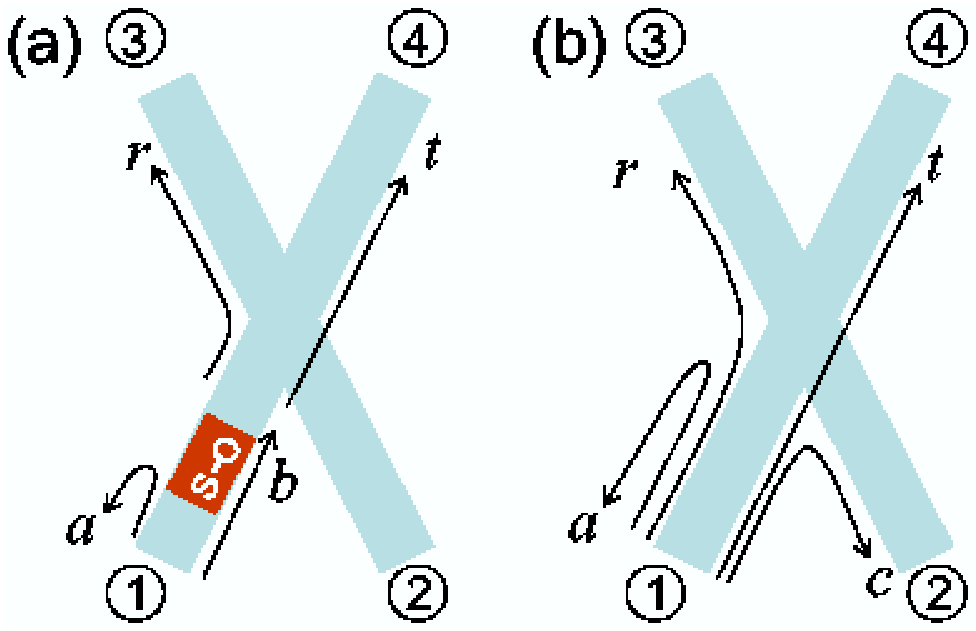, width=0.40\textwidth}
\end{center}
\caption{
Two cases for a beam-splitter with backscattering.
(a) Back-scattering in arm $1$ only,
induced by the {\em s-o} region, with $a$ the backscattering amplitude,
and $r$ and $t$ the original amplitudes of the beam-splitter.
(b) Backscattering present in all leads (amplitude $a$), with also
cross-backscattering ($c$). Here $r$ and $t$ are the modified amplitudes
satisfying the normalization $|a|^2+|c|^2+|r|^2+|t|^2=1$. This case will be
discussed in the Appendix  \ref{app-back-bs}.
}
\label{figback}
\end{figure}

Here we examine how the presence of backscattering in the lead containing
the spin-orbit interaction (lead 1) affects our results for the current
fluctuations. A small backscattering can indeed be produced by the small
band offset between the regions of the lead with and without the Rashba
interaction; see Appendix \ref{app-john} where we calculate explicitly the transmission and
reflection coefficients for a model quantum wire with spin-orbit interaction
of both the Dresselhaus and Rahsba types.
The situation is depicted in Fig. \ref{figback}(a).

We emphasize that here the
{}\textquotedblleft left\textquotedblright\ leads (1,4) and
{}\textquotedblleft right\textquotedblright\ leads (2,3) are not equivalent
anymore because of the presence of the spin-orbit region in lead 1.
This, as we shall see below,
yields that $S_{33}\neq S_{44}$.
 The case in which backscattering is
present in both leads 1 and 2 was first addressed in Ref. {%
\onlinecite{gl-prl2003}}, and is discussed in the Appendix
\ref{app-back-bs} within a simple heuristic picture [see Fig.
\ref{figback}(b)]. For simplicity, we consider here only the case
of uncoupled bands (Sec. II-C).

We take into account the backscattering by adding a tunnel barrier with
reflection amplitude $a$ to lead 1, described by the scattering matrix%
\begin{equation}
\mathbf{s}_{\mathrm{back}}=\left(
\begin{array}{cc}
a & b \\
b & a%
\end{array}%
\right) .  \label{eq86}
\end{equation}%
In order to combine $\mathbf{s}_{\mathrm{back}}$ with the
beam-splitter scattering matrix $\mathbf{s}$ [Eq. (\ref{eq40})],
we match the coefficients of the wavefunction going out of
$\mathbf{s}_{\mathrm{back}}$ with the corresponding ones going in
$\mathbf{s}$, and solve the resulting system of equations
eliminating the intermediate coefficients \cite{datta-book}. We
obtain the modified scattering matrix%
\begin{equation}
\mathbf{s}^{\prime }=\left(
\begin{array}{cccc}
a & 0 & br & bt \\
0 & 0 & t & r \\
br & t & ar^{2} & art \\
bt & r & art & at^{2}%
\end{array}%
\right) .  \label{eq87}
\end{equation}%
The probability $A=|a|^{2}$ of backscattering in lead 1 renormalizes the
transmission probabilities $R=|r|^{2}$ and $T=|t|^{2}$ by the factor $%
|b|^{2}=1-A$. It is interesting to note that the constraints of
unitarity introduce some backscattering in the leads 3 and 4.\
These correspond to the possibility for particles injected into leads
3 or 4 to scatter back into leads 3 and 4 after being reflected by
the tunnel barrier in lead 1. We note that the probabilities $R$
and $T$ are the ones of the original beam-splitter, and as such
satisfy the normalization condition $R=1-T$.

We proceed as before and introduce the spin rotation due to the Rashba
coupling by multiplying $s_{13}$ and $s_{14}$ by $\mathbf{U}_{so}$, Eqs. (%
\ref{eq41}) and (\ref{eq42}). The spin rotation has no effect on $s_{11}$
even if the backscattering occurs after the Rashba region, because the spin
of a particle with reversed momentum rotates in the reversed direction. We
can now use the general formulas
Eqs. (\ref{eq45}) and (\ref{eq46}) to evaluate the
autocorrelation noise in lead 3. Introducing the renormalized probability $%
R^{\prime }=R(1-A)$, we find%
\begin{equation}
S_{33}^{S}(\theta _{so})=\frac{e^{2}}{h\nu }\left[ TR+R^{\prime
}(1-R^{\prime })+2TR^{\prime }\cos (\theta _{so})\delta _{\epsilon
_{1},\epsilon _{2}}\right] ,  \label{backS}
\end{equation}%
\begin{equation}
S_{33}^{T_{e_{z}}}=\frac{e^{2}}{h\nu }\left[ TR+R^{\prime }(1-R^{\prime
})-2TR^{\prime }\delta _{\epsilon _{1},\epsilon _{2}}\right] ,
\label{backTe}
\end{equation}%
\begin{equation}
S_{33}^{T_{u_{\uparrow ,z}}}(\theta _{so})=\frac{e^{2}}{h\nu }\left[
TR+R^{\prime }(1-R^{\prime })-2TR^{\prime }\cos ^{2}(\theta _{so}/2)\delta
_{\epsilon _{1},\epsilon _{2}}\right] ,  \label{backTu}
\end{equation}%
with the average current in lead 3%
\begin{equation}
I_{3}=V\frac{e^{2}}{h}(1-AR).  \label{b-cur3}
\end{equation}

The factor $2TR$ found in the previous case without backscattering, Eqs. (%
\ref{eq53})--(\ref{eq55}), splits into two contributions,
$TR+R^{\prime }(1-R^{\prime })$ with the renormalized probability
$R^{\prime }$. The backscattering adds a contribution related to
the partition noise created by the tunneling barrier. Partition
noise (shot noise) corresponds to fluctuations arising from the
fact the barrier splits randomly the incident electron flow into
transmitted and backscattered flows. On the other hand, the
backscattering reduces the transmission probability $R^{\prime }$
for electrons injected in lead 1, and therefore can decrease the
{}\textquotedblleft beam-splitter noise\textquotedblright\
proportional to $R(1-R)$. For instance, the noise for entangled
triplets with same energies is increased from zero (in the absence
of backscattering) to a finite value. Similarly, the noise for
electrons with different energies is increased in the range $0\leq
A\leq \mathrm{max}\{2-1/T,0\}$, with a maximal increase of
$(T-1/2)^{2}\,(e^{2}/h\nu )$. However, the maximal value for the
noise (obtained in the case of singlets with $\epsilon
_{1}=\epsilon _{2}$ and $\theta _{so}=2n\pi ,\,n\in \mathbb{Z}$)
is not changed, as it corresponds to the maximal value reached in
the case of bunching of bosonic-like particles. In the case of
spin pairs injected along the $
Oy$ axis, the result for the singlet remains the same, while $%
S_{33}^{T_{u,y}}=S_{33}^{T_{e,z}}$ and $S_{33}^{T_{e,y}}(\theta
)=S_{33}^{S}(\pi -\theta _{so})$.

\begin{figure}[tbp]
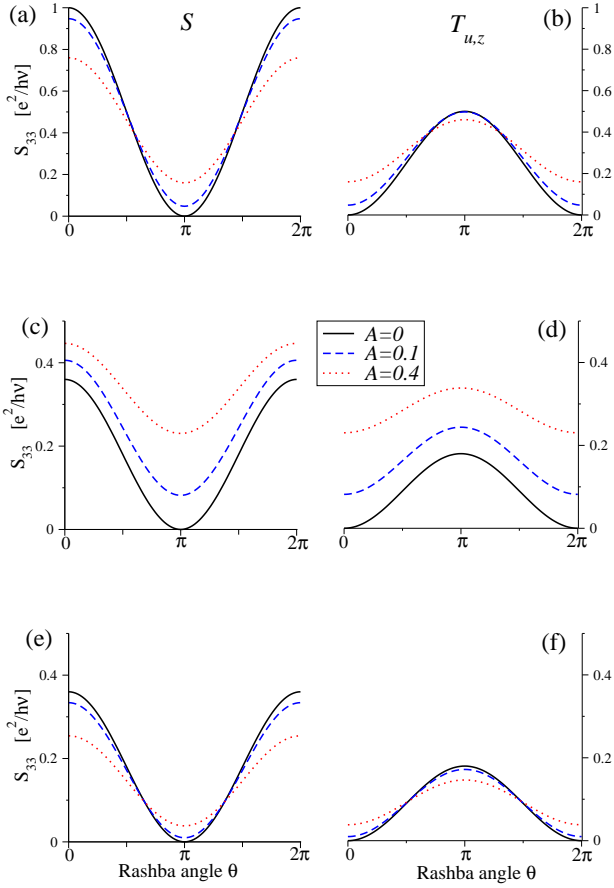

\epsfig{file=t05.eps, width=0.45\textwidth}\vspace{1cm}
\par
\epsfig{file=t09.eps, width=0.45\textwidth} \vspace{1cm}
\par
\epsfig{file=t01.eps, width=0.45\textwidth} \caption{Noise in the
presence of backscattering in the lead with \emph{s-o}
interaction, for backscattering probabilities given by $A=0$, $0.1$, and $%
0.4 $. The transmission probabilities are $R=1-T=0.5$ (a,b); $R=0.9$ (c,d);
and $R=0.1$ (e,f). The left (right) column shows the case of the singlet $S$
(unentangled triplet $T_{u,z}$) with $\protect\epsilon _{1}=\protect\epsilon %
_{2}$.
}
\label{fig4}
\end{figure}

Figure \ref{fig4} shows the auto-correlation noise $S_{33}$ in the presence
of backscattering in lead 1, of the singlet $S$ and for the unentangled
triplet $T_{u,z}$ in the case of equal energies, $\epsilon _{1}=\epsilon
_{2} $. We recall that the noise of the entangled triplet $T_{e,z}$ is
independent of $\theta _{so}$, as are $S$ and $T_{u,z}$ for different
energies. We see that the backscattering can either increase or decrease the
noise, depending on the rotation angle $\theta _{so}$ as well as the
beam-splitter properties (defined by $R$). However, in all cases the
backscattering reduces the visibility (amplitude) of the $\theta _{so}$ oscillations. In
the absence of backscattering ($A=0$), the bunching of singlets for $\theta
_{so}=0$ is maximal ($S_{33}=e^{2}/h\nu $) for a symmetric beam-splitter
(i.e., $T=R=0.5$); this comes from the fact that the randomness is maximal
in this configuration. In the presence of backscattering, the effective
transmission is decreased via $R\rightarrow R^{\prime }(1-A)$, so that one
moves away from this symmetric point, the randomness of the beam-splitter is
reduced, and the noise is suppressed. Similarly, the perfect anti-bunching
seen at $\theta _{so}=\pi $ (and at $\theta _{so}=0$ for the triplets) no
longer happens because of the residual partition noise related to the
backscattering and the noise is given by $S_{33}^{S}(\theta _{so}=\pi
)=S_{33}^{T_{u_{\uparrow ,z}}}(\theta _{so}=0)=S_{33}^{T_{e_{z}}}(\epsilon
_{1}=\epsilon _{2})=(e^{2}/h\nu )AR(1-AR)$ (see Appendix \ref{app-heur},
where we present a heuristic derivation of these results via simple
calculations of expectation values of number operators).

In the case of an asymmetric beam-splitter with a large value of $R>3/(4-A)$
(e.g., $R=0.9$), the noise for the singlets is increased by the
backscattering for all $\theta _{so}$. For $\theta _{so}=0$, one can
understand this by noting that the backscattering reduces the value of $%
R^{\prime }$ which therefore gets closer to the symmetric value of $0.5$
related to the maximal noise. On the other, in the case of small $R$ (e.g., $%
R=0.1$), the reduction of $R^{\prime }$ moves even further away from the
maximal $0.5$ value, and the noise is reduced for small $\theta _{so}$.
Around $\theta _{so}=\pi $ it increases from zero, because of the additional
partition noise. For the triplet, the maximal value is reached at $\theta
=\pi $, where electrons reach the beam-splitter with opposite spins, and
hence behave classically. The value $S_{33}=0.5e^{2}/h\nu $ at $\theta
_{so}=\pi $ for $A=0$ is modified in the presence of backscattering; it is
then increased when $R>1/(2-A)$.

For the cross-correlations of the noise between leads 3 and 4, we find%
\begin{equation}
S_{34}^{S}(\theta )=-\frac{e^{2}}{h\nu }\left[ TR+T^{\prime }R^{\prime
}+2TR^{\prime }\cos (\theta )\delta _{\epsilon _{1},\epsilon _{2}}\right] ,
\label{backS-auto}
\end{equation}%
\begin{equation}
S_{34}^{T_{e_{z}}}=-\frac{e^{2}}{h\nu }\left[ TR+T^{\prime }R^{\prime
}-2TR^{\prime }\delta _{\epsilon _{1},\epsilon _{2}}\right] ,
\label{backTe-auto}
\end{equation}%
\begin{equation}
S_{34}^{T_{u_{\uparrow ,z}}}(\theta )=-\frac{e^{2}}{h\nu }\left[
TR+T^{\prime }R^{\prime }-2TR^{\prime }\cos ^{2}(\theta /2)\delta _{\epsilon
_{1},\epsilon _{2}}\right] ,  \label{backTu-auto}
\end{equation}%
with $T^{\prime }=T(1-A)$. These expressions are very similar to the ones
for the auto-correlations, Eqs. (\ref{backS-auto})--(\ref{backTu-auto}). The
main difference is the negative sign of these correlations, which comes from
the fact that $S_{33}+S_{34}=0$ when $A=0$. The other difference is that the
term $R^{\prime }(1-R^{\prime })$ is replaced by $R^{\prime }T^{\prime }$,
which is easily understandable since the cross-correlation involves $\langle
n_{3}\rangle \langle n_{4}\rangle =R^{\prime }T^{\prime }$ for a particle
injected in lead 1, while the auto-correlation involves $\langle
n_{3}\rangle (1-\langle n_{3}\rangle )=R^{\prime }(1-R^{\prime })$, see
Appendix \ref{app-heur}.

\subsection{Spin-polarized case}

We now turn to the case of injection from a spin-polarized Fermi
liquid lead with a continuous energy spectrum. We start from the
standard expression for the noise \cite{butt}, which corresponds
to the continuous version of Eqs.~(\ref{eq37}-\ref{eq39})
\begin{eqnarray}
S_{\alpha \beta } &=&\frac{e^{2}}{h}\mathrm{Re}\int d\epsilon \,\sum_{\gamma
\delta uv}A_{\gamma \delta }^{uv}(\alpha ,\epsilon ,\epsilon )A_{\delta
\gamma }^{vu}(\beta ,\epsilon ,\epsilon )\times   \notag \\
&&f_{\gamma u}(\epsilon )\left[ 1-f_{\delta v}(\epsilon )\right] ,
\label{S-Fermi}
\end{eqnarray}%
where $f_{\alpha u}(\epsilon )=\langle a_{\alpha u}^{\dagger
}a_{\alpha u}\rangle $ and $u=(\sigma ,n)$. Here, again, $\sigma
=\uparrow ,\downarrow $ denotes the spin components and $n=a,b$
the band index. At zero temperature, for the scattering matrix of
the beamsplitter with a two-band spin-orbit active region in one
of its incoming arms (lead 1), Eqs.~(\ref{eq41}) and (\ref{eq42}),
we obtain for the shot noise in the outgoing leads (say, lead 3),
\begin{eqnarray}
S_{33} &=&\frac{e^{2}}{h}\int d\epsilon \,T\left( 1-T\right) \sum_{\sigma
\sigma ^{\prime }nm}|U_{\sigma n,\sigma ^{\prime }m}^{cc}|^{2}
\big\{f_{1\sigma n}(\varepsilon )[1-  \notag \\
&&
f_{2\sigma ^{\prime }m}(\varepsilon )]+ f_{2\sigma n}( \varepsilon)
\left[1-f_{1\sigma ^{\prime }m}(\varepsilon )\right] \big\},
\label{long-result}
\end{eqnarray}%
where $U_{\sigma n,\sigma ^{\prime }m}^{cc}$ denote the matrix elements of  $%
\mathbf{U}_{so}^{cc}$ given in Eq.~(\ref{eq36}).

\subsubsection{Injection into one subband}

We consider a (non-equilibrium) spin-polarized injection into leads $1$
and $2$ in subband $a$ only, which we model with
\begin{eqnarray}
\mu _{1\uparrow a}=\mu _{2\uparrow a} &=&\varepsilon _{F}+eV,  \label{mu1} \\
\mu _{1\downarrow a}=\mu _{2\downarrow a} &=&\varepsilon _{F}+\frac{1-p}{1+p}%
eV,  \label{mu2} \\
\mu _{1\sigma b}=\mu _{2\sigma b} &=&\varepsilon _{F},\quad \quad (\sigma
=\uparrow ,\downarrow ),  \label{mu3} \\
\mu _{3\sigma n}=\mu _{4\sigma n} &=&\varepsilon _{F},\quad \quad (\sigma
=\uparrow ,\downarrow ;n=a,b).  \label{mu4}
\end{eqnarray}
The degree of spin-polarized is controlled by the parameter $0\leq p \leq1$.  For full
polarization $p=1$ there is no voltage drop for spin-down electrons in Eq. (\ref{mu2}),
which therefore do not contribute to transport.
Note that here we only inject electrons in channel $a$, see Eqs. (\ref{mu3}) and (\ref{mu4}).
We then obtain for the current in the outgoing leads
\begin{equation}
I_{3}=I_4=\frac{e^{2}}{h}V\frac{2}{1+p},  \label{current2}
\end{equation}%
which gets halved for full polarization, $p=1$, as compared to the unpolarized case $p=0$.
For the auto-correlation noise we find
\begin{eqnarray}
S_{33} &=&\frac{2e^{2}}{h}T(1-T)\left[ \frac{2p}{1+p}|U_{\uparrow
a;\downarrow a}^{cc}|^{2}+|U_{\uparrow a;\downarrow
b}^{cc}|^{2}+|U_{\uparrow a;\uparrow b}^{cc}|^{2}\right.   \notag \\
&&\left.+ \frac{1-p}{1+p}(|U_{\downarrow a;\uparrow
b}^{cc}|^{2}+|U_{\downarrow a;\downarrow b}^{cc}|^{2})\right] eV.
\label{noise2}
\end{eqnarray}%
Note that $S_{33}=S_{44}=-S_{34}$.  We define the reduced Fano
factor $f_{p}\equiv F/T(1-T)\equiv S_{33}/2eI_{3}T(1-T)$, and
using Eq.~(\ref{eq36}) and $|\xi |^{2}=1$ we find
\begin{equation}
f_{p}=\frac{p}{4}\left( \cos ^{2}\frac{\theta _{d}}{2}+1-2\cos \frac{\theta
_{d}}{2}\cos \theta _{so}\right) +\frac{1}{2}\sin ^{2}\frac{\theta _{d}}{2}.
\label{fano2}
\end{equation}%
This agrees with Ref.~\onlinecite{bel} for $\beta =0$ and $p=1$.
Notice, however, that if
$p<1$ there is an additional correction, i.e., the last term in Eq.~(\ref{fano2}), which
is independent of $p$ and thus there is a small noise power even for the unpolarized case
$p=0$.
This is related to the presence of the additional empty channel $b$, which allows
for fluctations in the outgoing leads in the case of finite
 {\it s-o} induced interband coupling, $\theta_d \neq 0$.
In the strictly 1D case (i.e., single channel), this effect disappears as
$\theta _{d}=0$ and Eq.(\ref{fano2}) reduces to the expression
\begin{equation}
f_{p}=\frac{p}{2}\left( 1-\cos \theta _{so}\right) =p\sin ^{2}\frac{\theta
_{so}}{2}.  \label{rfano2a}
\end{equation}
This effect also vanishes in the case where both bands are
occupied, studied in the next section.

\subsubsection{Injection into both leads}

\label{bothleads}

We now consider the more realistic case of injection into both
subbands $a$ and $b$ and model it with
\begin{eqnarray}
\mu _{1\uparrow }=\mu _{2\uparrow } &=&\varepsilon _{F}+eV,  \label{mu1b} \\
\mu _{1\downarrow }=\mu _{2\downarrow } &=&\varepsilon _{F}+\frac{1-p}{1+p}%
eV,  \label{mu2b} \\
\mu _{3\sigma }=\mu _{4\sigma } &=&\varepsilon _{F},\quad \quad (\sigma
=\uparrow ,\downarrow ),  \label{mu3b}
\end{eqnarray}%
where $\mu_{i,\sigma,a}=\mu_{i,\sigma,b}=\mu_{i,\sigma}$.
We obtain the current
\begin{equation}
I_{3}=I_4=\frac{2e^{2}}{h}V\frac{2}{1+p},  \label{current3}
\end{equation}%
which is twice the current in the case of injection into a
single channel given by Eq.(\ref{current2}).
The Fano factor reads
\begin{equation}
F=T(1-T)\frac{p}{2}\sum_{n,m=a,b}\left\vert U^{cc}_{\uparrow n,\downarrow
m}\right\vert ^{2}.  \label{fano2e}
\end{equation}%
Using Eq. (\ref{eq36}), we obtain the reduced Fano factor $f_{p}\equiv
F/T(1-T)$,
\begin{equation}
f_{p}=\frac{p}{2}\left( 1-\cos \frac{\theta _{d}}{2}\cos \theta _{so}\right).  \label{fano2er}
\end{equation}%
This vanishes in the case of unpolarized injection, $p=0$, showing
that the interband coupling in lead 1 in itself does not give rise
to additional noise for two filled channels. For an uncoupled
channel ($\theta _{d}=0$), $f_{p}$ yields the usual form valid for
single channels, i.e. Eq.~(\ref{rfano2a}). The above results were
derived for a quantization axis along the $z$ direction. For other
directions, one can derive similar but distinct formulas. Finally,
we note that measuring such Fano factors enables one to quantify the
degree of spin-polarization $p$ of the reservoirs along different
quantization axis.

\subsection{Coherent injection into multiple discrete states}
\label{multiple-inj}

Here we generalize our previous results by studying the injection
of electrons into leads with a discrete spectrum. We show that
even in the case of injection into many different discrete states
of the lead, we can observe two-particle coherence, e.g., bunching
and antibunching, and thus detect entangled states.  We will
identify six asymptotic regimes separated by the relative
magnitude of the level spacing $\delta$ in the leads, the energy
mismatch $\Delta$ of the injected electrons, and the energy
broadening $\gamma$ of the injected electrons.  It turns out that
in four of the six regimes, we obtain (asymptotically) full
two-particle interference.  Only in two regimes, namely when the
energy mismatch $\Delta$ exceeds both $\delta$ and $\gamma$, we
obtain no two-particle interference.

Electron injection into leads 1 and 2 is assumed to be coherent
but with a finite width $\gamma$ in energy, such that several
energy levels of the lead can be filled.  This represents a
generalization of both the injection into single discrete levels
as discussed in this paper and in earlier works \cite{ble,egd} and
of recent investigations of the problem in the continuum limit\cite{Samu04}
which will be a special case of the following discussion.
\begin{figure}
\begin{center}
\epsfig{file=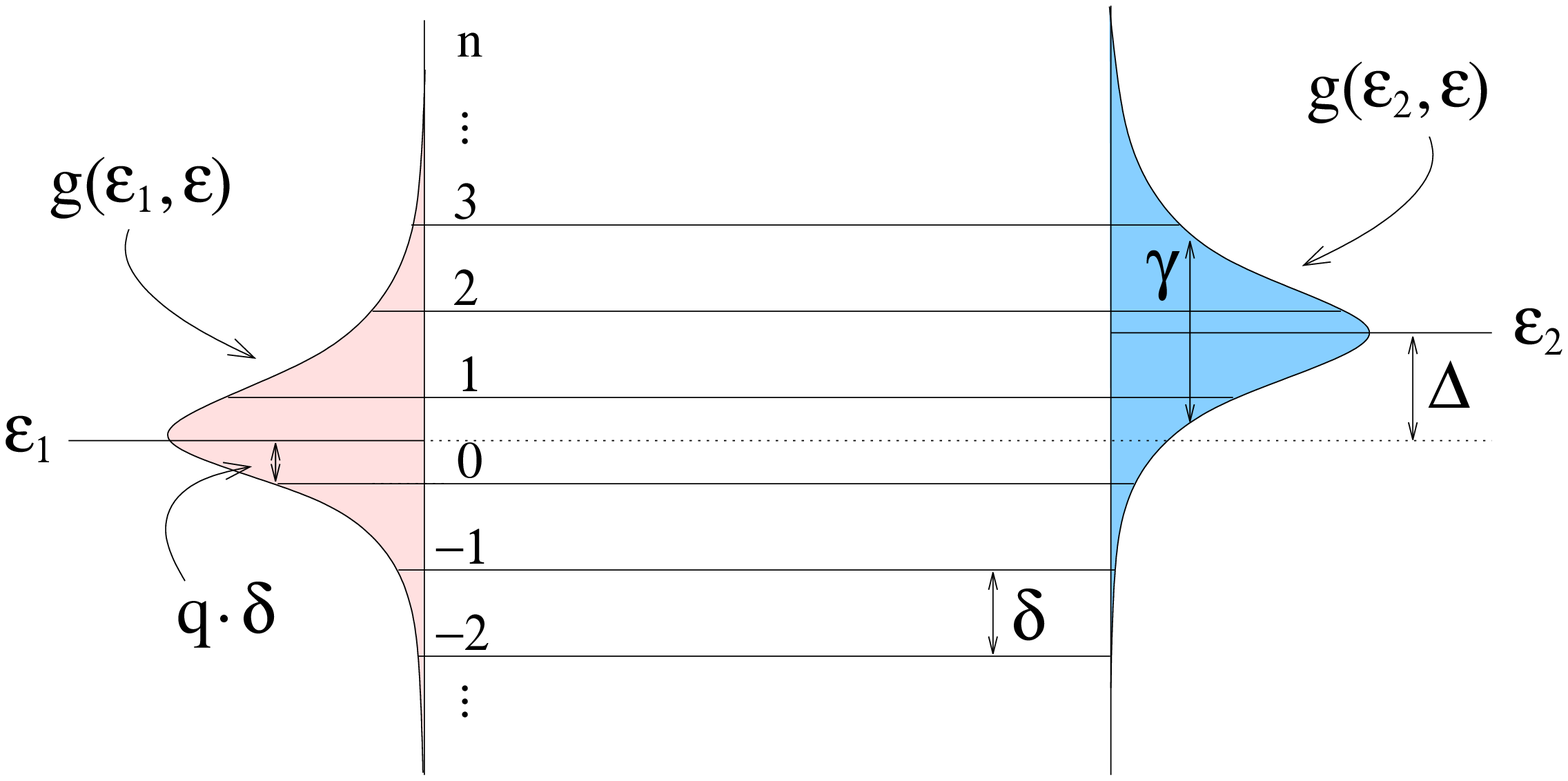, width=0.45\textwidth}
\end{center}
\caption{\label{discrete} Injection of electrons into equidistant
discrete lead states $n=0, \pm 1, \pm 2,\ldots$ with spacing
$\delta$ at center energies $\varepsilon_{1,2}$ separated by
$\Delta=\varepsilon_{2}-\varepsilon_{1}$ with distributions
$g(\varepsilon_{1,2},\varepsilon)$ of width $\gamma$. The
distributions $g(\varepsilon_{1,2},\varepsilon)$ are not drawn to
scale, as their normalizations depend on $q$.}
\end{figure}

We assume here injection into leads with equidistant levels
$\varepsilon_n = n\delta + q\delta$, where $n=0,\pm 1,\pm
2,\ldots$ and $0\le q<1$ is a fixed fractional offset.  The
assumption of equidistance simplifies our calculations because it
allows us to perform the discrete sums explicitly (Appendix \ref{sums}).
Also, we choose our discrete index $n$ to run from $-\infty$ to
$+\infty$, this is done for convenience and merely means that the
spectrum is well described within an energy band a few times
$\gamma$ wide. We do not expect our results and conclusions to be
significantly altered in cases of non-equidistant energy levels in
the leads.

\begin{figure}
\begin{center}
\epsfig{file=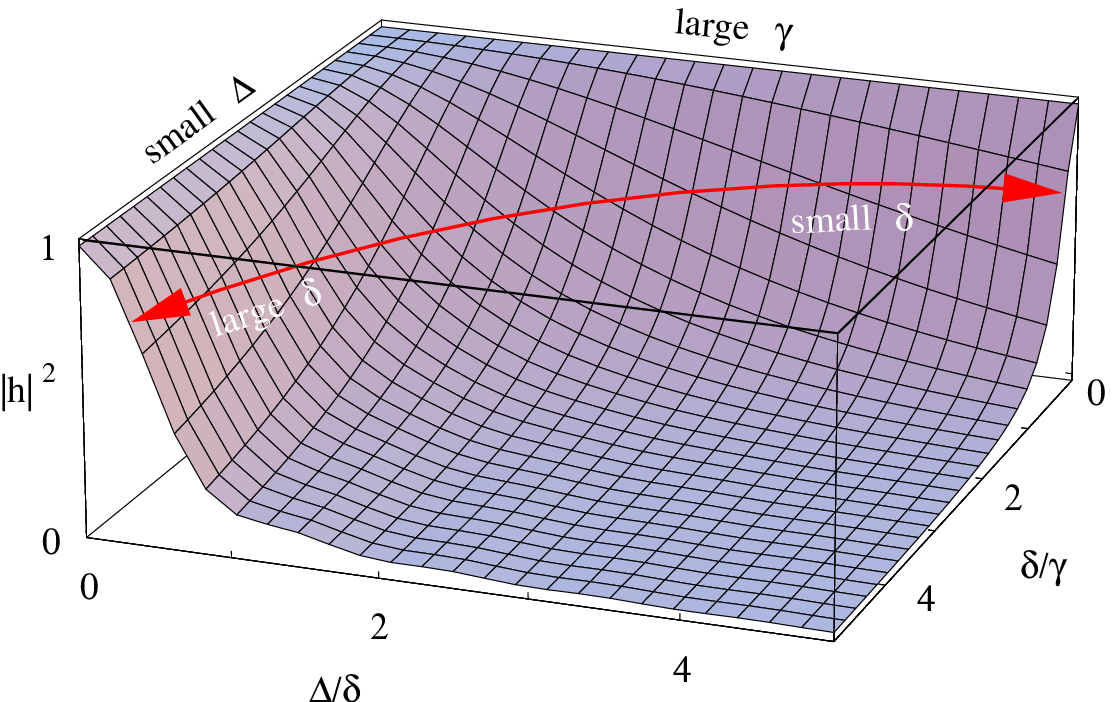,width=0.45\textwidth}
\end{center}
\caption{\label{hfunc}The function $|h(\Delta,\delta,\gamma)|^2$
as expressed in Eq.~(\ref{E10}), plotted versus the dimensionless
quantities $\Delta/\delta$ and $\delta/\gamma$, where
$\Delta=\varepsilon_2-\varepsilon_1$ denotes the mean energy
difference between the two injected electrons, $\gamma$ the width
of their energy distributions, and $\delta$ the level spacing in
the leads. The case of matching energies
$\varepsilon_1\approx\varepsilon_2$ or $\Delta\ll \delta,\gamma$
(Sec.~\ref{matching}) corresponds to the edge indicated as ``small
$\Delta$''; here we find $|h|^2=1$, irrespective of the ratio
$\gamma/\delta$. The case of a broad energy distribution (fast
injection) $\gamma\gg\Delta,\delta$ (Sec.~\ref{largestscale})
corresponds to the edge indicated as ``large $\gamma$''; along
this edge $|h|^2=1$, regardless of $\Delta/\delta$.  The red arrow
follows a line of constant $\gamma$ and $\Delta$ with variable
$\delta$. For large $\delta$, we find again $|h|^2=1$ for all
$\Delta/\gamma$, whereas for small $\delta$ (continuum limit,
Sec.~\ref{continuum}), the limit $|h|^2=0$ or $|h|^2=1$ is
reached, depending on whether $\Delta/\gamma\gg 1$ or
$\Delta/\gamma\ll 1$.}
\end{figure}

Injection of an electron with spin $\sigma=\uparrow,\downarrow$
into lead $\alpha$ and centered about the energy $\varepsilon$ is
described by the creation operator
\begin{equation}
  \label{E1}
  c^\dagger_{\alpha \sigma}(\varepsilon) = \sum_{n=-\infty}^\infty
     g(\varepsilon,\varepsilon_n)
     a^\dagger _{\alpha \sigma} (\varepsilon_n),
\end{equation}
where $a^\dagger _{\alpha \sigma} (\varepsilon_n)$ creates an
electron with the sharp energy $\varepsilon_n$, cf.\
Eq.~(\ref{eq37}) and below. The weight function $g$ will be
assumed to have the Breit-Wigner form
\begin{equation}
  g(\varepsilon,\varepsilon^\prime) =
   \frac{g_0(\varepsilon)}{\varepsilon -\varepsilon^\prime +i\gamma},
\label{E2}
\end{equation}
with the normalization condition
$\sum_{n=-\infty}^\infty|g(\varepsilon,\varepsilon^\prime)|^2=1$.
Unlike for the case of weak tunneling from, e.g., a quantum dot
\cite{recher-suk-loss}, we assume here that the single-electron
states are filled with probability $1$, but with uncertain energy.
This can be achieved with time-dependent tunnel barriers. Note
that for widely spaced levels $\delta\gg\gamma$ and away from
symmetric injection into two adjacent levels ($q=1/2$), injection
thus takes place into the nearest level with probability $\approx
1$. The two-particle injected states we are interested in are
\begin{equation}
  \left.\begin{array}{c}|S\rangle\\|T_{e_i}\rangle\end{array}\right\}
    = \frac{1}{\sqrt{2}}\left(
       c^\dagger_{1\uparrow}(\varepsilon_1)
       c^\dagger_{2\downarrow}(\varepsilon_2) \mp
       c^\dagger_{1\downarrow}(\varepsilon_1)
       c^\dagger_{2\uparrow}(\varepsilon_2)
      \right),
\label{E3}
\end{equation}
the singlet and entangled triplet states with single-particle
energies centered around $\varepsilon_1$ and $\varepsilon_2$ and
smeared over a width $\gamma$. Using (\ref{E1}), these states can
be expressed in terms of the states with sharp energies
Eq.~(\ref{eq43}) as $|S,T_{e_i}\rangle =
\sum_{\varepsilon_1^\prime,\varepsilon_2^\prime}
   g(\varepsilon_1,\varepsilon_1^\prime)g(\varepsilon_2,\varepsilon_2^\prime)
   |S,T_{e_i}\rangle_{\varepsilon_1^\prime,\varepsilon_2^\prime}$.
Using the normalization condition, we find that the average
current in the outgoing leads of the beamsplitter is unaffected by
the spread in energy, i.e., $I_3=I_4=-e/h\nu$. For the Fano factor
$F=S_{33}/2eI_3$, we find
\begin{equation}
  \label{E5}
  F_{S,T_{e}} = T(1-T)\left( 1\pm |h(\varepsilon_1,\varepsilon_2)|^2\right).  
\end{equation}
The discrete overlap function $h$ is given by
\begin{equation}
  \label{E6}
  h(\varepsilon_1,\varepsilon_2) =
   \sum_\varepsilon g(\varepsilon_2,\varepsilon)^*
                    g(\varepsilon_1,\varepsilon).
\end{equation}
The discrete summations required to evaluate the function $h$ are
carried out in Appendix \ref{sums} and lead to the main result of
this section ($\Delta\equiv \varepsilon_2-\varepsilon_1$),
\begin{equation}
  \label{E10}
  |h|^2 = \frac{\gamma^2}{(\Delta/2)^2+\gamma^2}
          \frac{\cosh^2 (2\pi\gamma/\delta) - \cos^2(\pi\Delta/\delta)}
          {\sinh^2 (2\pi\gamma/\delta)}.
\end{equation}
Note that $|h|^2$ appears in Eq. (\ref{E5}) for the Fano factor
and generalizes the Kronecker delta for the case of injection into
a single level. The function $0<|h(\Delta,\delta,\gamma)|^2\le 1$
determines the amount of two-particle interference that can be
observed (see Fig.~\ref{hfunc}). Note that as a consequence of the
normalization of $g(\varepsilon,\varepsilon_n)$, the overlap
function $|h|^2$ is \textit{independent} of the energy offset $q$
(see Appendix \ref{sums}). For $|h|^2=1$, full bunching and
antibunching of singlets and entangled triplets can be expected
which is ideal for the purpose of discriminating these entangled
states from unentangled two-particle states.  If $|h|^2\rightarrow
0$, then no interference, i.e., no bunching or antibunching will
be observed because the two wavepackets centered around
$\varepsilon_1$ and $\varepsilon_2$ have no overlap in energy
space.

Since $|h(\Delta,\delta,\gamma)|^2$ depends on three parameters
with the dimension of an energy, there are $3!=6$ different
asymptotic regimes. We will now discuss all six cases, grouped
into three sections with $\Delta$, $\gamma$, and $\delta$ as the
smallest energy, respectively. We also note that since $|h|^2$ is
a dimensionless number, it is in fact only a function of
\textit{two} dimensionless energy ratios, e.g., $\Delta/\delta$
and $\delta/\gamma$, (see Fig.~\ref{hfunc}) as can easily be seen
from Eq.~(\ref{E10}).

\subsubsection{Matching energies $\varepsilon_1 = \varepsilon_2$
($\Delta\ll\gamma,\delta$)}\label{matching}

If $\Delta =\varepsilon_1 - \varepsilon_2 = 0$, then $|h|^2=1$ for
arbitrary values of the other three parameters $\delta$, $\gamma$,
and $q$. In other words, if the two electrons are injected with
energy distributions whose centers coincide, then the interference
is always maximal.  This result persists for finite $\gamma$ and
$\delta$ as long as $\Delta \ll \gamma, \delta$; in this case, the
correction is $O(\Delta^2)$.  For $\Delta\ll\gamma\ll\delta$, we
find $|h|^2 \simeq 1-\pi^2\Delta^2/3\delta^2$. In the case
$\Delta\ll\delta\ll\gamma$, the correction is $|h|^2 \simeq
1-\Delta^2/4\gamma^2$.

\subsubsection{Sharp energies $\gamma \ll \delta, \Delta$}
\label{sharp} In the limit of sharp energies, $\gamma \ll \delta,
\Delta$, we obtain
\begin{equation}
  \label{E11}
  |h|^2 = \frac{\delta^2}{\pi^2 \Delta^2}\sin^2\pi\frac{\Delta}{\delta}
               +O(\gamma^2)
\end{equation}
from Eq.~(\ref{E10}). The limit $\gamma \ll \delta \ll \Delta$
describes the situation where the two electrons are injected into
two different discrete states and thus $|h|^2 \rightarrow 0$. In
the other limit, $\gamma \ll \Delta \ll \delta$, the two electrons
are injected into the same discrete level and $|h|^2 =
1-\pi^2\Delta^2/3\delta^2+O(\Delta^4/\delta^4,\gamma^2)$. The
limit $\gamma \ll \delta, \Delta$ discussed here describes
injection into single discrete levels discussed throughout this
paper and in earlier works \cite{ble,egd,gl-prl2003} where
$\varepsilon_{1,2}$ are multiples of $\delta$ and $|h|^2
\rightarrow \delta_{\varepsilon_1,\varepsilon_2}$ is given by a
Kronecker delta.

\subsubsection{Continuum limit $\delta\ll\Delta,\gamma$}
\label{continuum} In the continuum limit, i.e., the asymptotic
case where the single-electron level spacing in the leads becomes
small compared to all other relevant energy scales, we find a
Lorentzian \cite{Samu04}
\begin{equation}
  \label{E12}
  |h|^2 \simeq \frac{\gamma^2}{(\Delta/2)^2 + \gamma^2},
\end{equation}
with the usual special limits $|h|^2 \rightarrow 0$ for
$\delta\ll\gamma\ll \Delta$ and $|h|^2 \rightarrow 1$ for
$\delta\ll\Delta\ll \gamma$.

\subsubsection{Discussion in terms of largest energy scale}
\label{largestscale} Above, we have discussed all six asymptotic
cases, grouped according to the smallest energy scale.
Alternatively, we can discuss the asymptotic regimes characterized
by the largest energy scale.  In the \textit{detuned case}
$\Delta\gg\delta,\gamma$, we never see two-particle interference
since $|h|^2\simeq (4\gamma^2/\Delta^2) f(\Delta;\delta,\gamma)
\rightarrow 0$ where $f(\Delta;\delta,\gamma)$ is a function which
is bounded for fixed $\delta$ and $\gamma$. In the case of a
\textit{wide distribution}, as e.g.\ effected by a \textit{fast
injection} into the lead with injection time $\hbar/\gamma$, we
are in the regime $\gamma\gg\delta,\Delta$, and we always find
that $|h|^2$ exponentially approaches $1$. In the
\textit{single-level case} $\delta\gg\gamma,\Delta$, there is only
one level to fill and we obtain $|h|^2\rightarrow 1$ irrespective
of the relative magnitude of $\Delta$ and $\gamma$.

\section{Summary}

We have carried out a thorough study of current and noise for spin-polarized and spin-entangled electrons in
a beam-splitter geometry, including a local spin-orbit interaction (Rashba and Dresselhaus) in one of the
incoming arms.
We have considered incoming leads with one or two channels, as well as backscattering effects.
The channels can be coupled via the {\it s-o} interaction for incoming energies near the band-crossing.
We have found that the spin-orbit interaction is a useful mechanism to coherently rotate spin states.
Such rotation can be used to modulate noise signals, thus providing unique signatures of spin-polarization
and spin-entanglement.

For spin-polarized electrons, noise measurements can give a direct
measure of the degree of polarization along different directions.
For electron pairs, the coupling between the channels can play an
important role. For pairs with incoming energies near the
band-crossing injected into one of the channels, we find an
additional modulation due to the coherent transfer of electrons
between the two channels. In this case, noise measurement allows
us to distinguish all the different triplets states defined along
the $y$ direction, in addition to the singlet. Furthermore, for
equal strengths the combined effect of the Rashba and Dresselhaus
interactions can partially cancel out. In this case, the spin and
orbital degrees of freedom are separable, the interband coupling
essentially disappears, and the propagation of spin states is
robust against scattering off non-magnetic impurities.

We have also considered the influence of backscattering in the
beam-splitter with a single channel. The main effect is an
additional contribution related to the partition noise due to the
tunnel barrier describing the backscattering. This reduces the
visibility of the oscillations in the shot-noise as the {\em s-o}
rotation angle $\theta_{so}$ is varied. It also reduces the
maximal noise value found for perfect antibunching of singlets.

We have generalized earlier results for the shot noise of
entangled electrons by allowing the injection of wave packets,
i.e., coherent superpositions of discrete momentum eigenstates
(plane waves). We have found a general analytical formula for the
two-particle interference visibility $|h|^2$ in terms of all three
relevant energy  scales $\Delta$, $\gamma$, and $\delta$.  Our new
result contains and generalizes both the discrete single-level
case and the continuum case.

Finally, we have developed a simple heuristic picture for the
noise based on number operators in the different leads and the
relevant transmission and reflection probability amplitudes in the
beamsplitter. Within this picture we can more intuitively
re-derive some of the formulas for the noise -- previously derived
within the rigorous scattering formalism (Sec. III) --  in the
presence of spin-orbit interaction and backscattering in the incoming leads.

\acknowledgements

The authors acknowledge useful discussions with S. Erlingsson. This
work was supported by NCCR Nanoscale Science, CNPq,
EU-Spintronics, Swiss NSF, DARPA, ARO, and ONR.

\appendix

\section{Boundary conditions at $x=L$}

Here we show in some detail that both the wave function and the
velocity operator acting on it are continuous at the exit $(x=L)$
of the \textit{s-o} region in lead 1.

\subsection{Continuity of \ $\Psi (x,y)$.}

The continuity of the wavefunction at $x=L$ is trivially satisfied by the
two-channel state
\begin{eqnarray}
\Psi (x,y) &=&\left(
\begin{array}{c}
\cos \left( \theta _{d}/2\right) e^{-i\theta _{so}/2}+e^{i\theta _{so}/2} \\
\xi \left[ \cos \left( \theta _{d}/2\right) e^{-i\theta
_{so}/2}-e^{i\theta _{so}/2}\right]
\end{array}%
\right) \frac{1}{2}e^{i\left( k_{c}+k_{so}\right) x}\phi _{a}(y)+  \notag \\
&&\left(
\begin{array}{c}
-i\sin \left( \theta _{d}/2\right) e^{-i\theta _{so}/2} \\
i\xi \sin \left( \theta _{d}/2\right) e^{-i\theta _{so}/2}%
\end{array}%
\right) \frac{1}{2}e^{i\left( k_{c}-k_{so}\right) x}\phi _{b}(y),  \label{A1}
\end{eqnarray}%
describing the electron state within the \emph{s-o} region, $0\leq x\leq L$,
and the state
\begin{equation}
\Phi (x,y)=\left(
\begin{array}{c}
A \\
B%
\end{array}%
\right) \frac{1}{2}e^{i\left( k_{c}+k_{so}\right) x}\phi _{a}(y)+\left(
\begin{array}{c}
C \\
D%
\end{array}%
\right) \frac{1}{2}e^{i\left( k_{c}-k_{so}\right) x}\phi _{b}(y),  \label{A2}
\end{equation}%
valid for $x\geq L$ in lead 1, if we choose $A$, $B$,\ $C$, and $D$, equal
to the corresponding components of $\Psi (L,y)$.

\subsection{Continuity of the current flow.}

The continuity of the (non-diagonal) velocity operator\cite{bc}
\begin{equation}
\hat{v}_{so}=\left(
\begin{array}{cccc}
\frac{\hbar }{im}\partial _{x} & \frac{\beta +i\alpha }{\hbar } & 0 & 0 \\
\frac{\beta -i\alpha }{\hbar } & \frac{\hbar }{im}\partial _{x} & 0 & 0 \\
0 & 0 & \frac{\hbar }{im}\partial _{x} & \frac{\beta +i\alpha }{\hbar } \\
0 & 0 & \frac{\beta -i\alpha }{\hbar } & \frac{\hbar }{im}\partial _{x}%
\end{array}%
\right)  \label{A3}
\end{equation}%
acting on the wavefuctions at $x=L$
\begin{equation}
\hat{v}_{so}\Psi (x,y)|_{x\rightarrow L^{-}}=\hat{v}_{so}\Phi
(x,y)|_{x\rightarrow L^{+}},  \label{A4}
\end{equation}%
assures current conservation. The left-hand side of (\ref{A4}) yields

\begin{widetext}
\begin{eqnarray}
\hat{v}_{so}\Psi (x,y)|_{x\rightarrow L^{-}}= &&\frac{\hbar }{4m}\left[
\left(
\begin{array}{c}
\left( k_{c}-\Delta /2+k_{so}\right) e^{-i\Delta L/2} \\
\xi \left( k_{c}-\Delta /2+k_{so}\right) e^{-i\Delta L/2} \\
\left( k_{c}-\Delta /2-k_{so}\right) e^{-i\Delta L/2} \\
-\xi \left( k_{c}-\Delta /2-k_{so}\right) e^{-i\Delta L/2}%
\end{array}%
\right) +\left(
\begin{array}{c}
\left( k_{c}+\Delta /2+k_{so}\right) e^{i\Delta L/2} \\
\xi \left( k_{c}+\Delta /2+k_{so}\right) e^{i\Delta L/2} \\
-\left( k_{c}+\Delta /2-k_{so}\right) e^{i\Delta L/2} \\
\xi \left( k_{c}+\Delta /2-k_{so}\right) e^{i\Delta L/2}%
\end{array}%
\right) \right] e^{ik_{c}L}+  \notag \\
&&\frac{\hbar }{2m}\left(
\begin{array}{c}
\left( k_{2}-k_{so}\right) e^{ik_{2}L} \\
-\xi \left( k_{2}-k_{so}\right) e^{ik_{2}L} \\
0 \\
0%
\end{array}%
\right) ,  \label{A5}
\end{eqnarray}
\end{widetext}

or

\begin{widetext}%
\begin{eqnarray}
\hat{v}_{so}\Psi (x,y)|_{x\rightarrow L^{-}} &=&\frac{\hbar }{2m}\left(
\begin{array}{c}
\left( k_{c}+k_{so}\right) \left[ \left( \cos (\theta _{d}/2)+i\frac{\Delta
}{\left( k_{c}+k_{so}\right) }\sin (\theta _{d}/2)\right) e^{-i\theta
_{so}/2}+e^{i\theta _{so}/2}\right]  \\
\xi \left( k_{c}+k_{so}\right) \left[ \left( \cos (\theta _{d}/2)+i\frac{%
\Delta }{\left( k_{c}+k_{so}\right) }\sin (\theta _{d}/2)\right) e^{-i\theta
_{so}/2}-e^{i\theta _{so}/2}\right]
\end{array}%
\right) e^{i\left( k_{c}+k_{so}\right) L}\phi _{a}(y)+  \notag \\
&&\frac{\hbar }{2m}\left(
\begin{array}{c}
-i\left( k_{c}-k_{so}\right) \left( \sin (\theta _{d}/2)+i\frac{\Delta }{%
\left( k_{c}-k_{so}\right) }\cos (\theta _{d}/2)\right) e^{i\theta _{so}/2}
\\
-i\xi \left( k_{c}-k_{so}\right) \left( \sin (\theta _{d}/2)+i\frac{%
\Delta }{\xi \left( k_{c}-k_{so}\right) }\cos (\theta
_{d}/2)\right)
e^{i\theta _{so}/2}%
\end{array}%
\right) e^{i\left( k_{c}-k_{so}\right) L}\phi _{b}(y).  \label{A6}
\end{eqnarray}%
\end{widetext}
where we used $k_{2}-k_{so}=k_{c}+k_{so}.$ On the other hand, it is
straightforward to show that
\begin{widetext}
\begin{eqnarray}
\hat{v}_{so}\Phi (x,y)|_{x\rightarrow L^{+}} &=&\frac{\hbar }{2m}\left(
k_{c}+k_{so}\right) \left(
\begin{array}{c}
\left( \cos (\theta _{d}/2)e^{-i\theta _{so}/2}+e^{i\theta _{so}/2}\right)
\\
\xi \left( \cos (\theta _{d}/2)e^{-i\theta _{so}/2}-e^{i\theta
_{so}/2}\right)%
\end{array}%
\right) e^{i\left( k_{c}+k_{so}\right) L}\phi _{a}(y)+  \notag \\
&&\frac{\hbar }{2m}\left( k_{c}-k_{so}\right) \left(
\begin{array}{c}
-i\sin (\theta _{d}/2)e^{i\theta _{so}/2} \\
-i\xi \sin (\theta _{d}/2)e^{i\theta _{so}/2}%
\end{array}%
\right) e^{i\left( k_{c}-k_{so}\right) L}\phi _{b}(y).  \label{A7}
\end{eqnarray}
\end{widetext}Again, assuming $\Delta \ll k_{c}\sim k_{F}$ we can drop the
terms proportional to $\Delta /\left( k_{c}\pm k_{so}\right) $ in Eq. (\ref%
{A6}) thus arriving at the desired equality $\hat{v}_{so}\Psi
(x,y)|_{x\rightarrow L^{-}}=\hat{v}_{so}\Phi (x,y)|_{x\rightarrow L^{+}}$
which assures current conservation.

\section{Transport through quantum wire at $|\protect\alpha|=\protect\beta$}\label{app-john}

As an example of a backscattering mechanism in quasi-1D channels,
we calculate here the transmission and reflection coefficients for
a quantum wire of length $a$ in the presence of spin-orbit
coupling of both the Rashba and Dresselhaus types with equal
strengths $|\alpha |=\beta $. The wire is attached to leads which
are modeled as semi-infinite quantum wires without
spin-orbit interaction. Again we choose the axis of the wire along the $x$%
-direction. The central region $0\leq x\leq a$ is characterized by
an effective mass $m_{2}$ and a confining potential $V_{2}(y)$,
whereas the
leads ($x<0$ and $x>a$) have effective mass $m_{1}$ and confining potential $%
V_{1}(y)$, in general $m_{1}\neq m_{2}$, $V_{1}\neq V_{2}$. For
definiteness let us consider the case $\alpha =-\beta $. Here the
spin is independent of the momentum along the wire thus being a
good quantum number. The problem of matching the wave functions at
the interfaces separates then into these two
spin directions. Without loss of generality, we will concentrate on the $+$%
-state as defined in Eq.~(\ref{eigenstate}). Moreover, to analyze
ballistic transport through this arrangement we will use the
approximation that the reflected and transmitted electrons have
the same subband index as the incoming ones. Thus, the incoming
and reflected parts of the wave function at $x<0$ read
\begin{equation}
\psi (x,y)=\left( \lambda e^{ikx}+Ae^{-ikx}\right) \phi
_{n}^{1}(y)\,,
\end{equation}%
and for the transmitted part at $x>a$ we have
\begin{equation}
\psi (x,y)=Ce^{ikx}\phi _{n}^{1}(y)\,.
\end{equation}%
Here $k$ is the wave vector of the incoming particle along the
wire axis
with energy $\varepsilon =\varepsilon _{n}^{1}+(\hbar k)^{2}/2m_{1}$, where $%
\varepsilon _{n}^{1}$ is the subband energy of the transverse wave function $%
\varphi _{n}^{1}(y)$ according to the potential $V_{1}(y)$. In the
above equations, $\lambda $ is the amplitude of the incoming wave,
and $A$ and $C$ are the amplitudes of tne reflected and
transmitted wave, respectively. Note that the above form of the
transmitted and reflected part of the wave function are a
restricted ansatz containing the aforementioned approximation that
the subband index is the same as in the incoming part. For the
wave function in the region with spin-orbit coupling ($0\leq x\leq
a$) we use the general ansatz
\begin{eqnarray}
\psi (x,y) & = & \sum_{j}\Bigl[ \left(
B_{+}^{j}e^{iq_{+}^{j}x}+B_{-}^{j}e^{iq_{-}^{j}x}\right)\times \nonumber\\
 & & \qquad\qquad\phi
_{j}^{2}(y)e^{i\sqrt{2}\alpha m_{2}y}\Bigr]
\end{eqnarray}%
with
\begin{equation}
q_{\pm }^{j}=-\sqrt{2}\alpha m_{2}\pm \sqrt{\frac{2m_{2}}{\hbar
^{2}}\left( \varepsilon +\frac{2m_{2}\alpha ^{2}}{\hbar
^{2}}-\varepsilon _{j}^{2}\right) }\,,
\end{equation}%
and $\varepsilon _{j}^{2}$ is the subband energy of the transverse
wave function $\varphi _{j}^{2}$ in the absence of spin-orbit
coupling. The above ansatz contains all subbnads, and the
corresponding wave vectors $q_{\pm }^{j}$ will acquire imaginary
parts for large enough subband energies. However, since this
region of the setup is finite ($0\leq x\leq a$), the wave
functions remain normalizable.

The continuity conditions on the wave functions at the interface
$x=0$ lead to
\begin{eqnarray}
\left( \lambda +A\right) \phi _{n}^{1}(y) &=&\sum_{j}\Bigl[ \left(
B_{+}^{j}+B_{-}^{j}\right)\times \nonumber\\
 & & \qquad\phi _{j}^{2}(y)e^{i\sqrt{2}\alpha m_{2}y}%
\Bigr] \,,  \label{match1.1} \\
\frac{m_{2}}{m_{1}}k\left( \lambda -A\right) \phi _{n}^{1}(y)
&=&\sum_{j}\Bigl[\left(\eta _{j}B_{+}^{j}-\eta
_{j}B_{-}^{j}\right)\times
\nonumber\\
 & & \qquad\phi _{j}^{2}(y)e^{i\sqrt{2}\alpha m_{2}y}\Bigr]
\label{match1.2}
\end{eqnarray}%
with
\begin{equation}
\eta _{j}=\sqrt{\frac{2m_{2}}{\hbar ^{2}}\left( \varepsilon +\frac{%
2m_{2}\alpha ^{2}}{\hbar ^{2}}-\varepsilon _{j}^{2}\right) }\,,
\end{equation}%
and at $x=a$ one has
\begin{eqnarray}
Ce^{ika}\phi _{n}^{1}(y) &=&\sum_{j}\Bigl[ \left(
B_{+}^{j}e^{iq_{+}^{j}a}+B_{-}^{j}e^{iq_{-}^{j}a}\right) \times\nonumber\\
 & & \qquad\phi
_{j}^{2}(y)e^{i\sqrt{2}\alpha m_{2}y}\Bigr] \,,  \label{match1.3} \\
\frac{m_{2}}{m_{1}}kCe^{ika}\phi _{n}^{1}(y) &=&\sum_{j}\Bigl[
\left( \eta _{j}B_{+}^{j}e^{iq_{+}^{j}a}-\eta
_{j}B_{-}^{j}e^{iq_{-}^{j}a}\right)\times
\nonumber\\
 & & \qquad\phi _{j}^{2}(y)e^{i\sqrt{2}\alpha m_{2}y}\Bigr] \,.  \label{match1.4}
\end{eqnarray}%
Multiplying the above by $(\phi _{j}^{2}(y))^{\ast
}e^{-i\sqrt{2}\alpha m_{2}y}$ and integrating over the transverse
direction $y$ gives
\begin{eqnarray}
\left( \lambda +A\right) S_{jn} &=&B_{+}^{j}+B_{-}^{j}\,,
\label{match2.1}
\\
\frac{m_{2}}{m_{1}}k\left( \lambda -A\right) S_{jn} &=&\eta
_{j}B_{+}^{j}-[\eta _{j}B_{-}^{j}\,,  \label{match2.2} \\
Ce^{ika}S_{jn}
&=&B_{+}^{j}e^{iq_{+}^{j}a}+B_{-}^{j}e^{iq_{-}^{j}a}
\label{match2.3} \\
\frac{m_{2}}{m_{1}}kCe^{ika}S_{jn} &=&\eta
_{j}B_{+}^{j}e^{iq_{+}^{j}a}-\eta _{j}B_{-}^{j}e^{iq_{-}^{j}a}\,,
\label{match2.4}
\end{eqnarray}%
where we have defined the overlap integrals
\begin{equation}
S_{jn}=\int dy(\phi _{j}^{2}(y))^{\ast }e^{+i\sqrt{2}\alpha
m_{2}y}\phi _{n}^{1}(y)\,.
\end{equation}%
Eliminating the quantities $A$ and $C$ yields
\begin{eqnarray}
2\frac{m_{2}}{m_{1}}k\lambda S_{jn} &=&\left(
\frac{m_{2}}{m_{1}}k+\eta
_{j}\right) B_{+}^{j}\nonumber\\
 & & +\left( \frac{m_{2}}{m_{1}}k-\eta _{j}\right)
B_{-}^{j}\,,  \label{match3.1} \\
0 &=&\left( \frac{m_{2}}{m_{1}}k-\eta _{j}\right)
B_{+}^{j}e^{iq_{+}^{j}a}\nonumber\\
 & & +\left( \frac{m_{2}}{m_{1}}k+\eta _{j}\right)
B_{-}^{j}e^{iq_{-}^{j}a}\,,  \label{match3.2}
\end{eqnarray}%
or, solving for $B_{\pm }^{j}$,
\begin{eqnarray}
\left(
\begin{array}{c}
B_{+}^{j} \\
B_{-}^{j}%
\end{array}%
\right)  & =  & \frac{\left(
\begin{array}{rr}
\left( \frac{m_{2}}{m_{1}}k+\eta _{j}\right) e^{iq_{-}^{j}a} & -\left( \frac{%
m_{2}}{m_{1}}k-\eta _{j}\right)  \\
-\left( \frac{m_{2}}{m_{1}}k-\eta _{j}\right) e^{iq_{+}^{j}a} & \left( \frac{%
m_{2}}{m_{1}}k+\eta _{j}\right)
\end{array}%
\right) }{\left( \frac{m_{2}}{m_{1}}k+\eta _{j}\right)
^{2}e^{iq_{-}^{j}a}-\left( \frac{m_{2}}{m_{1}}k-\eta _{j}\right)
^{2}e^{iq_{+}^{j}a}} \times\nonumber\\
 & & \qquad\left(
\begin{array}{c}
2\frac{m_{2}}{m_{1}}k\lambda S_{jn} \\
0%
\end{array}%
\right) \,.  \label{match4}
\end{eqnarray}%
Moreover, multipying Eqs.~(\ref{match1.1})-(\ref{match1.4}) by
$(\phi _{n}^{1}(y))^{\ast }$ and integrating over $y$ gives
\begin{eqnarray}
\lambda +A &=&\sum_{j}\left[ \left( B_{+}^{j}+B_{-}^{j}\right) S_{jn}^{\ast }%
\right] \,,  \label{match5.1} \\
\frac{m_{2}}{m_{1}}k\left( \lambda -A\right)  &=&\sum_{j}\left[
\left( \eta _{j}B_{+}^{j}-\eta _{j}B_{-}^{j}\right) S_{jn}^{\ast
}\right] \,,
\label{match5.2} \\
Ce^{ika} &=&\sum_{j}\Bigl[ \left(
B_{+}^{j}e^{iq_{+}^{j}a}+B_{-}^{j}e^{iq_{-}^{j}a}\right)\times\nonumber\\
 & & \qquad S_{jn}^{\ast }%
\Bigr] \,  \label{match5.3} \\
\frac{m_{2}}{m_{1}}kCe^{ika} &=&\sum_{j}\Bigl[ \left( \eta
_{j}B_{+}^{j}e^{iq_{+}^{j}a}-\eta
_{j}B_{-}^{j}e^{iq_{-}^{j}a}\right)\times
\nonumber\\
  & & \qquad S_{jn}^{\ast }\Bigr] \,.  \label{match5.4}
\end{eqnarray}%
Inserting the above expressions for $B_{\pm }^{j}$ into
Eq.~(\ref{match5.1}) yields the following result for the
reflection amplitude,
\begin{widetext}
\begin{equation}
\frac{A}{\lambda }=-1+\sum_{j}\frac{\left( \left(
\frac{m_{2}}{m_{1}}k+\eta _{j}\right) e^{-i\eta _{j}a}-\left(
\frac{m_{2}}{m_{1}}k-\eta _{j}\right)
e^{i\eta _{j}a}\right) 2\frac{m_{2}}{m_{1}}k|S_{jn}|^{2}}{\left( \frac{m_{2}%
}{m_{1}}k+\eta _{j}\right) ^{2}e^{-i\eta _{j}a}-\left( \frac{m_{2}}{m_{1}}%
k-\eta _{j}\right) ^{2}e^{i\eta _{j}a}}\,,  \label{match6.1}
\end{equation}
\end{widetext}
and from Eq.~(\ref{match5.3}) one finds for the transmission
amplitude
\begin{equation}
\frac{C}{\lambda }=\sum_{j}\frac{e^{-ika-i\sqrt{2}\alpha m_{2}a}4\eta _{j}%
\frac{m_{2}}{m_{1}}k|S_{jn}|^{2}}{\left( \frac{m_{2}}{m_{1}}k+\eta
_{j}\right) ^{2}e^{-i\eta _{j}a}-\left( \frac{m_{2}}{m_{1}}k-\eta
_{j}\right) ^{2}e^{i\eta _{j}a}}\,.  \label{match6.2}
\end{equation}%
It is worthwhile to note that using
Eqs.~(\ref{match5.2}),(\ref{match5.4}) instead of
Eqs.~(\ref{match5.1}),(\ref{match5.3}) leads to an identical
expression for the transmission amplitude and to an equivalent
result for the reflection amplitude,
\begin{widetext}
\begin{equation}
\frac{A}{\lambda }=1-\sum_{j}\frac{\left( \left(
\frac{m_{2}}{m_{1}}k+\eta _{j}\right) e^{-i\eta _{j}a}+\left(
\frac{m_{2}}{m_{1}}k-\eta _{j}\right)
e^{i\eta _{j}a}\right) 2\eta _{j}|S_{jn}|^{2}}{\left( \frac{m_{2}}{m_{1}}%
k+\eta _{j}\right) ^{2}e^{-i\eta _{j}a}-\left(
\frac{m_{2}}{m_{1}}k-\eta _{j}\right) ^{2}e^{i\eta _{j}a}}\,.
\label{match6.3}
\end{equation}
\end{widetext}
This is indeed the same as Eq.~(\ref{match6.1}) as one can see using $%
\sum_{j}|S_{jn}|^{2}=1$. Thus,
Eqs.~(\ref{match2.1})-(\ref{match2.4}) (from
which (\ref{match4}) was obtained) are consistent with Eqs.~(\ref{match5.1}%
)-(\ref{match5.4}). This is a nontrivial finding since our
original ansatz for the wave function was a restricted one
containing an approximation, and it is not a priori clear that
such an ansatz would lead to a consistent system of equations.

\section{Heuristic picture of the noise in a beam-splitter \label{app-heur}}

One can obtain a simple physical picture of the expressions for the current
noise by considering the fluctuations of the number operator in the outgoing
leads when one injects a pair of particles in leads 1 and 2 (i.e., $\langle
n_{1}\rangle =\langle n_{2}\rangle =1$). For instance,
\begin{equation*}
S_{33}\left( \frac{e^{2}}{h\nu }\right) ^{-1}\sim \,\langle \Delta
n_{3}^{2}\rangle =\langle n_{3}^{2}\rangle -\langle n_{3}\rangle ^{2},
\end{equation*}%
\begin{equation*}
S_{34}\left( \frac{e^{2}}{h\nu }\right) ^{-1}\sim \,\,\langle \Delta
n_{3}\Delta n_{4}\rangle =\langle n_{3}n_{4}\rangle -\langle n_{3}\rangle
\langle n_{4}\rangle .
\end{equation*}%
By considering classical, Fermi or Bose particles we can derive formulas
which are in direct correspondence with the results for the current
fluctations for electron pairs with spin previously obtained \ in Sec. III
using the rigorous scattering formalism. For clarity we discuss separately
the different configurations, illustrated in Fig. \ref{figback}.
We first consider a symmetric beam-splitter
with backscattering and no local \emph{s-o} effect in lead 1; see Fig. \ref{figback}(b).
We then consider the case of backscattering in the \emph{s-o} lead 1 only, shown in
Fig. \ref{figback}(a).
Finally, we return to the \emph{s-o} rotation in a beam-splitter with no
backscattering.

\subsection{Backscattering at the beam-splitter \label{app-back-bs}}

In realistic experiments\cite{liu} the beam-splitter is not perfect, and can
actually contain a significant amount of backscattering in all input leads
(i.e., $s_{11},s_{22}\neq 0$),  as well as {}\textquotedblleft
cross-backscattering\textquotedblright\ between the input leads ($s_{12}\neq
0$).
The most symmetric scattering
matrix corresponding to this situation is%
\begin{equation*}
\mathbf{s}=\left(
\begin{array}{cccc}
a & c & r & t \\
c & a & t & r \\
r & t & a & c \\
t & r & c & a%
\end{array}%
\right) ,
\end{equation*}%
where all backscattering amplitudes are the same $%
s_{11}=s_{22}=s_{33}=s_{44}=a$ and so are the cross-backscattering
amplitudes $s_{12}=s_{34}=c$. Defining $A=|a|^{2},C=|c|^{2},R=|r|^{2}$, and $%
T=|t|^{2}$, the unitarity of $\mathbf{s}$ imposes $A+C+R+T=1$ and $c=-ar/t$,
so that $a$ and $c$ are not independent. To have independent $a$ and $c$,
one must drop the requirment of symmetry between input and ouput, as is
considered in section \ref{back-body} and \ref{app-back-one} for the case of
backscattering in the \emph{s-o} lead only. We now calculate the expectation
values of the number operators in lead 3 and 4, by simply considering the
probabilities for the different scattering configurations. Let $P(3)$, $%
P(3,3)$, and $P(3,4)$ denote the probabilities of finding one and two
electrons in lead 3, and one electron in each lead 3 and 4, respectively. We
need to determine $\langle n_{3}\rangle =P(3)+2P(3,3)$, $\langle
n_{3}^{2}\rangle =P(3)+4P(3,3)=\langle n_{3}\rangle +2P(3,3)$, and $\langle
n_{3}n_{4}\rangle =P(3,4)$.

For classical particles, we have $\langle n_{3}\rangle
_{C}=|s_{13}|^{2}(|s_{21}|^{2}+|s_{22}|^{2}+|s_{24}|^{2})+
|s_{23}|^{2}(|s_{11}|^{2}+|s_{12}|^{2}+|s_{14}|^{2})+2|s_{13}|^{2}|s_{23}|^{2}=R+T
$, $\langle n_{3}^{2}\rangle _{C}=R+T+2RT,$ and $\langle n_{3}n_{4}\rangle
_{C}=|s_{13}|^{2}|s_{24}|^{2}+|s_{14}|^{2}|s_{23}|^{2}=R^{2}+T^{2}$. We find
the auto- and cross-correlations%
\begin{eqnarray}
\langle \Delta n_{3}^{2}\rangle _{C} &=&T(1-T)+R(1-R)  \notag \\
&=&(A+C)(R+T)+2RT,  \label{nC}
\end{eqnarray}%
\begin{equation}
\langle \Delta n_{3}\Delta n_{4}\rangle _{C}=-2RT.  \label{1}
\end{equation}%
As we shall see in App. \ref{app-back-one} below, these results correspond to the
current noise $S_{33}$ and $S_{34}$ for electrons behaving classically
(i.e., with different energies and/or with opposite spins). The result (%
\ref{nC}) consists of the sum of two terms corresponding to the partition
noise for electrons coming from lead 2 and 1, respectively. One can simply
add these contributions because the classical particles are independent. In
the second equality we can recognize a partition noise term $(A+C)(R+T)$ in
addition to the usual {}\textquotedblleft beam-splitter
noise\textquotedblright\ $\sim 2RT$.

The situation is different for quantum particles obeying Fermi or Bose
statistics, in which case one must first add or substract the amplitudes
before building the probabilities for indistiguishable events. One can
satisfy unitarity by choosing, for convenience, $\Re (r^{\ast }t)=\Re
(a^{\ast }c)=0$, which yields $|r^{2}\pm t^{2}|=R\mp T$ and $|a^{2}\pm
c^{2}|=A\mp C$. For spinless fermions, one has $P(3,3)=0$ and $\langle
n_{3}\rangle
_{F}=|s_{13}s_{24}-s_{14}s_{23}|^{2}+|s_{13}s_{21}-s_{11}s_{23}|^{2}+|s_{13}s_{22}-s_{12}s_{23}|^{2}=R+T
$, $\langle n_{3}^{2}\rangle _{F}=\langle n_{3}\rangle _{F}=R+T$, and $%
\langle n_{3}n_{4}\rangle _{F}=|s_{13}s_{24}-s_{14}s_{23}|^{2}=(R+T)^{2}$.
The correlations read%
\begin{eqnarray*}
\langle \Delta n_{3}^{2}\rangle _{F} &=&T(1-T)+R(1-R)-2RT \\
&=&(A+C)(R+T),
\end{eqnarray*}%
\begin{equation*}
\langle \Delta n_{3}\Delta n_{4}\rangle _{F}=0.
\end{equation*}%
For the autocorrelation $\langle \Delta n_{3}^{2}\rangle _{F}$, we see that
the zero value found in the absence of backscattering becomes finite, which
is a consequence of the partition noise created by the additionnal
backscattering channels. One can obtain this result by substracting from the
classical result the forbidden case with two electrons in lead 3, $\langle
\Delta n_{3}^{2}\rangle _{F}=\langle \Delta n_{3}^{2}\rangle _{C}-2P(3,3)$.
The cross-correlations, on the other hand, remains zero. These results for
fermionic particles correspond to electrons with equal energies in a triplet
state.

For spinless bosons, we must double $P(3,3)$ thus obtaining  $\langle
n_{3}\rangle _{B}=R+T$, $\langle n_{3}^{2}\rangle _{B}=R+T+4RT$, and $%
\langle n_{3}n_{4}\rangle _{B}=(R-T)^{2}$. The correlations are%
\begin{eqnarray}
\langle \Delta n_{3}^{2}\rangle _{B} &=&T(1-T)+R(1-R)+2RT  \notag \\
&=&(A+C)(R+T)+4RT,  \label{heur-auto-B}
\end{eqnarray}%
\begin{equation}
\langle \Delta n_{3}\Delta n_{4}\rangle _{B}=-4RT,  \label{heur-cross-B}
\end{equation}%
and correspond to electrons in a singlet pair with equal energies. We
recognize again the sum of a partition-type noise and the beam-splitter
noise $4RT$.

We can establish a connection between the above results and those of Ref. {%
\onlinecite{gl-prl2003}} in which the case with no cross-backscattering $%
s_{12}=0$ was studied. In that work, the back-scattering was introduced in
the same way as in Section \ref{back-body}, namely, by taking the
probabilities $R$ and $T$ from the original beam-splitter and adding a
tunnel barrier with reflection probability $R_{B}$. Since the
cross-backscattering does not play a direct role for the noise in the output
leads, one can identify $A+C\rightarrow R_{B}$, $R\rightarrow R(1-R_{B})$,
and $T\rightarrow T(1-R_{B})$, which establishes the equivalence of Eq. (\ref%
{heur-auto-B}),(\ref{heur-cross-B}) and Eqs. (5),(6) of Ref. {%
\onlinecite{gl-prl2003}}.

\subsection{Backscattering in one lead \label{app-back-one}}

We consider here the case of backscattering in lead 1 only, which was
discussed in section \ref{back-body}. The problem is no longer symmetric and
the transmission probabilities $T$ and $R$ are not equivalent anymore.
See Appendix B where we calculate explicitly the transmission and
reflection coefficients for a model quantum wire with spin-orbit interaction
of both the Dresselhaus and Rahsba types.
We recall that $R$ and $T$ here are the original quantities before adding the
backscattering channel, and therefore satisfy $T=1-R$. We calculate the
fluctuations of the number operator in lead 3 $n_{3}$ by the same procedure
as in Appendix \ref{app-back-bs}.

For classical particles we find $\langle n_{3}\rangle _{C}=1-AR,$ $\langle
n_{3}^{2}\rangle _{C}=1-AR+2TR^{\prime }$, which yields%
\begin{equation}
\langle \Delta n_{3}^{2}\rangle _{C}=TR+R^{\prime }(1-R^{\prime }).
\label{numC}
\end{equation}%
This result corresponds to the current noise expressions (\ref{backS}),(\ref%
{backTe}), and (\ref{backTu}) with $\epsilon _{1}\neq \epsilon _{2}$. Hence,
in this configuration electrons with distinct energies are not affected by
the (anti-)symmetrization related to their spin state, and behave,
effectively, like classical particles.

For fermions, one finds $\langle n_{3}\rangle _{F}=\langle n_{3}^{2}\rangle
_{F}=1-AR$, which gives%
\begin{equation}
\langle \Delta n_{3}^{2}\rangle _{F}=TR+R^{\prime }(1-R^{\prime
})-2TR^{\prime }.  \label{numF}
\end{equation}%
This result corresponds to the case of triplets (\ref{backTe}), (\ref{backTu}%
) with $\epsilon _{1}=\epsilon _{2}$ and $\theta _{so}=0$.

For bosons we can proceed likewise to find $\langle n_{3}\rangle _{B}=1-AR,$
$\langle n_{3}^{2}\rangle _{B}=1-AR+4TR^{\prime }$. The result%
\begin{equation}
\langle \Delta n_{3}^{2}\rangle _{B}=TR+R^{\prime }(1-R^{\prime
})+2TR^{\prime },  \label{numB}
\end{equation}%
corresponds to the case of singlets with $\epsilon _{1}=\epsilon _{2}$ and $%
\theta _{so}=0$. This can also be found by doubling the probability to have
electrons in the same lead 3, $\langle \Delta n_{3}^{2}\rangle _{B}=\langle
\Delta n_{3}^{2}\rangle _{C}+2P(3,3).$

We can also rewrite the results above as
\begin{equation*}
\langle \Delta n_{3}^{2}\rangle _{C}=AR(1-AR)+2TR^{\prime },
\end{equation*}%
\begin{equation*}
\langle \Delta n_{3}^{2}\rangle _{F}=AR(1-AR)=\langle n_{3}\rangle
(1-\langle n_{3}\rangle ),
\end{equation*}%
\begin{equation*}
\langle \Delta n_{3}^{2}\rangle _{B}=AR(1-AR)+4TR^{\prime }.
\end{equation*}%
This shows how the backscattering induces a partition noise $\sim \langle
n_{3}\rangle (1-\langle n_{3}\rangle )$, in addition to renormalizing the
beam-splitter noise $\sim TR^{\prime }$ via $R^{\prime }=R(1-A).$

\subsection{Spin-orbit rotation}

We can incorporate some of the effects of \emph{s-o}\textbf{\ }induced
rotation within the heuristic scheme presented above. We neglect
backscattering for simplicity. Considering the result for the singlet [Eq. (%
\ref{eq53})] with $\epsilon _{1}=\epsilon _{2}$, $S_{33}^{S}=(e^{2}/h\nu
)\,2TR(1+\cos \theta )$, we see that the \emph{s-o} angle $\theta _{so}$
{}\textquotedblleft interpolates\textquotedblright\ from the bosonic
behavior ($\theta _{so}=0$) to the fermionic behavior ($\theta _{so}=\pi $).
For the angle $\theta _{so}=\pi /2$, we actually recover the same result as
in the classical case ($\epsilon _{1}\neq \epsilon _{2}$) -- although one
must emphasize that any locally-rotated state is still maximally entangled,
and therefore is not classical. In order to consider particles that
effectively behave in an intermediate way between bosons and fermions, we
consider particles that follow intermediate statistics, i.e., anyons that
acquire a finite phase $e^{i\theta }$ upon antisymmetrization
 (with $\theta _{so}=0$ for bosons
and $\theta _{so}=\pi $ for fermions). Indeed, one can easily recover the
formula%
\begin{equation}
\langle \Delta n_{3}^{2}\rangle _{A}=2TR(1+\cos \theta ),  \label{numA}
\end{equation}%
corresponding to a Rashba-rotated singlet pair with $\epsilon _{1}=\epsilon
_{2}$ [Eq. (\ref{eq53})] by the following calculation for anyons%
\begin{eqnarray}
\langle n_{3}\rangle _{A} &=&P(3)+2P(3,3)=\left\vert s_{13}s_{24}+e^{i\theta
}s_{14}s_{23}\right\vert ^{2}  \notag \\
&&+2\left\vert s_{13}s_{23}\frac{1+e^{i\theta }}{\sqrt{2}}\right\vert
^{2}=R+T=1,  \label{n3}
\end{eqnarray}%
\begin{eqnarray}
\langle n_{3}^{2}\rangle _{A} &=&P(3)+4P(3,3)=\langle n_{3}^{2}\rangle
_{A}+4\left\vert s_{13}s_{23}\frac{1+e^{i\theta }}{\sqrt{2}}\right\vert ^{2}
\notag \\
&=&1+2TR(1+\cos \theta ).  \label{n32}
\end{eqnarray}%
The factor $(1+e^{i\theta })/\sqrt{2}$ interpolates between the Pauli
exclusion principle and the bosonic bunching occuring for two particles in
the same lead 3.

\section{Evaluation of the noise matrix element \label{app-gen-form}}

Here we sketch the derivation of general formulas for the relevant matrix
elements in the noise calculation for injected electron pairs. From the
noise definition [Eq. (\ref{eq39})] it is clear that we need to evaluate
objets of the form
\begin{eqnarray}
D &=&\langle 0|a_{\mu ,\sigma _{\mu }}(\varepsilon _{\mu })a_{\nu ,\sigma
_{\nu }}(\varepsilon _{\nu })a_{\alpha ,\sigma }^{\dagger }(\varepsilon
)\times  \notag \\
&&\times a_{\beta ,\sigma ^{\prime }}(\varepsilon ^{\prime })a_{\gamma
,\sigma _{\gamma }}^{\dagger }(\varepsilon _{\gamma })a_{\eta ,\sigma _{\eta
}}^{\dagger }(\varepsilon _{\eta })|0\rangle  \label{A8}
\end{eqnarray}%
and
\begin{eqnarray}
Q &=&\langle 0|a_{\mu ,\sigma _{\mu }}(\varepsilon _{\mu })a_{\nu ,\sigma
_{\nu }}(\varepsilon _{\nu })a_{\alpha ,\sigma }^{\dagger }(\varepsilon
)a_{\beta ,\sigma ^{\prime }}(\varepsilon ^{\prime })\times  \notag \\
&&\times a_{\alpha ^{\prime },\sigma ^{\prime \prime }}^{\dagger
}(\varepsilon ^{\prime \prime })a_{\beta ^{\prime },\sigma ^{\prime \prime
\prime }}(\varepsilon ^{\prime \prime \prime })a_{\gamma ,\sigma _{\gamma
}}^{\dagger }(\varepsilon _{\gamma })a_{\eta ,\sigma _{\eta }}^{\dagger
}(\varepsilon _{\eta })|0\rangle .  \label{A9}
\end{eqnarray}%
Wick's theorem tells us that we can express the above matrix elements in
terms of all possible pairings of the fermionic operators. Here the possible
non-zero pairings involve only one creation and one destruction operator.
For the first case we have:
\begin{widetext}
\begin{eqnarray}
D &=&+\langle 0|a_{\alpha ,\sigma }^{\dagger }(\varepsilon )a_{\beta ,\sigma
^{\prime }}(\varepsilon ^{\prime })|0\rangle \langle 0|a_{\nu ,\sigma _{\nu
}}(\varepsilon _{\nu })a_{\gamma ,\sigma _{\gamma }}^{\dagger }(\varepsilon
_{\gamma })|0\rangle \langle 0|a_{\mu ,\sigma _{\mu }}(\varepsilon _{\mu
})a_{\eta ,\sigma _{\eta }}^{\dagger }(\varepsilon _{\eta })|0\rangle  \notag
\\
&&-\langle 0|a_{\alpha ,\sigma }^{\dagger }(\varepsilon )a_{\beta ,\sigma
^{\prime }}(\varepsilon ^{\prime })|0\rangle \langle 0|a_{\nu ,\sigma _{\nu
}}(\varepsilon _{\nu })a_{\eta ,\sigma _{\eta }}^{\dagger }(\varepsilon
_{\eta })|0\rangle \langle 0|a_{\mu ,\sigma _{\mu }}(\varepsilon _{\mu
})a_{\gamma ,\sigma _{\gamma }}^{\dagger }(\varepsilon _{\gamma })|0\rangle
\notag \\
&&+\langle 0|a_{\nu ,\sigma _{\nu }}(\varepsilon _{\nu })a_{\alpha ,\sigma
}^{\dagger }(\varepsilon )|0\rangle \langle 0|a_{\beta ,\sigma ^{\prime
}}(\varepsilon ^{\prime })a_{\gamma ,\sigma _{\gamma }}^{\dagger
}(\varepsilon _{\gamma })|0\rangle \langle 0|a_{\mu ,\sigma _{\mu
}}(\varepsilon _{\mu })a_{\eta ,\sigma _{\eta }}^{\dagger }(\varepsilon
_{\eta })|0\rangle  \notag \\
&&-\langle 0|a_{\nu ,\sigma _{\nu }}(\varepsilon _{\nu })a_{\alpha ,\sigma
}^{\dagger }(\varepsilon )|0\rangle \langle 0|a_{\beta ,\sigma ^{\prime
}}(\varepsilon ^{\prime })a_{\eta ,\sigma _{\eta }}^{\dagger }(\varepsilon
_{\eta })|0\rangle \langle 0|a_{\mu ,\sigma _{\mu }}(\varepsilon _{\mu
})a_{\gamma ,\sigma _{\gamma }}^{\dagger }(\varepsilon _{\gamma })|0\rangle
\notag \\
&&+\langle 0|a_{\mu ,\sigma _{\mu }}(\varepsilon _{\mu })a_{\alpha ,\sigma
}^{\dagger }(\varepsilon )|0\rangle \langle 0|a_{\nu ,\sigma _{\nu
}}(\varepsilon _{\nu })a_{\gamma ,\sigma _{\gamma }}^{\dagger }(\varepsilon
_{\gamma })|0\rangle \langle 0|a_{\beta ,\sigma _{z}^{\prime }}(\varepsilon
^{\prime })a_{\eta ,\sigma _{\eta }}^{\dagger }(\varepsilon _{\eta
})|0\rangle  \notag \\
&&-\langle 0|a_{\mu ,\sigma _{\mu }}(\varepsilon _{\mu })a_{\alpha ,\sigma
}^{\dagger }(\varepsilon )|0\rangle \langle 0|a_{\beta ,\sigma ^{\prime
}}(\varepsilon ^{\prime })a_{\gamma ,\sigma _{\gamma }}^{\dagger
}(\varepsilon _{\gamma })|0\rangle \langle 0|a_{\nu ,\sigma _{\nu
}}(\varepsilon _{\nu })a_{\eta ,\sigma _{\eta }}^{\dagger }(\varepsilon
_{\eta })|0\rangle .  \label{A10}
\end{eqnarray}
\end{widetext}The first two terms are zero since they both have a
destruction operator acting on the vacuum state. Hence, we have
\begin{eqnarray}
D &=&\delta _{\nu \alpha }\delta _{\sigma _{\nu }\sigma }\delta
_{\varepsilon _{\nu }\varepsilon }(\delta _{\beta \gamma }\delta _{\mu \eta
}\delta _{\sigma ^{\prime }\sigma _{\gamma }}\delta _{\sigma _{\mu }\sigma
_{\eta }}\delta _{\varepsilon ^{\prime }\varepsilon _{\gamma }}\delta
_{\varepsilon _{\mu }\varepsilon _{\eta }}  \notag \\
&&-\delta _{\beta \eta }\delta _{\mu \gamma }\delta _{\sigma ^{\prime
}\sigma _{\eta }}\delta _{\sigma _{\mu }\sigma _{\gamma }}\delta
_{\varepsilon ^{\prime }\varepsilon _{\eta }}\delta _{\varepsilon _{\mu
}\varepsilon _{\gamma }})+  \notag \\
&&\delta _{\mu \alpha }\delta _{\sigma _{\mu }\sigma }\delta _{\varepsilon
_{\mu }\varepsilon }(\delta _{\nu \gamma }\delta _{\beta \eta }\delta
_{\sigma _{\nu }\sigma _{\gamma }}\delta _{\sigma ^{\prime }\sigma _{\eta
}}\delta _{\varepsilon _{\nu }\varepsilon _{\gamma }}\delta _{\varepsilon
^{\prime }\varepsilon _{\eta }}  \notag \\
&&-\delta _{\beta \gamma }\delta _{\nu \eta }\delta _{\sigma ^{\prime
}\sigma _{\gamma }}\delta _{\sigma _{\nu }\sigma _{\eta }}\delta
_{\varepsilon ^{\prime }\varepsilon _{\gamma }}\delta _{\varepsilon _{\nu
}\varepsilon _{\eta }}).  \label{A11}
\end{eqnarray}%
Similarly, we find
\begin{widetext}%
\begin{eqnarray}
Q &=&+\delta _{\alpha \beta }\delta _{\alpha ^{\prime }\beta ^{\prime
}}\delta _{\sigma \sigma ^{\prime }}\delta _{\sigma ^{\prime \prime }\sigma
^{\prime \prime \prime }}\delta _{\varepsilon \varepsilon ^{\prime }}\delta
_{\varepsilon ^{\prime \prime }\varepsilon ^{\prime \prime \prime }}[\delta
_{\nu \gamma }\delta _{\mu \eta }\delta _{\sigma _{\nu }\sigma _{\gamma
}}\delta _{\sigma _{\mu }\sigma _{\eta }}\delta _{\varepsilon _{\nu
}\varepsilon _{\gamma }}\delta _{\varepsilon _{\mu }\varepsilon _{\eta
}}-\delta _{\mu \gamma }\delta _{\nu \eta }\delta _{\sigma _{\mu }\sigma
_{\gamma }}\delta _{\sigma _{\nu }\sigma _{\eta }}\delta _{\varepsilon _{\mu
}\varepsilon _{\gamma }}\delta _{\varepsilon _{\nu }\varepsilon _{\eta }}]
\notag \\
&&-\delta _{\alpha ^{\prime }\beta ^{\prime }}\delta _{\mu \eta }\delta
_{\sigma ^{\prime \prime }\sigma ^{\prime \prime \prime }}\delta _{\sigma
_{\mu }\sigma _{\eta }}\delta _{\varepsilon ^{\prime \prime }\varepsilon
^{\prime \prime \prime }}\delta _{\varepsilon _{\mu }\varepsilon _{\eta
}}[\delta _{\nu \gamma }\delta _{\beta \alpha }\delta _{\sigma _{\nu }\sigma
_{\gamma }}\delta _{\sigma ^{\prime }\sigma }\delta _{\varepsilon _{\nu
}\varepsilon _{\gamma }}\delta _{\varepsilon ^{\prime }\varepsilon }-\delta
_{\nu \alpha }\delta _{\beta \gamma }\delta _{\sigma _{\nu }\sigma }\delta
_{\sigma ^{\prime }\sigma _{\gamma }}\delta _{\varepsilon _{\nu }\varepsilon
}\delta _{\varepsilon ^{\prime }\varepsilon _{\gamma }}]  \notag \\
&&+\delta _{\alpha ^{\prime }\beta ^{\prime }}\delta _{\mu \gamma }\delta
_{\sigma ^{\prime \prime }\sigma ^{\prime \prime \prime }}\delta _{\sigma
_{\mu }\sigma _{\gamma }}\delta _{\varepsilon ^{\prime \prime }\varepsilon
^{\prime \prime \prime }}\delta _{\varepsilon _{\mu }\varepsilon _{\gamma
}}[\delta _{\nu \eta }\delta _{\beta \alpha }\delta _{\sigma _{\nu }\sigma
_{\eta }}\delta _{\sigma ^{\prime }\sigma }\delta _{\varepsilon _{\nu
}\varepsilon _{\eta }}\delta _{\varepsilon ^{\prime }\varepsilon }-\delta
_{\nu \alpha }\delta _{\beta \eta }\delta _{\sigma _{\nu }\sigma }\delta
_{\sigma ^{\prime }\sigma _{\eta }}\delta _{\varepsilon _{\nu }\varepsilon
}\delta _{\varepsilon ^{\prime }\varepsilon _{\eta }}]  \notag \\
&&-\delta _{\alpha ^{\prime }\beta ^{\prime }}\delta _{\mu \alpha }\delta
_{\sigma ^{\prime \prime }\sigma ^{\prime \prime \prime }}\delta _{\sigma
_{\mu }\sigma }\delta _{\varepsilon ^{\prime \prime }\varepsilon ^{\prime
\prime \prime }}\delta _{\varepsilon _{\mu }\varepsilon }[\delta _{\nu \eta
}\delta _{\beta \gamma }\delta _{\sigma _{\nu }\sigma _{\eta }}\delta
_{\sigma ^{\prime }\sigma _{\gamma }}\delta _{\varepsilon _{\nu }\varepsilon
_{\eta }}\delta _{\varepsilon ^{\prime }\varepsilon _{\gamma }}-\delta _{\nu
\gamma }\delta _{\beta \eta }\delta _{\sigma _{\nu }\sigma _{\gamma }}\delta
_{\sigma ^{\prime }\sigma _{\eta }}\delta _{\varepsilon _{\nu }\varepsilon
_{\gamma }}\delta _{\varepsilon ^{\prime }\varepsilon _{\eta }}]  \notag \\
&&-\delta _{\nu \alpha }\delta _{\mu \eta }\delta _{\sigma _{\nu }\sigma
}\delta _{\sigma _{\mu }\sigma _{\eta }}\delta _{\varepsilon _{\nu
}\varepsilon }\delta _{\varepsilon _{\mu }\varepsilon _{\eta }}[\delta
_{\beta \gamma }\delta _{\beta ^{\prime }\alpha ^{\prime }}\delta _{\sigma
^{\prime }\sigma _{\gamma }}\delta _{\sigma ^{\prime \prime \prime }\sigma
^{\prime \prime }}\delta _{\varepsilon ^{\prime }\varepsilon _{\gamma
}}\delta _{\varepsilon ^{\prime \prime \prime }\varepsilon ^{\prime \prime
}}-\delta _{\beta \alpha ^{\prime }}\delta _{\beta ^{\prime }\gamma }\delta
_{\sigma ^{\prime }\sigma ^{\prime \prime }}\delta _{\sigma ^{\prime \prime
\prime }\sigma _{\gamma }}\delta _{\varepsilon ^{\prime }\varepsilon
^{\prime \prime }}\delta _{\varepsilon ^{\prime \prime \prime }\varepsilon
_{\gamma }}]  \notag \\
&&+\delta _{\nu \alpha }\delta _{\mu \gamma }\delta _{\sigma _{\nu }\sigma
}\delta _{\sigma _{\mu }\sigma _{\gamma }}\delta _{\varepsilon _{\nu
}\varepsilon }\delta _{\varepsilon _{\mu }\varepsilon _{\gamma }}[\delta
_{\beta \eta }\delta _{\beta ^{\prime }\alpha ^{\prime }}\delta _{\sigma
^{\prime }\sigma _{\eta }}\delta _{\sigma ^{\prime \prime \prime }\sigma
^{\prime \prime }}\delta _{\varepsilon ^{\prime }\varepsilon _{\eta }}\delta
_{\varepsilon ^{\prime \prime \prime }\varepsilon ^{\prime \prime }}-\delta
_{\beta \alpha ^{\prime }}\delta _{\sigma ^{\prime }\sigma ^{\prime \prime
}}\delta _{\varepsilon ^{\prime }\varepsilon ^{\prime \prime }}\delta
_{\beta ^{\prime }\eta }\delta _{\sigma ^{\prime \prime \prime }\sigma
_{\eta }}\delta _{\varepsilon ^{\prime \prime \prime }\varepsilon _{\eta }}]
\notag \\
&&-\delta _{\nu \alpha }\delta _{\mu \alpha ^{\prime }}\delta _{\sigma _{\nu
}\sigma }\delta _{\sigma _{\mu }\sigma ^{\prime \prime }}\delta
_{\varepsilon _{\nu }\varepsilon }\delta _{\varepsilon _{\mu }\varepsilon
^{\prime \prime }}[\delta _{\beta \eta }\delta _{\beta ^{\prime }\gamma
}\delta _{\sigma ^{\prime }\sigma _{\eta }}\delta _{\sigma ^{\prime \prime
\prime }\sigma _{\gamma }}\delta _{\varepsilon ^{\prime }\varepsilon _{\eta
}}\delta _{\varepsilon ^{\prime \prime \prime }\varepsilon _{\gamma
}}-\delta _{\beta \gamma }\delta _{\beta ^{\prime }\eta }\delta _{\sigma
^{\prime }\sigma _{\gamma }}\delta _{\sigma ^{\prime \prime \prime }\sigma
_{\eta }}\delta _{\varepsilon ^{\prime }\varepsilon _{\gamma }}\delta
_{\varepsilon ^{\prime \prime \prime }\varepsilon _{\eta }}]  \notag \\
&&+\delta _{\mu \alpha }\delta _{\nu \eta }\delta _{\sigma _{\mu }\sigma
}\delta _{\sigma _{\nu }\sigma _{\eta }}\delta _{\varepsilon _{\mu
}\varepsilon }\delta _{\varepsilon _{\nu }\varepsilon _{\eta }}[\delta
_{\beta \gamma }\delta _{\sigma ^{\prime }\sigma _{\gamma }}\delta
_{\varepsilon ^{\prime }\varepsilon _{\gamma }}\delta _{\beta ^{\prime
}\alpha ^{\prime }}\delta _{\sigma ^{\prime \prime \prime }\sigma ^{\prime
\prime }}\delta _{\varepsilon ^{\prime \prime \prime }\varepsilon ^{\prime
\prime }}-\delta _{\beta \alpha ^{\prime }}\delta _{\sigma ^{\prime }\sigma
^{\prime \prime }}\delta _{\varepsilon ^{\prime }\varepsilon ^{\prime \prime
}}\delta _{\beta ^{\prime }\gamma }\delta _{\sigma ^{\prime \prime \prime
}\sigma _{\gamma }}\delta _{\varepsilon ^{\prime \prime \prime }\varepsilon
_{\gamma }}]  \notag \\
&&-\delta _{\mu \alpha }\delta _{\nu \gamma }\delta _{\sigma _{\mu }\sigma
}\delta _{\sigma _{\nu }\sigma _{\gamma }}\delta _{\varepsilon _{\mu
}\varepsilon }\delta _{\varepsilon _{\nu }\varepsilon _{\gamma }}[\delta
_{\beta \eta }\delta _{\sigma ^{\prime }\sigma _{\eta }}\delta _{\varepsilon
^{\prime }\varepsilon _{\eta }}\delta _{\beta ^{\prime }\alpha ^{\prime
}}\delta _{\sigma ^{\prime \prime \prime }\sigma ^{\prime \prime }}\delta
_{\varepsilon ^{\prime \prime \prime }\varepsilon ^{\prime \prime }}-\delta
_{\beta \alpha ^{\prime }}\delta _{\sigma ^{\prime }\sigma ^{\prime \prime
}}\delta _{\varepsilon ^{\prime }\varepsilon ^{\prime \prime }}\delta
_{\beta ^{\prime }\eta }\delta _{\sigma ^{\prime \prime \prime }\sigma
_{\eta }}\delta _{\varepsilon ^{\prime \prime \prime }\varepsilon _{\eta }}]
\notag \\
&&+\delta _{\mu \alpha }\delta _{\nu \alpha ^{\prime }}\delta _{\sigma _{\mu
}\sigma }\delta _{\sigma _{\nu }\sigma ^{\prime \prime }}\delta
_{\varepsilon _{\mu }\varepsilon }\delta _{\varepsilon _{\nu }\varepsilon
^{\prime \prime }}[\delta _{\beta \eta }\delta _{\sigma ^{\prime }\sigma
_{\eta }}\delta _{\varepsilon ^{\prime }\varepsilon _{\eta }}\delta _{\beta
^{\prime }\gamma }\delta _{\sigma ^{\prime \prime \prime }\sigma _{\gamma
}}\delta _{\varepsilon ^{\prime \prime \prime }\varepsilon _{\gamma
}}-\delta _{\beta \gamma }\delta _{\sigma ^{\prime }\sigma _{\gamma }}\delta
_{\varepsilon ^{\prime }\varepsilon _{\gamma }}\delta _{\beta ^{\prime }\eta
}\delta _{\sigma ^{\prime \prime \prime }\sigma _{\eta }}\delta
_{\varepsilon ^{\prime \prime \prime }\varepsilon _{\eta }}].  \label{A12}
\end{eqnarray}%
\end{widetext}With the help of Eqs. (\ref{A11}) and (\ref{A12}) we can
systematically determine all the relevant matrix elements appearing in the
noise calculation for a particular type of injected electron pair (singlet,
triplets, Bell states, etc.).

\section{Discrete sums for $h(\varepsilon_1,\varepsilon_2)$}
\label{sums}

We choose $\varepsilon_1$ as our reference energy and define
$\Delta=\varepsilon_2-\varepsilon_1$. We further assume an
equidistant discrete single-particle spectrum of the leads,
$\varepsilon = \varepsilon_1 + (q + n) \delta$, where $n$ is an
integer, $q$ a real number between $0$ and $1$, and $\delta$ is
the level spacing of the leads, see Fig.~\ref{discrete}. We obtain
$h(q) = g_0(q-\frac{\Delta}{\delta}) g_0(q) h_0(q)$ with the
discrete sums
\begin{widetext}
\begin{eqnarray}
  h_0(q) &=& \sum_{n=-\infty}^\infty \frac{1}{(n\delta + q\delta -\Delta +i\gamma)
                                        (n\delta + q\delta -i\gamma)}
   = \frac{-2\pi}{\delta(\Delta -2i\gamma)}\frac{\sin \pi(q+i\frac{\gamma}{\delta})
    \left(\cos \pi (q-3i\frac{\gamma}{\delta})
      -\cos \pi(q+i\frac{\gamma}{\delta}-2\frac{\Delta}{\delta})\right)}
  {(\cosh 2\pi\frac{\gamma}{\delta}-\cos 2\pi q)
    (\cosh 2\pi\frac{\gamma}{\delta}-\cos 2\pi(q-\frac{\Delta}{\delta}))},\nonumber\\
\label{E7a}\\
  \frac{1}{g_0(q)^2} &=& \sum_{n=-\infty}^\infty \frac{1}{(n\delta+q\delta)^2+\gamma^2}
   = \frac{\pi}{\gamma\delta} \frac{\sinh 2\pi
   \frac{\gamma}{\delta}}{\cosh  2\pi\frac{\gamma}{\delta} - \cos 2\pi q}.\label{E7b}
\end{eqnarray}
\end{widetext}
Taking the modulus squared of the complex function $h$, we obtain
$|h(q)|^2=g_0(q)^2g_0(q-\Delta/\delta)^2|h_0(q)|^2=A B(q)$, with a
manifestly $q$-independent factor
\begin{equation}
  \label{E8}
  A = \frac{\gamma^2}{(\Delta/2)^2+\gamma^2}
          \frac{1}{2\sinh^2 2\pi\gamma/\delta},
\end{equation}
and an apparently $q$-dependent factor
\begin{equation}
  \label{E9}
  B(q) = \frac{|\cos\pi(q-3i\gamma/\delta)-\cos\pi(q-2\Delta/
  \delta+i\gamma/\delta)|^2}{\cosh 2\pi\gamma/\delta -\cos 2\pi(q-\Delta/\delta)}.
\end{equation}
Inspection of Eq.~(\ref{E9}) shows that the dependence of $B(q)$
on the offset $q$ drops out completely, and we are left with
$B\equiv B(q)=\cosh(4\pi\gamma/\delta)-\cos(2\pi\Delta/\delta) =
\cosh^2(2\pi\gamma/\delta)-\cos^2(\pi\Delta/\delta)$, where we
have used $\cos^2x=(1+\cos x)/2$ and $\cosh^2x=(1+\cosh x)/2$.
Combining $A$ and $B$, we finally obtain Eq.~(\ref{E10}). We note
that the cancellation of $q$ from the interference function
$|h|^2$ is a consequence of our normalization condition, i.e.,
that electrons are injected with unit probability.


\begin{thebibliography}{99}
\bibitem{mismatch} The band offset due to the gate-induced Rashba term
arises from the energy shift $\epsilon _{R}\equiv
\hbar^{2}k_{R}^{2}/2m$ and is very small since $\epsilon _{R}\ll
\varepsilon _{F}$. We can estimate the transmission amplitude from
$t=2\left( 1+\epsilon _{R}/\varepsilon _{F}\right)
^{1/4}/[1+(1+\epsilon _{R}/\varepsilon _{F})^{1/2}]$ (see Refs. [
\onlinecite{molen,larsen}]). Typically \cite{nitta} $\epsilon
_{R}/\varepsilon _{F}=0.1$,
 which yields $|t|^{2}=\allowbreak 0.99943.$
\bibitem{delay} It is crucial that subsequent pairs of entangled
electrons are well separated in time and space. This requires that
the \textquotedblleft delay time\textquotedblright\ $\tau _{d}$
between two electrons within an entangled pair an electron pair is
negligible compared to the \textquotedblleft transit
time\textquotedblright\ $\tau _{t}$ for an electron to cross the
beam-splitter structure. For typical parameters we find
\cite{review} $\tau _{d}\sim 0.6$ ps  $\ll \tau _{t}\sim 10-100$
ps for the entanglers of Refs. \onlinecite{recher-suk-loss} and
\onlinecite{saraga-loss}.
\end{thebibliography}
\end{document}